\documentclass[preprint]{aastex}
\usepackage{natbib,subfig}
\usepackage{subfig}

\newcommand{\sgra}{Sgr A$\textrm{*}$}

\newcommand   {\mic}   {\mbox{$\mu$m}}

\renewcommand {\deg}   {\mbox{$^\circ$}}

\newcommand   {\kms}   {\mbox{km\,s$^{-1}$}}
\renewcommand {\ga}    {\mbox{\rlap{\hbox{\lower5pt\hbox{$\sim$}}}\hbox{$>$}}}
\renewcommand {\la}    {\mbox{\rlap{\hbox{\lower5pt\hbox{$\sim$}}}\hbox{$<$}}}

\begin{document}
\pagenumbering{arabic} 
\def\kms {\hbox{km{\hskip0.1em}s$^{-1}$}} 

\def\msol{\hbox{$\hbox{M}_\odot$}}
\def\lsol{\hbox{$\hbox{L}_\odot$}}
\def\kms{km s$^{-1}$}
\def\Blos{B$_{\rm los}$}
\def\etal   {{\it et al.}}                     
\def\psec           {$.\negthinspace^{s}$}
\def\pasec          {$.\negthinspace^{\prime\prime}$}
\def\pdeg           {$.\kern-.25em ^{^\circ}$}
\def\degree{\ifmmode{^\circ} \else{$^\circ$}\fi}
\def\ut #1 #2 { \, \textrm{#1}^{#2}} 
\def\u #1 { \, \textrm{#1}}          
\def\nH {n_\mathrm{H}}

\def\ddeg   {\hbox{$.\!\!^\circ$}}              
\def\deg    {$^{\circ}$}                        
\def\le     {$\leq$}                            
\def\sec    {$^{\rm s}$}                        
\def\msol   {\hbox{$M_\odot$}}                  
\def\i      {\hbox{\it I}}                      
\def\v      {\hbox{\it V}}                      
\def\dasec  {\hbox{$.\!\!^{\prime\prime}$}}     
\def\asec   {$^{\prime\prime}$}                 
\def\dasec  {\hbox{$.\!\!^{\prime\prime}$}}     
\def\dsec   {\hbox{$.\!\!^{\rm s}$}}            
\def\min    {$^{\rm m}$}                        
\def\hour   {$^{\rm h}$}                        
\def\amin   {$^{\prime}$}                       
\def\lsol{\, \hbox{$\hbox{L}_\odot$}}
\def\sec    {$^{\rm s}$}                        
\def\etal   {{\it et al.}}                     

\def\xbar   {\hbox{$\overline{\rm x}$}}         

\def\la{\lower.4ex\hbox{$\;\buildrel <\over{\scriptstyle\sim}\;$}}
\def\ga{\lower.4ex\hbox{$\;\buildrel >\over{\scriptstyle\sim}\;$}}
\def\refitem{\par\noindent\hangindent\parindent}
\oddsidemargin = 0pt \topmargin = 0pt \hoffset = 0mm \voffset = -17mm
\textwidth = 160mm  \textheight = 244mm
\parindent 0pt
\parskip 5pt

\slugcomment{revised}
\shorttitle{Sgr A*}
\shortauthors{}

\title{Simultaneous Multi-Wavelength Observations of Sgr A*\\ 
during 2007 April 1-11}
\author{F. Yusef-Zadeh\footnote{Dept. of Physics and Astronomy, Northwestern  University, Evanston, Il. 
60208},
H. Bushouse
\footnote{STScI, 3700 San Martin Drive, Baltimore, MD  21218},
M. Wardle
\footnote{Department of Physics and Engineering, Macquarie University, Sydney NSW 2109, Australia},
C. Heinke
\footnote{Dept. of Physics, University of Alberta, Room \#238 CEB, 11322-89
Avenue, Edmonton AB T6G 2G7, Canada},
D. A. Roberts
\footnote{Adler Planetarium and Astronomy Museum, 1300
South Lake Shore Drive, Chicago, IL 60605},
C.D. Dowell\footnote{Cal Tech, Jet
Propulsion Laboratory, Pasadena, CA 91109},
A. Brunthaler
\footnote{Max-Planck-Institut f\"ur
Radioastronomie, Auf dem Huegel 69, 53121 Bonn, Germany},
M. J. Reid
\footnote{Harvard-Smithsonian CfA, 60 Garden Street,            
Cambridge, MA 02138},
C. L. Martin
\footnote{Oberlin College, Dept. of Physics and Astronomy, Professor 110 
N.  St.,Oberlin, OH 44074},
D. P. Marrone
\footnote{National Radio
Astronomy Observatory; University of Chicago, 5640 South Ellis Avenue,
Chicago IL 60637},\,
D.  Porquet
\footnote{Observatoire astronomique de Strasbourg, Universit\'e
de Strasbourg, NRS, INSU, 11 rue de l'Universit\'e, 67000 Strasbourg, 
France},
N. Grosso$^{11}$, 
K. Dodds-Eden
\footnote{Max-Plank-Institut f\"ur Extraterrestrische Physik 1312, D-85471, 
Garching, Germany},\, 
G. C. Bower\
\footnote{Radio Astronomy Lab, 601 Campbell Hall,
University of California, Berkeley, CA 94720},\, 
H. Wiesemeyer
\footnote{Institut de RadioAstronomie Millimetrique, 300 rue de la Piscine,
Domaine Universitaire 38406 Saint Martin d'Heres, France, on leave to
IRAM Granada, Spain},\,
A.  Miyazaki
\footnote{Mizusawa VLBI Observatory, National Astronomical 
Observatory of Japan, Mizusawa, Oshu, Iwate 023-0861, Japan},\,\,
S.  Pal\ 
\footnote{School of Physics, University of Western Australia,  35 Stirling Highway,
Crawley, WA, 6009, Australia},\, 
S. Gillessen$^{12}$, 
A. Goldwurm
\footnote{Service d'Astrophysique / IRFU / DSM, CEA Saclay, Bat. 709, 91191, 
Gif-sur-Yvette Cedex, France and
AstroParticule \& Cosmologie (APC) / Universit{\'e} Paris VII / CNRS / CEA /
Observatoire de Paris Bat. Condorcet, 10, rue Alice Domon et L{\'e}onie Duquet,
75205 Paris Cedex 13, France},
G. Trap$^{18}$, and 
H. Maness$^{13}$}

\begin{abstract} 
We report the detection of variable emission from Sgr A* in almost all wavelength bands (i.e. centimeter, 
millimeter, submillimeter, near-IR and X-rays) during a multi-wavelength observing campaign. Three new 
moderate flares are detected simultaneously in both near-IR and X-ray bands. The ratio of X-ray to near-IR 
flux in the flares is consistent with inverse Compton scattering of near-IR photons by submillimeter 
emitting relativistic particles which follow scaling relations obtained from size measurements of Sgr A*. 
We also find that the flare statistics in near-IR wavelengths is consistent with the probability of flare 
emission being inversely proportional to the flux. At millimeter wavelengths, the presence of flare 
emission at 43 GHz (7mm) using VLBA with milli-arcsecond spatial resolution 
indicates the first direct evidence that 
hourly time scale flares are localized within the inner 30$\times$70 Schwarzschild radii of Sgr A*. We also 
show several cross correlation plots between near-IR, millimeter and submillimeter light curves that 
collectively demonstrate the presence of time delays between the peaks of emission up to three hours. The 
evidence for time delays at millimeter and submillimeter wavelengths  are consistent with the source of 
emission being optically 
thick initially followed by  a transition to  an  optically thin regime. In particular, there is 
an intriguing correlation between the optically thin near-IR and X-ray flare 
and optically thick radio flare at 43 GHz that occurred on 2007 April 4. This would be the first evidence 
of a radio flare emission at 43 GHz delayed with respect to the near-IR 
and X-ray flare emission. The time 
delay 
measurements support the expansion of hot self-absorbed 
synchrotron plasma blob and  weaken the hot spot model of flare emission.
In addition, a simultaneous fit to  43 and 84 GHz light curves, using an adiabatic 
expansion model  of hot plasma, appears to support a 
power law rather than a relativistic Maxwellian 
distribution of particles. 
\end{abstract}



\section{Introduction}
The black hole at the center of our own galaxy was first detected as the
radio source Sgr A* over 30 years ago \citep{balick74}.  It was
found to lie at the center of a cluster of young massive stars.
Submillimeter and far-infrared observations showed that \sgra\ is
encircled by a torus of gas approximately 10 light-years across, which
orbits with a speed of 100 \kms\ (e.g.\ Genzel \& Townes 1987).
The gravity required to hold onto this material implies a mass of
several million solar masses, although a portion of this is
contributed by the stars in the stellar cluster.  These measurements
suggested that \sgra\ could be a black hole.  More detail was
provided through studies of the light distribution of stars in the
cluster, as well as the motions of ionized and molecular gas clouds
orbiting \sgra. These measurements implied a mass of approximately
3--4 million times that of the sun \citep{genzel00,genzel87} lying within a third of a
light year of the radio source.  Recently, more precise measurements
of fast moving stars in close orbits around \sgra\ have
conclusively demonstrated that it has a mass of $\sim4\times 10^6$
\msol\  \citep{ghez05,eisenhauer05,ghez08,schoedel02,gillessen09} and that the size of
the radio source is about $\sim$4 times its Schwarzschild
radius at 230 GHz (R$_{s})$ \citep{doeleman08}.  This dark, massive object has also 
been uniquely
identified through the proper motion of the radio source, 
which show that \sgra\ must contain $>4\times10^{5}$ \msol\  \citep{reid04}.
Taken together, these measurements provide  strong
evidence that \sgra\ is a black hole with mass $\sim4\times10^6$
\msol. No other known category of astrophysical object can easily  fit so 
much mass into a sub-AU-size region.

This massive black hole is a hundred times closer to us than the next
nearest example, presenting an unparalleled opportunity to closely  
study the process by which gas is captured by black holes.  It is   
therefore the subject of intense scrutiny.
The energy radiated by \sgra\ is thought to be liberated from gas that
is falling into the black hole after being captured from the powerful
winds of members of its neighboring cluster of massive stars (e.g., 
Melia 1992).  The broad band spectrum of Sgr A* peaks at
submillimeter wavelengths \citep{zylka92,falcke98};
this is thought to be the dividing line between optically thick and
optically thin  emission at low and high frequencies,  respectively.  The
bolometric luminosity of Sgr A* $\sim 100$ \lsol\  is several orders of magnitudes below
that predicted  given its expected rate of capture of material from 
stellar winds, prompting a number of theoretical models to explain its
very low efficiency 
\citep{narayan95,liu01,yuan03,goldston05,liu04,falcke09}.

Now that the quiescent spectrum of emission from \sgra\ has been
characterized from radio to X-rays, attention has turned to
variability of emission in multiple wavelengths. 
 These measurements probe the structure and the 
physical parameters of the hot plasma in the vicinity of the black hole by measuring the 
time variations of its flux in different wavelength bands as well as their cross-correlation 
with each other.
Flaring activity on $<1-4$ hour time scale is seen in all wavelength bands
in which quiescent  emission has been detected.

Flaring X-ray emission from \sgra\ has been detected and has been argued to
originate within a few 
Schwarzschild radii of the $\sim4\times10^6$
\msol\ black hole \citep{baganoff03,goldwurm03,porquet03,belanger05}.
At near-IR (NIR) wavelengths  \citep{genzel03,yuan03,ghez04,hornstein07},
flare emission
from \sgra\ is shown to be due to optically thin  synchrotron emission, whereas the 
long-wavelength flaring activity in  
submillimeter, millimeter  and centimeter bands is due to optically thick 
synchrotron emission.
The exact  frequency at which the transition from 
optically thick to thin flare emission occurs is unknown.

A variety of mechanisms have been proposed to explain the origin of
the variability of \sgra.  Many of these models have considered
different energy distributions for the relativistic particles to
explain the origin of submillimeter emission
 \citep{markoff01,yuan02,melia02,liu02,yuan03,nayakshin03,
eckart04,eckart06a,eckart06b,zadeh06a,gillessen06,goldston05,liu06,falcke09};
Melia and Falcke (2001 and references therein).  The
direction that has been taken in the past in interpreting the flaring
activity of \sgra\ is within one of the established paradigms for the
accretion flow that have been developed based on the time-averaged
emission -- for example a thin accretion disk, a disk and jet,
outflow, an advection-dominated accretion flow, radiatively
inefficient accretion flow, accretion disk inflow/outflow solutions
\citep{melia92,yuan03,falcke00,falcke09,narayan98,blandford99} and
then the predicted spectrum is compared with the observed spectrum.

We have recently analyzed the NIR flaring of \sgra, which is produced
by synchrotron emission from a transient population of particles
produced within $\sim 10$ Schwarzschild radii of the massive black
hole \citep{genzel03,eckart06a,gillessen06}.  We argued that the $\sim
2-3$ hour duration of submillimeter flares could not be due to
synchrotron cooling when observed simultaneously with a NIR flare
(estimated to be $\sim 20$ minutes and $\sim 12$ hours at 1.6$\mu$m
[188 THz] and 850$\mu$m [350 GHz], respectively).  The decline in
submillimeter light curves was interpreted to be due to adiabatic
cooling associated with expansion of the emitting plasma
\citep{zadeh06a,zadeh06b} under the assumption that the same
accelerated population of particles is responsible for NIR and
submillimeter emission.  Time delays detected between peaks of flare
emission at radio, submillimeter and NIR/X-rays wavelengths are
consistent with this picture (e.g., Yusef-Zadeh et al. 2006b; 
Marrone et al.  2008; Yusef-Zadeh
et al.  2008; Meyer et al.  2008; Eckart et al.  2008).  However, the
lack of long simultaneous coverage have not placed strong constraints
in time delay measurements, especially between radio and NIR
wavelengths.  Simple modeling of the total and polarized intensity of
the hot expanding plasma provide predictions that can be tested
observationally by carrying out observational campaigns such as the
one we coordinated during April 2007 to examine the mechanisms for the
variability, with implications on the nature of the accretion flow.
The results presented here are the third in a series of papers that
came from the multi-wavelength observing campaign  that took place on 2009 April
\citep{porquet08,dodds-eden09}.  The results of soft $\gamma$-ray
observations are given separately \citep{trap09}.

The structure of this paper is as follows.  $\S3$ presents light
curves of all the useful data that were taken in this campaign,
following the observational details described in $\S2$.  In $\S4$ we
analyze the statistical properties of flare emission and the
corresponding spectral and power spectrum distributions at NIR
wavelengths, as well as cross-correlation analysis of light curves.
We then discuss in $\S5$ the origin of X-ray emitting flares and
provide observational support for the expanding hot plasma model of
flare emission.  The polarization results will be given elsewhere.
                                                   
\section{Observations and Data Reduction}
 
The primary purpose of observations made during 2007 April 1-11 
was  the coordination of 
several telescopes operating at many wavelengths to monitor the emission from Sgr A* and 
measure the time evolution of its spectrum. 
There were  a total of 13 observatories that participated in 
this campaign, including XMM-Newton, the Hubble Space Telescope (HST),
the International Gamma-Ray Astrophysics Laboratory INTEGRAL, 
the Very Large Array (VLA) of the National Radio Astronomy 
Observatory\footnote{The National
Radio Astronomy Observatory is a facility of the National
Science Foundation, operated under a cooperative agreement by
Associated Universities, Inc.} (NRAO), the Very Long Baseline Array (VLBA$^{18}$), 
the Caltech Submillimeter Observatory (CSO), the Very Large Telescope (VLT), 
the Submillimeter Array  \citep{Blundell04}, the 30m Pico 
Veleta Telescope of  
the Institute for Millimeter Radioastronomy (IRAM), the Submillimeter Telescope 
(SMT), the Nobeyama Millimeter Array (NMA), the Combined Array for 
Research in 
Millimeter-wave Astronomy (CARMA), and the Giant Meterwave Radio Telescope (GMRT). 
The campaign was organized by first obtaining observing time with
XMM-Newton (PI: D.\ Porquet) and HST (PI: F.\ Yusef-Zadeh),
 and then coordinating the 
ground-based 
facilities to the
allotted space-based schedules.

Figure 1 shows the schedule of all observations and their  rough durations.
Details  of  XMM-Newton and   VLT/NACO 
observations have  already been reported in  Porquet et al. (2008), and 
\citep{dodds-eden09}, respectively.
Summary of the results from VLT/VISIR and INTEGRAL observations has also been given 
by Trap et al.\ (2009). 
Briefly, XMM observations were carried out using  three observations for a total of 
230 ks blocks of time
during 2007 March 30 to April 4, and the  VLT/NACO observations took place
between  2007 April 1--6 using H (1.66 $\mu$m),  K (2.12$\mu$m),  
and L' (3.8$\mu$m) bands. 
INTEGRAL observations took place in parallel to XMM-Newton on  April 1 and 4 for 
a total effective exposure time of 212 ks for IBIS/ISGRI (20--100 keV) and 
46 ks for JEM-X 1 (3--20 keV).

\subsection {HST NICMOS: 1.45\mic\ and 1.70\mic}
We obtained 40 orbits of HST observations using the NICMOS camera 1,
with the orbits distributed over seven consecutive days between
2007 April 1--7.
The observations used the NICMOS F145M and F170M filters, with exposure 
times of 144 sec in each filter and readout samplings of $\sim$16 sec 
within each exposure.
Use of this pair of filters has several advantages.
First, they have no overlap in wavelength space and are therefore
suitable for spectral index measurements.
Second, they are well matched to one another in terms of throughput,
so that we can use identical exposure and readout times, thus producing
time series data that are evenly sampled in both filters. This has
great advantages for making periodicity measurements.   
Third, their relatively high throughput also allows us to use relatively
short exposure times, so that we can cycle back and forth between the
two filters fairly rapidly. The near-simultaneous observations then
allow us to make meaningful spectral index measurements of flare events.
This is especially true for the peaks of flare events, where the flux
from \sgra\ does not change rapidly over the course of several minutes
or more. 
The spectral index distribution can not be measured accurately,
however, during the rise and fall of flare events because the overall
flux is changing more rapidly than the cadence of our filter cycling
times.

The IRAF ``apphot'' routines were used to perform aperture photometry
of sources in the NICMOS \sgra\ field, including \sgra\ itself.
For stellar sources the measurement aperture was positioned on  
each source using an automatic centroiding routine.
This approach could not be used for measuring \sgra\ because its
signal is spatially overlapped by that of the orbiting star S0-2 and S17.
Therefore the photometry aperture for \sgra\ was positioned by 
using a constant offset from the measured location of S0-2 in each image.
The offset between S0-2 and \sgra\ was derived from the orbital
parameters given by \citet{ghez03}.
The position of \sgra\ was estimated to be 0.16$''$ south
and 0.01$''$ west of S0-2 at the time of the HST observations. To confirm 
the accuracy of the position of \sgra, two images of \sgra\ taken 
before and during a flare event were aligned and subtracted, which 
resulted in an image showing the location of the flare emission. 

We used a measurement aperture with a diameter of 3 detector
pixels, which corresponds to $\sim 0.13''$. This size was
chosen as a suitable compromise between wanting to maximize the fraction
of the \sgra\ PSF included in the aperture, while at the same time 
limiting the amount of signal coming from the wings of the PSF from the 
adjacent star S0-2.
We derived aperture correction factors by making measurements of a
reasonably well isolated star in the field through a series of apertures
of increasing size. The aperture corrections, which convert the fluxes
measured through our 3-pixel diameter aperture to a semi-infinite 
aperture, are 2.91 and 3.40 for the 1.45 and 1.70 \mic\ bands, respectively.
Absolute calibration of the photometric measurements was accomplished
using the latest calibrations for the NICMOS F145M and F170M 
obtained from STScI.

De-reddened fluxes were computed using the appropriate  extinction law 
for the Galactic center \citep{moneti01} and their extinction value 
of A(K)=3.3 mag.  These translate to extinction values for the NICMOS 
filter bands of A(F170M)=5.03 mag and A(F145M)=6.52 mag, which then 
correspond to correction factors of 103 and 406, respectively.  
Because there is some contribution from the neighboring star S0-2 within
our measurement aperture, we have determined the flux of \sgra\ when it 
is flaring by subtracting the mean flux level measured during
``quiescent'' episodes. The resulting net flux can therefore be safely
attributed to \sgra\ flares.

\subsection {VLA: 43  and 22 GHz}

We used a fast-switching technique to observe \sgra\ simultaneously
using the VLA D configuration at 43.3 GHz (7mm) and 22.4 GHz (13mm)
GHz.  These observations took place on 2007 April 1--4, each lasting
for $\sim$7 hours, using 8 VLA and 18 eVLA antennas. The two IFs were
separated by 50MHz in each observation, except for those on 2007 April
4, when the two IFs were centered at non-standard frequencies of
43.1851 and 43.5351 GHz, which corresponds to a frequency separation
of 350 MHz. The separation was used to carry out polarization
measurements, the result of which will be given elsewhere.

We cycled between \sgra\ and the fast-switching calibrator 17444-31166
(2.3 degrees away from \sgra) for 90 sec and 30 sec, respectively,
throughout the observation. On 2007 April 4, we also used the
fast-switching phase calibrator 17459-28204, which is weaker, but closer
($\sim 21'$) to \sgra.  3C286 was used as the flux calibrator and
NRAO530 was observed as a polarization and additional phase
calibrator.  The light curves at 43 GHz restricted data to a {\it uv}
range greater than 90  k$\lambda$ with full width at half point of
2.45$''\times1.3''$ (PA=$-4^0$).  We used NRAO530 for pointing every
30 minutes; the bootstrapped flux of NRAO530 at 43 GHz is
2.43$\pm0.05$ Jy.  
At 22 GHz, the strong continuum emission 
from
ionized gas associated with extended features surrounding \sgra\
overwhelmed the flux of \sgra\ itself, making the variability analysis
uncertain.  Therefore the 22 GHz data are not useful and are not
presented here.  In all the measurements presented here, we used only
antennas that had constant gain curves with similar values, thus many
of the eVLA antennas were not used.

In all cases, at least two and sometimes three phase calibrators were
used in order to ensure that amplitude variability or calibration
errors of one of the calibrators would not be introduced into the
light curve of \sgra.  In the case of multiple phase calibrators, the
same calibrator used to calibrate the gains of \sgra\ was used to
cross-calibrate the other calibrators.  In cases where calibrator
light curves are shown as alongside those of Sgr A*, they were
obtained from cross-calibration using one of the other phase
calibrators and {\sl not} from self-calibration.  Additionally, as a
check, all light curves of \sgra\ made using phase calibrations from
the principle phase calibrator (usually 17444-31166) were compared
against light curves of \sgra\ made using the other phase
calibrators.  These comparisons were used to identify bad data in the
calibrators and after editing and recalibration light curves
using different calibrators were consistent.

For the final light curves of \sgra, the data were calibrated using
the principle phase calibrator (17444-31166).  A phase
self-calibration was applied to \sgra\ before the determination of a
light curve.  No amplitude self-calibration was done to \sgra\ or any
backup phase calibrators, whose light curves are shown as a reference,
since amplitude calibration would remove time variation from the light
curves.  After phase self-calibration, large images were made and
found no confusing point sources above the rms noise (typically 2.5
mJy beam$^{-1}$ in a full run at 43 GHz) when the selected  uv data
were  greater than 90k$\lambda$. 

In order to derive the light curve in the visibility plane, the 
Astronomical Image Processing System (AIPS)
task DFTPL was used.  DFTPL plots the direct Fourier transform of a
vector averaged set of measured visibilities as a function of time.
Since we use this on data that has been phase self-calibrated using
the point source \sgra, the vector average gives the flux of \sgra.
Visibilities are averaged in bins with defined time widths, however,
since the number of visibilities in each bin varies, the error
associated with the average will not be constant and are derived at
each time range.

\subsection {VLBA: 43, 22, and 15 GHz}
We observed Sgr~A* with the 
VLBA  in two different experiments. One took place 
on 2007 April 1,  5  and 11 at 43 GHz  under program
BR124. All observations employed four 8 MHz bands in dual circular 
polarization each. 
These observations  were made at 43 GHz and involved 
rapid switching between Sgr~A* and the two background continuum 
sources J1748-291 and J1745-283. Sources were changed every 15 seconds in the 
sequence Sgr~A* -- J1748-291 -- Sgr~A* -- J1745-283 -- Sgr~A*, yielding an 
on-source time of $\sim$ 10 seconds. Before, in the middle, and after each 
observation 16 quasars were observed within $\sim$ 40 minutes. NRAO\,530 was 
also observed as fringe-finder. The total observing time including the quasars 
was 8 hours for each observation. The data were correlated with 16 spectral 
channels per frequency band and an integration time of 0.131 seconds. 
The  $\sim1$ hour  gaps in the light curves of Sgr A* are due to geodetic 
measurements.

In the second experiment (proposal code BB230),  the observations on 2007 April 2 
and 10 involved rapid 
switching between two 
frequencies on Sgr~A* . We observed 43 and 22 GHz on 2007 April 2 and 43 and 
14 GHz on April 10. We changed the receiver every 20 seconds, yielding an 
on-source time of $\sim$10 seconds for each frequency. The observations were 
interrupted three times by 20 minute observations of four different quasars, 
including the fringe-finder 3C\,345. All data were correlated with 16 spectral 
channels per band and an integration time of one second. The total observing 
time including the quasars was 6 hours for each observation.

The VLBI data were edited and calibrated using standard techniques in AIPS.
First, we applied the latest
values of the Earth's orientation parameters. A-priori amplitude calibration 
was applied using system temperature measurements and standard gain curves. 
We performed a ``manual phase-calibration'' using the data from NRAO\,530 or 
3C\,345 to remove instrumental phase offsets among the three  frequency bands. 
Then, we fringe fitted the data from Sgr~A* using only the five inner VLBA 
antennas (PT, KP, FD, OV, and LA). Then, we discarded all data with elevations
below 15$^\circ$  and performed one round of phase self-calibration on 
Sgr~A*. 
Finally, we divided the calibrated {\it uv}-data by a model described in 
Bower et al. (2004), i.e. an elliptical Gaussian component of 
0.71$\times$0.41 mas with a position angle of 78$^\circ$. Lightcurves were 
extracted form the uv-data with the AIPS task {\sc DFTPL}.



\subsection {CSO: 350$\mu$m, 450$\mu$m \& 850$\mu$m}

Nightly observations of \sgra\  were made at three wavelengths over the 
period 2007 April 1--6 UT, using the SHARC-II camera.  The observations 
on April 1--5 were made with the SHARP imaging polarimeter module 
installed  \citep{li08}, using the 350 $\micron$ half-wave plate.  
The observing bands (selected by a cryogenic filter) were 450 $\micron$ 
on April 1--3 and 350 $\micron$ on April 4--5.  The observations on April 6 
were made at 850 $\micron$ with SHARP removed from the optical path.
 
In this paper, we report only the total intensity measurements.  Because
SHARP is a dual-polarization instrument, and because the
observations were made over cycles of half-wave plate angles that fully
modulate the polarization, our total intensity results are insensitive to
the polarization of the source.
                                                                                           
Except for the stepping of the polarimeter half-wave plate between
integrations, the observing and analysis method was similar to
past CSO observations \citep{zadeh06a,zadeh08a}.  We used Lissajous
scans with typical full amplitudes of 100$''$ in both azimuth and
elevation.  The instantaneous field of view is $57''\times57''$ in polarimeter 
mode and is $154''\times58''$ without the polarimeter.
 
The measured beam sizes were 8.4$''$ at 350 $\micron$, 10.1$''$
at 450 $\micron$, and 18.8$''$ at 850 $\micron$.
We used the Dish Surface Optimization System (DSOS) on April 1--5.  Three quadrants
were working fully during the run, but the fourth quadrant of the system
was available for only part of the run.  For observations at the elevation
of \sgra, we expect no significant effect on the results from the
non-operational quadrant.  Any change in the beam FWHM due to the status of
the DSOS quadrant was less than 3\%.
                                                                                           
Mauna Kea weather conditions were good overall during the April 1--6
campaign.  Occasional thin cirrus was observed visually and on satellite
photos on April 3, 4, and 6; otherwise skies were clear.  Local humidity
was $<$ 30\% during the observations.  The wind speed for April 1 was
noticeably high (roughly 30 mph), but less than 20 mph on the other
nights.  The zenith atmospheric opacity at 225 GHz was marginal for
observations at 450 $\micron$ on April 1 ($\tau_{225} \approx\ 0.07$),
as well as at 850 $\micron$ on April 6 ($\tau_{225} \approx\ 0.15$), and 
rising in both cases.  Atmospheric opacity on April 2--5
was excellent and relatively steady, ranging from 0.03 to 0.06 at 225 GHz.
                                                                                           
In producing the light curves for this paper, we reconsidered
the image registration and absolute calibration for all of our CSO
observations of \sgra\ from 2004 September through 2008 May.  The absolute
pointing of the images is based on hourly measurements of point-like calibration
sources, such as planets, and the pointing model for the telescope.  This
procedure appears to average down in a reasonable manner.
The 350 $\micron$ and 450 $\micron$ position that
we measure for the variable component of \sgra\ is within 0.3$''$ of
the nominal position of $\alpha_{2000} = 17^h:45^m:40^s.03, \delta_{2000} =
-29^0:00':28''.1$.  The agreement at 850 $\micron$, at which the telescope 
beam size is larger, is somewhat worse at 0.9$''$.
 
At any particular point in time, the telescope pointing model has only
$\sim 2''$ accuracy.  Therefore, we shifted the individual observations
to align with the average of all the observations, using the bright dust emission
in the images as the reference.  Subsequent photometry of \sgra\ assumes a
fixed position and beam size.
                                                                                           
Minor changes have been made to the absolute calibration scale factor and
\sgra\ ``zero point'', including data which have been published in the past
\citep{zadeh06a,zadeh08a,marrone08b}.
For the scale factor, we have adopted the following
brightness temperatures for calibration at 350, 450, and 850 $\micron$,
respectively: Callisto (128, 122, 120 K), Neptune (61, 66, 81 K), and
Uranus (64, 70, 86 K), arranged in decreasing order of usage and with an
estimated 10\% uncertainty.  These brightness temperatures are not
significantly different from our past assumptions.
The zero point relates
to the difficulty of measuring the total flux of \sgra\ with $\sim 10''$
resolution because of confusion from surrounding dust emission.  We estimate an
additive uncertainty of 1 Jy in our measurements of the absolute flux at
350 $\micron$ and 850 $\micron$, and an additive uncertainty of 0.5 Jy at
450 $\micron$.  To be consistent with the results published in this paper, the
August--September 2004 measurements at 850$\micron$ reported by \citet{zadeh06a}
should be shifted upwards by $\sim$0.2 Jy; the 450 $\micron$
measurements for the same period are essentially unchanged.  The 850 
$\micron$ measurements for July 2006, reported by \citet{zadeh08a} and
\citep{marrone08b}, should be shifted upwards by $\sim$0.5 Jy.  The 350
$\micron$ results for July 2005, reported by \citep{marrone08b}, should
be shifted upwards by $\sim$0.4 Jy; the 450 $\micron$ and 850 $\micron$
results for the same period are essentially unchanged.

\subsection {SMA: 230 GHz}


The SMA observed Sgr~A* 
on 
the nights of 1, 3, 4, and 5 April 2007, typically covering
the interval 1200--1830~UT. On the first three nights the array was
tuned to observe 231.9 (221.9)~GHz in the upper (lower) sideband,
while on the last night the frequency was tuned to 246.0
(241.0)~GHz. The array was in its ``compact-north'' configuration,
resulting in angular resolution of approximately 3\arcsec. All eight
antennas were used except on April 4, when one was lost to an
instrument problem. The SMA polarimetry system
\citep{marrone08a} was used in these observations to
convert the linearly polarized SMA feeds to circular polarization sensitivity, 
which prevents confusion between linear polarization and
total intensity variations.
                                                                                       
The data were gain calibrated using the quasar J1733$-$130, which was
observed approximately every 10 minutes. The absolute flux density
scale was derived from observations of Callisto and has an uncertainty
of 10\%. To remove the effects of the extended emission that surrounds
Sgr~A*, only projected baselines longer than 20~k$\lambda$ were used
in the light curve determination. Flux density measurements were made
by applying the quasar gains to the Sgr~A* data, removing the average
phase on Sgr~A* in each light curve interval via phase
self-calibration, which reduces the effect of baseline errors and
phase drifts on the measurement, and fitting a central point source to
the calibrated visibilities. Errors in the flux density account for
thermal noise, as well as the time-variable uncertainty in the gain,
which is estimated from the data themselves.

\subsection {IRAM-30m Telescope: 240 GHz}

Observations with the IRAM-30m telescope at Pico Veleta, Spain, were
carried out on 2007 April 1-4. Because of the low elevation of Sgr A*
at Pico Veleta, gain drifts due to atmospheric fluctuations are the
most severe limitation to accurate flux monitoring. For the same
reason, accurate peak-up is important if flux variations are to be
measured that are small with respect to the quiescent flux. To
account for both requirements, we alternated between Sgr B2 and Sgr
A* with the following procedure, applying a wobbling secondary mirror
to remove the 240 GHz  emission from the atmospheric and from extended
($>70"$) source structure. First, we pointed at Sgr B2 and measured the
position of its point-source component by fitting simultaneously a
Gaussian and a linear baseline (for a refined removal of extended
emission) to the pointing subscans taken in on-the-fly mode (two in
azimuth direction, two in elevation). The positional correction was
entered and the procedure repeated to recover the correct flux. We
used either the azimuth or elevation subscan, depending on where the
flux was larger (and thus a better peak-up was provided). Then the
antenna was moved to Sgr A*, where the same procedure was repeated.
Thus, for each time sample, there are two data points representing
the flux of Sgr A*, one from the pointing, and another one from the
peaked up pointing. Both results were used if the pointing correction
was sufficiently small. Error estimates were made by comparing the
results of subscans in the same direction. To avoid effects due to
instrumental and atmospheric gains drifts, only scaled  fluxes of Sgr A*
were retained for further analysis, using Sgr B2 as a non-variable
flux reference. Data reduction was done with the MOPSIC software
package \footnote{http://www.iram.es/IRAMES/mainWiki/CookbookMopsic}
The average Sgr B2 flux density is estimated to be 38.0$\pm1.2$ Jy
and was derived using the HII region G10.62-0.38 as absolute flux
reference. The beam FWHM is 11$''^2$.


\subsection {SMT: 250 GHz}
Observations were undertaken at the SMT located at
3200m altitude on Mount Graham in eastern Arizona
\cite{1999PASP..111..627B} 
using the 250~GHz channel of the facility's
four color bolometer \cite{1990LIACo..29..265K}.  
This  bolometer was used to observe at 250 GHz with a
broad band, ranging between 200 and 290 GHz on 2007 April 1--4. 
We made use of the telescope's beam switching mode, chopping horizontally 
$\pm 2\arcmin$
with the subreflector at a rate of 2~Hz along with an
``off-on-on-off'' observing mode that shifted the position of the
telescope every 10 seconds to remove any asymmetries in the
observations due to the chopping.  Jupiter, Saturn, and Mars were used
for focus and pointing references confirming the telescope's typical
half power beam width at these frequencies of $30 \arcsec$ and
pointing accuracy of $2 \arcsec$.  While Jupiter was used to set the
gain of the bolometer and skydips to find the atmospheric opacity, NRAO
530, 1757-240  and G34.3 were also observed throughout the observations as
secondary calibrators to check the stability and repeatability of the
measurements.  Finally, after splitting the data into 80 second
increments (consisting of two iterations of the 40 second long
``off-on-on-off'' observing mode), the raw data were reduced using a
version of the standard GILDAS NIC reduction program customized for
the four color bolometer. Because a single   calibrator was not used 
continuously during the first  
two days of observations, the flux variation of \sgra\ was uncertain and 
thus the data are not presented here.


\subsection {NMA: 150 \&  230 GHz}

Interferometric NMA observations 
were carried out simultaneously
at 90 and 102 GHz in the 3-mm band, and simultaneously at
134 and 146 GHz in the 2-mm band with bandwidth of 1024 MHz
on 2007 April 1--4. 
The 2 and 3 mm flux densities  are  measured to be 1.8$\pm$0.4 and
2.0$\pm$0.3 Jy, respectively. 
The light curves of data from April 3 and 4 are presented using five and 
six  antennas, respectively. Th weather was bad on April 1 and 2 so we 
discarded the data on the first days of observations.
1744-31 (J2000) was used as the phase  calibrator and the data was binned 
every 3-4 minutes. 
The flux measurements of \sgra\ were estimated by fitting 
a point source model in the {\it uv}  plane restricted 
to distances $>$ 20 k$\lambda$, in order to suppress the contamination
from extended components surrounding \sgra. 
The FWHM of the synthesized  beam in the 2mm observation 
on 2007, April 4 is  6$\arcsec\times1.4\arcsec$. 
We used 3C279 as a passband calibrator and Neptune
as the primary flux calibrator.

\subsection {CARMA: 94 GHz}
Interferometric CARMA observations were done 
to  observe Sgr A* on 2007, April 2-5.  
Observations were made at 94 GHz using nine 6m diameter BIMA and six 10m diameter  OVRO antennas with the
exception of observations on 2007, April 3  which did not include any OVRO
antennas. In all days, 
Uranus was used as the primary flux calibrator, 1744-312 as the complex 
gain calibrator and 1751+096 was used as a passband calibrator.  
The weather was poor for observing at 94 GHz on the first half of 2007, April 3 and the 
second half of 2007 April 4. We did not include the data during these times. 
Five 
frequency windows, each 469 MHz wide, were used at frequencies from 94 to 
100 GHz.  We used NRAO530 (1730-130) to cross calibrate 1744-312, in 
order to independently track the amplitude stability of 1744-312. 
 All 
calibration was done using MIRIAD package and calibrated visibility data 
for each day were read into AIPS and the DFTPL task was used to extract 
light curves for the source.

\subsection {GMRT: 1.28 GHz}

We observed Sgr A* using Giant Meterwave Radio Telescope (GMRT) in 1280, 610 and 325 MHz 
frequencies with central observation time on MJD 54195.0, 54191.1 and 54190.1 (5.0 April, 1.1 
April and 31.1 March 2007 UT) respectively. GMRT\footnote{www.gmrt.ncra.tifr.res.in} consists of 
thirty fully steerable parabolic antenna array, where fourteen antennas are randomly distributed 
in 1 km area and rest of the sixteen antennas are placed in three arms, spread over 25 km 
area, forming nearly a shape like `{\it Y}'. The diameter of each antenna is forty five meter. 
Observation band-width in each frequency was 32 MHz and integration time was 16.9 second. The 
source was observed for 6.1, 4.2 and 7.0 hr in 1280, 610 and 325 MHz respectively. We have done 
flux calibration using 3C286 and 3C48 and used \cite{baa77} for setting flux density scale. 
J1830-360 was used as phase calibrator. The bad data and radio frequency interferences (RFIs) 
are eliminated from the data set and the source is self-calibrated. The original data has 
channel width of 125 KHz in the spectral line mode. To take care of effect of the band width 
smearing in low frequency, we did not averaged all the channels after calibration but averaged 
32, 16 and 8 channels in 1280, 610 and 325 MHz respectively (forming effective channel width of 
4, 2 and 1 MHz in 1280, 610 and 325 MHz). The images are corrected for the beam-shape.
Because of the strong emission from the nonthermal emission surrounding 
Sgr A*,  the light curve of Sgr A* could have not been obtained reliably at  these low frequencies 
330, 630 and 1280 MHz. This is mainly due to the  instantaneous elongated beam shape which contains 
extended structures surrounding Sgr A*. 

\section {Light Curves: Individual Telescopes}

The results of XMM and VLT observations in X-ray and NIR 
have already been 
presented elsewhere
\citep{porquet08,dodds-eden09}. A detailed account of INTEGRAL observations 
are  given elsewhere (Trap et al. 2009). 
To present all the data that were taken during this  campaign, 
we include the XMM and VLT light curves again here
and briefly review the results of these observations that have 
already been published elsewhere.  

\subsection {NICMOS Photometric Measurements}

Figure 2a shows the observed variability of \sgra\ in the NICMOS 
1.45 and 1.70\mic\ bands, where there is good agreement between 
the two bands.
The observed ``quiescent'' emission levels of \sgra\  in the 1.45 
and 1.70\mic\ bands are $\sim32$ and $\sim38.5$ mJy, respectively,
but some fraction of this total signal is due to the neighboring 
star S0-2. 
During flare events, the emission is seen to increase by anywhere
from a few percent to 25\%\ above these levels. In spite 
of the somewhat lower signal-to-noise ratio for the 1.45\mic\ data,
due to the somewhat lower sensitivity of the NICMOS detector and 
increased effects of extinction, the flare activity is still easily 
detected in this band. In order to confirm that the observed variability 
of \sgra\ is not due to either instrumental or data reduction effects,
we have compared the \sgra\ light curves to that of the star S0-2 and
to a region of background emission with the NICMOS images, as shown in
Figure 2b.  The photometric measurements for \sgra\ show obvious signs 
of variability in six of the seven  windows  of HST observations, while 
the 
corresponding light curves of S0-2 and the background remain quite stable.

The panels of Figure 3a-e present detailed light curves of \sgra\ and, for
comparison, S0-2 for each 
of the seven HST observing windows. These plots show the time-ordered
measurements in the 1.45 and 1.70\mic\ bands, where we have now subtracted
the mean ``quiescent'' flux level, leaving the net variations in emission
for both \sgra\ and S0-2. All light curves are aperture and extinction
corrected.
Each observing window consists of 5 to 7 HST orbits, with  each orbit
covering $\sim46$ minutes.  We have identified flaring activity in at 
least one orbit in each of the seven observing windows. These activities are 
identified in the light curves with labels designating the day (1--7)
and the flare even within the day (A--C).
A typical flare event lasts between 10 and 40 minutes. 
The  amplitudes  and durations of the events are similar to what was
found in our earlier HST observations \citep{zadeh06a}.


To examine the short time scale variability in more detail, 
the 1.70\mic\ light curves are shown in Figure 4 with a sampling of 64 seconds.
The Sgr A* and S0-2   light curves are qualitatively similar to 
those 
in Figure 3, except for the finer sampling and we show only the 1.70\mic\ band 
because the 1.45\mic\ data do not have sufficient signal-to-noise in this
shorter integration period.
There are 16 identified flaring events, all of which are shown 
in 45min periods  in two panels of eight flares.  
One type of fast fluctuation that we have detected is generally associated 
with the rise or fall of bright flares, or at the peaks of 
bright flare emission, as seen for the flares 1A, 2A, and 5A.
Similar minute time-scale variability has  also been detected by 
Dodds-Eden et al. (2009). 
Another type of fluctuation is the point-to-point variability seen during 
some of the quiescent phases of low-level of activity, such as flares 1B,
1C, 2B, 3A, 4B, 6A, and 7A.

\subsection {VLT NIR and Mid-IR Observations}

The VLT observations used multiple bands to observe \sgra\ on 2007 April 1--7,
using the two instruments NACO (NIR) and VISIR (mid-IR). 
The results of these observations, which included the identification of seven
flaring events are discussed in detail by Dodds-Eden et al. (2009).   The brightest 
flare  detected  at 3.80$\mu$m  coincides with  the brightest X-ray flare on 
April 4. Figure 5 shows a composite light curve of VLT observations with 
labeled flares  using 3.8$\mu$m, 2.12$\mu$m and 1.66$\mu$m NIR bands.  
No NIR spectral index measurements are available for the detected flares. 
However, a 3$\sigma$ upper limit of 57 mJy 
is placed at 11.88$\mu$m for the bright 3.8$\mu$m flare on April 4 with a peak 
flux density  of $\sim$30 mJy (see also Trap et al. 2009).  The brightest NIR flare detected in this 
campaign consists of a cluster of 
overlapping flares that last for about two hours. 
The second  brightest  flare detected by the VLT is  identified as \#6 in Figure 5. 
This flare precedes the bright NICMOS flare 5A (April 5), as shown 
in Figure 3e. These flares are  components of another   period of flaring 
activity lasting for about two  hours.
                                                                                                                               
\subsection {X-ray  Flaring Activity}

The X-ray light curves between 2 and 10 keV
with a time binning of 144s are shown in Figure 6.
A total of five flares were observed: one in 2007 April 2 (labeled $\#$1)
with a peak X-ray luminosity L$_{2-10keV}=3.3\times10^{34}$ erg s$^{-1}$
and four on 2007 April 4 (labeled \#2, \#3, \#4, \#5) with peak L$_{2-10keV}$=24.6, 
6.1, 6.3, and 8.9$\times10^{34}$ erg s$^{-1}$, respectively \citep{porquet08}.
For the first time, within a time interval of roughly half a day, an
enhanced
incidence rate of X-ray flaring was observed,
with a bright flare ($\#$2, with a duration of 2900 s) followed by three
flares
of more moderate amplitude (\#3, \#4, \#5, with durations of 300,
1300, and 800\,s respectively).
An enhanced rate of X-ray flares,
although with lower amplitudes, was also  reported  in Belanger et al.\ (2005) 
when one moderate and two  weak flares  were detected within a period of eight hours. 
These  rates of X-ray activity \citep{porquet08,belanger05} are  clearly higher than 
the typical  duty cycle of one X-ray flare a day \citep{baganoff-03}.
The brightest  event on 2007 April 4 
represents the second-brightest X-ray flare from Sgr A* after the X-ray flare with 
$\Gamma$ =2.2 $\pm0.3$ 
on  2002, October 3, 
on record with a peak amplitude of about 100 times above the 2--10\,keV quiescent
luminosity\footnote{No detection was made using  INTEGRAL in the 20--40
keV
and 40--100 keV energy bands, leading to 3$\sigma$ upper limits of
2.63 and 2.60$\times10^{35}$ ergs s$^{-1}$, respectively (Trap et al. 2009).}
This bright X-ray flare exhibits similar light-curve shape (i.e.,nearly
symmetrical), duration ($\sim$3 ks)
and spectral characteristics to the very bright flare observed on 2002, October 3
with 
XMM-Newton \citep{porquet03}.  Its measured spectral parameters, assuming
an absorbed
power law model including the effects dust scattering, are
N$_{\rm H}= 12.3^{+2.1}_{-1.8} \times 10^{22}$ cm$^{-2}$ and 
$\Gamma$ =2.3 $\pm0.3$ 
where the quoted errors are 
at the 90\% confidence level.
Therefore, the two brightest X-ray flares observed so far from Sgr A*
exhibited similar soft spectra $\Gamma\sim 2.2-2.3$.
The spectral parameter fits of the sum of the three following moderate
flares, while lower
(N$_{\rm H}= 8.8^{+4.4}_{-3.2} \times 10^{22}$  cm$^{-2}$ and $\Gamma$
=1.7$^{+0.7}_{-0.6}$),
are compatible within the error bars with those of the bright flares.
However, fixing 
the column density  at the value found for the brightest flare
(i.e. N$_{\rm H}= 12.3 \times 10^{22}$ cm$^{-2}$)
leads to a larger photon index value
for the sum of these moderate flares, i.e. $\Gamma =2.1\pm$0.4.



\subsection {43 GHz Time Variability: VLA}

Figure 7a,b shows light curves measured during April 1--4 at 43 GHz with the VLA, 
using 87sec and 300 sec sampling, respectively.  The light curve of the phase 
calibrator 17444-31165, which itself is cross calibrated by NRAO530, is flat and 
is shown at the bottom of each panel in Figure 7a. Since NRAO530 is not the 
primary calibrator, it provides a second check on instrumental stability and that 
its light curve was flat also. 


The light 
curves of \sgra\ show variations on a variety  of time scales from as short as 
30 min to longer than five hours at 43 GHz. 
The fluctuations on time scales of several hours $\sim5-6$ hours can be 
seen in 
Figure 7a,b. The slow  flux  variation  over 5-6 hours could, in principle,  
result from the contamination of the 
emission by an asymmetric distribution of extended structures surrounding \sgra\ 
especially when a compact configuration of the VLA is used. However, the 
contamination of flux by 
extended emission is minimal for {\it uv} data $> 90k\lambda$ (or 2.3$''$) 
and the 
variability on several hour time scale is  intrinsic to \sgra.  
Previous high resolution data taken with a wide configuration of the VLA 
have also shown the presence of flux variation 
of Sgr A* on such time scales (Yusef-Zadeh  et al.\ 2006a,b and 2008).
The contamination 
of extended emission was  clearly seen  at low elevations in the  {\it uv} data $<90 
\rm k\lambda$ at 43 GHz, and  our 22GHz data taken simultaneously with 43 GHz data on  
2007 April 1-4  were useless for time variability analysis because of the limited {\it 
uv} range (i.e., $<$ 70k$\lambda$). 


Most of the 
power of the 43 GHz fluctuations in four consecutive days of observations appears to 
fall in a range between 30 minutes and few  hours, as best shown all
light curves of Figure 7b. For example, fluctuations with $\sim$1h time 
scale are detected at a level of 200 mJy in the April 1 and April 2 light curves 
centered near 13h and 11:15h UT, respectively. The light curve of April 4 shows 
largest flux variations at a level of $\sim$40\%\ are seen to 
increase flux density from 1.1 Jy at 9h UT to 1.6 Jy near  15h UT.
Another interesting feature of the April 4 light curve is 
the presence of multiple weak fluctuations at  a level of 50 mJy on a time scale of 
$\sim$20-30 minutes.  
Figure 7c shows the 
light curves of April 4 for simultaneous observations at frequencies of 43.1851 GHz and 
43.5351 GHz with a 30sec  sampling time. The 
frequency separation 
between these light curves is 345 MHz.
We note at least five 20--30 minute fluctuations that are seen in both 
light curves. A more detailed account of the power spectrum analysis of the time variability of 
Sgr A* in radio wavelengths will be given  elsewhere. 

\subsection {14, 22 and 43 GHz Time Variability: VLBA}

Figure 8a shows 43 GHz light curves based on VLBA observations on April 1, 5 and 11, 
with a 60 sec sampling time. Figure 8b shows the light curves at 22 GHz and 43 GHz on 
April 2 whereas Figure 8c shows the light curves at 15 GHz and 43 GHz on April 10. 
The flux density of \sgra\ show variations on several hour time scales 
in these VLBA observations at multiple frequencies. 
These light curves show the first measurements of the flux variation of \sgra\ on a 
VLBA (milli-arcsecond) scale at several frequencies.  

Fluctuations in phase coherence and amplitude errors could 
produce significant changes in flux on short 
timescales. 
However, it is unlikely that calibration errors are similar at two
frequencies, thus the flux variation on $\sim$5-hour time scale (Fig.\ b,c) 
is intrinsic to Sgr A*.

                              
\subsubsection {43 GHz Light Curve: VLBA and VLA Comparison}

Because VLBA and VLA measurements on April 1 and 2 are taken simultaneously at 
43 GHz, we compared the two light curves, as shown in Figure 9a,b with a 300 
sec sampling time, respectively. The comparison of the light curves examines 
directly the localization of flaring events at radio wavelengths.  The largest 
fluctuations in both VLA and VLBA light curves appear to agree with each 
other.  Peaks with 
hourly time scale  durations occur in both light curves near 
13h UT on April 1, as seen  in  Figure 9a. Similarly, the slow decreasing trend in
the flux of Sgr A* over few hours is seen in the light curves of 2007 April 2 at 
43 GHz using both 
the VLA and VLBA, as shown in Figure 9b.  
The behavior of the light curves  on hourly time scales measured with VLBA 
provides the first 
direct  evidence that flaring activity arises from the innermost region of 
\sgra\ on milliarcsecond (mas) scales. The size of the flare emission  is 
dominated by interstellar scattering. 
The general 
agreement between the VLA and 
VLBA light curves  imply that flaring region  that has been detected
is unresolved with the VLA.

There are  also  discrepancies between the two light curves. 
One  is the different values  of ``average  levels'' of flux taken  in the 
light curves measured with the VLA and VLBA. 
In all the measurements shown in Figures 7a and 8a,b  the 
average-level of VLBA flux  appears to be lower than than
that of the VLA by $\sim200$ mJy. 
The second discrepancy  is  the flux variations 
do not agree  with each other
on small time scales in VLA and VLBA  light curves. 
The 
uncertainty in the absolute flux density calibration of
\sgra\ at 43 GHz using VLBA and VLA  could easily 
explain the first  discrepancy. 
It is possible that 
the emission from \sgra\ could be contaminated by extended emission
from the surrounding medium, as measured with 
the VLA, even though we have selected data with {\it uv} $>$ 100k$\lambda$. 
This could explain why the VLA and VLBA light curves 
do not agree with each other on 10-15 minute time scales. Lastly, it is possible that 
these discrepancies   could be explained by 
a core-halo structure of emission from Sgr A* in which the halo component is   resolved 
out in VLBA observations. 
Future simultaneous VLA observations using  its most extended array configuration and 
VLBA should be able  to examine closely  the reason for these discrepancies.  

\subsection{CSO  350$\mu$m, 450$\mu$m, 850$\mu$m Light Curves}

Figure 10 shows the light curves at three submillimeter wavelengths.
The data have been smoothed to increase the signal-to-noise ratio
with a sampling time of $\sim$6.5 minutes. 
As at radio wavelengths, the flux of \sgra\ appears to be varying  
on hourly time scales. The largest \sgra\ increase is detected 
at the beginning of the observation  near 13:30 UT on 2007 April 1.
 These light curves show evidence  for 
hourly and intraday  variability at 450$\mu$m,
at a level of 14\%. 
The mean daily flux of Sgr A* at 450$\mu$m is $\sim3\pm0.25$ Jy. 

Figure 10b shows some of the first variability of Sgr A* at 350$\mu$m. 
The flux increase on 2007, April 4 over 5 hours is about 50\% of the initial flux 
of Sgr A*. This steady increase of flux density over several  hours is seen to continue at 
90 GHz (see section 3.10). 
Figure 10c shows the light curve at 850\mic\ on 2007, April 6. unlike 
the other submillimeter light curves 
shown here, this light curve appears to show
time variability on a time scale of $\sim10$ minutes as seen near 13h:20m UT. 
Such sharp variations, though with low signal-to-noise values,  at  
850\mic\ at such a short time scale resembles 
the recent light curve obtained with a  different instrument (LABOCA of APEX)  
\citep{eckart08}.  
The reality of such a short time scale variation needs to be 
confirmed. 

\subsection{SMA: 230 GHz Light Curves}

Figure 11 shows the light curves taken from four days of observations 
with the SMA at 230 GHz. 
The 2007 April 1 data shows an asymmetric profile indicating 
a duration of possibly $\sim4$ hours considering that there is a gap
between 16h and 17h:30m UT. Similar submillimeter 
characteristics have been seen 
recently at 850$\mu$m \citep{zadeh08a}.
The 2007 April 3 light curve shows an emission peak near 14h UT, before
a slow decay that lasts for about 4 hours.  
The light curve obtained with SMT on the same day and at the same wavelength showed
the rising part of the light curve suggesting that the duration of the flare on this day 
 could be as long as 8 hours. 
The 2007 April 4 light curve shows a typical profile of submillimeter flare emission,
except for a dip in the flux at a level of 100 mJy near 14h UT. 
The April 5 data shows a light curve with multiple peaks as the
light curve decays. The typical time scale for this variation is between $\sim$1-2 hours. 
The overall percentage of flux variation during 6 hours of observations is between 
10\%\ and 30\%. 

\subsection{SMT 250 GHz Light Curves}

The SMT light curves of Sgr A* and calibrators (in blue) for 2007,  April 1-4
are shown in Figure 12. 
Because SMT and SMA observed Sgr A* at the same wavelength considering 
the broad bandwidth of the SMT, we compared the SMT light curves with 
those of SMA on April 1, 3 and 4. An increase in flux of $\sim$1 Jy 
in the rising part of the light curve is seen 
between 10h UT and 14h UT on 2007, April 3. 
This increase is similar  to the decrease in the 
flux of 
Sgr A* found in the  decaying  part of the SMA light curve,
as seen in Figure 11. 

The low spatial resolution of the SMT results in a higher background level for the 
\sgra\ light curve. The discrepancy in the zero level flux of Sgr A* using SMT and SMA 
on April 3 and 4   
is due to the fact that the emission from Sgr A* using SMT is contaminated by 
3.5$\pm0.2$ Jy of flux from extended features. If this flux is subtracted 
from 
the April 3 data and combined with the SMA light curve, the duration 
of the variability is estimated to be $\sim$6 hours.

The 2007, April 1 shows the most dramatic flux variation of $\sim$2 Jy 
The reality of this feature can not be confirmed as 
different  calibrators were used at the beginning of the 
observation. However, a  3.6$\pm0.2$  Jy subtraction 
from the SMT data  matches well with  the SMA data, thus 
suggests that the sudden rise 11h UT is likely to be real. 


\subsection{IRAM 240 GHz Light Curves}
Figure 13 shows the light curves of the two days of IRAM
observations on  2007 April 3 and 4. There is no evidence for any 
flux variations in these observations. 
The April 4 light curve overlaps with the biggest NIR/X-ray  flare detected during 
this campaign. However, there is no indication that the 240 GHz flux density changed 
by more than  1.5$\pm0.5$ Jy between 5h and 7h UT, during which the 
bright NIR/X-ray flare took place.
 
\subsection{NMA 140 GHz and 230 GHz Light Curves}
Figure 14 shows the NMA light curves of Sgr A* and 1744-312 based on two 
days of 
observations under excellent weather conditions. The flux 
of the calibrator remains 
flat during these observations, 
whereas the flux of Sgr A* increases by $\sim$0.5 Jy 
at 90 GHz on April 3.
The April 4 light curve shows a slight increase at 146 GHz before 
decaying strongly by more than 1 Jy. The duration of the flare 
is roughly 2 hours.   

\subsection{CARMA Light Curves}
Fig 15 shows the light curves of \sgra\ and the calibrator 1733-130 at 94 GHz 
taken for four days of observations on 
 2007, April 2-5. There is flux variation on short and long time scales in all days 
of observations. 
There is concern  on the variation of the calibrator evident in almost  all days of 
observations. 
Due to  this  uncertainty, we compared the light curves with other 94 GHz and 43 GHz 
measurements 
and we believe the large  scale variation may reflect the intrinsic variable emission 
from \sgra. However, the flux variation of \sgra\ on short time scale may not be valid.

\subsection{GMRT Light Curves}
As we observe 
Sgr A* at long wavelengths, the light curve of \sgra\ may be contaminated 
by the extended nonthermal emission surrounding  \sgra\ as well as 
by interstellar scattering which becomes more important at long wavelengths. 
 In  order to avoid the 
contamination by extended emission, 
we used the {\it uv} data at the highest  elevation as 
well as restricted the {\it uv} distribution  be greater than 80k$\lambda$. 
Also, interstellar scintillation is expected to operate on longer 
time scales  than hourly time scales  that we are sensitive to.
A flux variation  of $\sim$80 mJy over four hours
was detected. Given the limited resolution of the GMRT data at this frequency,  
it was not clear if  this variation  reflects 
the flux of Sgr A* 
at 1.28 GHz or an artifact of  the contamination of extended flux.  It is clear that higher 
resolution data are  needed to separate Sgr A* from the extended features in its vicinity.


\section {Analysis}

\subsection {NIR Flare Statistics}

Given the ability of HST to produce continuous observations over many 45 min
orbital visibility periods, along with its long-term photometric stability,
the NIR NICMOS data provide an excellent way to investigate the flare
strength distribution over many flare episodes.
Figure 16 shows a histogram of the NICMOS 1.70\mic\ net flare emission for
the 7 days of data obtained in this campaign. 
The net flare emission is measured by  first  subtracting the background  emission for each day 
before the excess flux above the background is selected. Thus, the selected data 
points do not sample the peak flare emission but rather the flux associated with flaring 
activity.  
The peak of values centered at a net flux of zero represents the emission
from \sgra\ during ``quiescent'' periods. The positive half of the histogram,
on the other hand, shows a tail of flare emission events extending out to
$\sim$10 mJy.
The ``quiescent'' distribution is best fitted with a Gaussian, which is
expected from the level of random noise in the observations.
The tail of flare emission can be fitted with a power-law distribution having
an index of -1.19$\pm0.27$ and a low-energy cutoff at S$_\nu = 1$ mJy.
The dotted line in the figure shows the result of simultaneous Gaussian and
power-law fits to these two components.

\citet{zadeh06a} reported that distribution of flare activity seen
in our more limited 2004 NICMOS observations could be fitted by two simultaneous
Gaussians profiles.  A reanalysis of those data, however, now show that a 
power-law distribution with a low-energy cutoff yields a good fit to the 2004
epoch data as well.
Figure 17 shows a histogram of the 2004 1.60\mic\ data, with Gaussian and
power-law fits to the two components shown by the broken lines.
The best power-law fit to these data has an index of -1.11$\pm0.13$, with a
low-energy cutoff of S$_\nu = 0.25$ mJy. This is remarkably consistent with the 
best-fit power-law index of the 2007 data. 
We note that the fraction of observing time that flare activity has been
detected in the 2004 and 2007 campaigns is $\geq 32\%$ and $\geq 37\%$,
respectively. 

The NIR flare histograms for the two epochs show that 
the probability of measuring flux $S_\nu$ at any instant is approximately
proportional to $1/S_\nu$.  Presumably this reflects the statistics of the flaring
behavior of Sgr A* at NIR wavelengths.  To explore this we
construct a simple phenomenological model for the flaring by 
 simulating a  light curve and 
then sample it to construct a
simulated histogram.  This model shows that the observed $1/S_\nu$
behavior arises quite naturally, but does constrain the statistics of
the flaring.

Our phenomenological model represents the flaring as a sequence of
100 Gaussian profiles occurring over 100 arbitrary time units, with
flare $i$ characterized by peak flux $S_i$, timing
of the peak $t_i$, and standard deviation of the flare $\sigma_i$, so
that the net light curve may be written as
\begin{equation}
S_\nu (t) = \sum_{i=1}^{100} S_i \exp\left(-\frac{(t-t_i)^2 }{
2\sigma_i^2}\right)\,. 
\end{equation}
The parameters $S_i$, $t_i$, and $\sigma_i$ are drawn randomly and
uniformly from the ranges $[0,1]$, $[0,100]$ and
$[0,\sigma_\mathrm{max}]$, respectively, and the resulting light curve 
is evenly sampled every 0.2 time units to create a flare histogram.
Note that $\sigma_\mathrm{max}$ is the only independent parameter of
this model, as increasing the number of flares and changing the maximum
flare amplitude can be accommodated by rescaling the flux and time units.
In addition, changing the sampling rate or the number of flares does
not affect the statistics, provided that the light curve has already
been adequately sampled (which is the case for our adopted sampling
rate of 50 per time unit).  We find that $\sigma_\mathrm{max} \la 0.5$
yields the observed $1/S_\nu$ behavior.  

A typical simulated light curve 
and the corresponding histogram for $\sigma_\mathrm{max}=0.5$ are given in 
Figures 18a and b, respectively. The slope of S$_{\nu}^{-1}$ is drawn on Figure 18b. 
Larger values of $\sigma_\mathrm{max}$ lead to significant overlap between
flares, tending to give a flatter dependence of the flux probability
on $S_\nu$.
This does not, of course, prove definitively that the flares behave as
given by equation 1. 
Other choices of functional form or
different statistics for $S_i$ may also yield the $1/S_{\nu}$
behavior of the histogram.  It does, however, seem to require that
the flare events do not significantly overlap each other.

\subsection {Spectral Index Distribution Between 1.45$\mu$m and 1.70$\mu$m}

We have constructed a log-log distribution of spectral index 
based on the NICMOS 1.45$\mu$m  and 1.70$\mu$m data.  Figure 19 shows 
the "color" distribution of all the data selected with signal-to-noise 
S/N$=$3. 
The diagonal line (in red) shows the spectral index of $\beta$=0.6,  where 
F$_{\nu}\propto\nu^{-\beta}$.   For comparison, $\beta$ of -4, -2, 2, and 4 are also 
plotted. 
This figure  shows a tendency for 
the spectral index of low flux values  to be steeper than 1, whereas the 
high flux values are represented by a flatter distribution of spectral index.
Because the data points used in making Figure 19 are not taken 
simultaneously  at the two different wavelengths, we attempted to
estimate spectral  index values of  adjacent data points, where the flux of
\sgra\ is not varying rapidly, such as during the fast rise or fall of 
individual flares. 
The 1.45 and 1.7$\mu$m  NICMOS images were acquired back-to-back in long
sequences, in which the exposures within each pair are separated in time
by about 2.5 minutes. The points shown in Fig. 19 represent all adjacent
pairs of measurements (adjacent meaning $\sim$2.5 minute separation) for which
the S/N in the individual measurements is greater than  3 (hence not all available
pairs from Fig. 3 are included). The fact that the overall Sgr A* flux
could be changing within that 2.5 minute time scale is a concern and is
why the spectral index values listed in Table 1 where taken from only those
measurements where we could see from the light curves that the overall
Sgr A* flux was not changing much on that $\sim$2.5 minute time scale. The
full light curves also indicate that the overall flux of Sgr A* does not
often change on such short time scales and therefore the number of
suspect measurements in Fig. 19 should be a relatively small fraction of
all measurements. Hence we only deduce general trends from that diagram.

We identified  five sets of data points associated with five different flares
during which the overall \sgra\ flux is not varying rapidly.
Table 1 shows the corresponding flux and spectral index values using data
sampled at 144 sec intervals.
The two brightest flares, 5A and 2A, have spectral indexes 0.73$\pm$0.16 and 
0.97$\pm0.27$, whereas the weaker flares have indexes steeper than $\beta$=1.5. 
These individual measurements are consistent with the spectral index trend shown 
in Figure 19.   
We also  find that the spectral index of the brightest flares
are  consistent with recent Keck measurements, which yield a spectral index 
of 0.6 \citep{hornstein07}. The spectral index  of low  flux values 
is also consistent with  VLT measurements, which show a steep spectrum for weak flares 
\citep{eisenhauer05,gillessen06}. These measurements suggest that the spectral index of 
flares varies with the NIR flare strength, support  earlier measurements by Gillessen et al. (2006) 
and disagree  with measurements by Hornstein et al. (2007) who claim a
constant spectral index in NIR wavelengths. 
The variation of spectral index with flare emission at NIR wavelengths  has important 
implications on the inverse Compton scattering mechanism of  X-ray and soft 
$\gamma$-ray emission from Sgr A* \citep{zadeh06a} as well as on the hypothesis 
that X-ray  emission is due to synchrotron mechanism  (Dodds-Eden et al. 2009).
It is possible that weak flares with a steep energy index of particles
are  associated with low-level activity of the accretion disk of Sgr A*, 
whereas the bright flares represent the hot magnetically-dominated events that are 
launched from the disk. Polarization characteristics of the weak and strong flares may 
constrain models of the flare emission. 
The correlation of the spectral index and flux has  been discussed in the context
of electron heating and cooling by a turbulent magnetic field \citep{bittner07}. 
In the synchrotron scenario, the higher value of the spectral  index 
at low NIR fluxes could be an
indication of  the cooling break.

\subsection{NIR Power Spectrum  Analysis}

Genzel et al. (2003) had reported a possible 17 min NIR periodicity with 
implications for the spin of the black hole. Our previous 2004 HST data 
\citep{zadeh06a} showed a marginal detection of power at 33$\pm$2 minutes. We 
investigated the power spectrum of flare data taken with the new NICMOS 
measurements. We created Lomb-Scargle periodograms \citep{scargle82} to search for 
periodicities in our unevenly-spaced NIR measurements.  We performed 1000 
simulations of each light curve, with the same sampling and variance as the data, 
and with simulated noise following a power-law ($P(f)\propto f^{-\delta}$) chosen 
to match the periodogram of the data as closely as possible (following Timmer \& 
Konig 1995, Mauerhan et al. 2005), typically with an index $\delta$ of 1.5 or 2.  
Artificial signals are seen at the 90 minute orbital period of HST, and the 144 
second filter switching cycle, and discounted.  For each point in a lightcurve, we 
identify the periodogram simulation at the nth (where n=99, 99.9) percentile of the 
distribution, and thus create lines below which n\% of the simulations fall. Figure 
20 shows the power spectrum as a solid line and the dotted lines show the the 
spectrum of the noise using power-law distributions. Only one HST observation shows 
any  power above the 99.9 percentile line, on 2007 April 4, near 2 hours.

The 99.9 percentile refers to the local distribution; however, the chance of 
getting a point above the 99.9 percentile line must be computed considering all 
trials (Benlloch et al. 2001).  We sample 158 frequencies above 10.8 minutes, the 
lower limit of our simulation software, and perform seven observations, so our 
total is 1106 observations, suggesting $\sim$1 peak above the 99.9 percentile line.  
We have two adjacent points above the 99.9 percentile line, but these points are 
probably not independent. Altering the index $\delta$ within a range consistent 
with the data does not change the strength of the signal.  We conclude that the 
significance of this possible periodicity is not particularly strong.

The lack of any significant power between  17 and 20 minutes supports the results from 
an earlier 
analysis of HST data in 2004 \citep{zadeh06a}. 
Recent analysis of data taken with the combined VLT and the Keck observations shows
no  significant power on short  time scales \citep{do08,meyer08}.


\section{Discussion}

\subsection{X-ray Flare Emission Mechanism}

As described in \S3.3, five X-ray flares were detected in the
present observing campaign.  The strongest X-ray flare (\#2) coincided
with a strong NIR flare observed with the VLT (Dodds-Eden et al.\
2003).  HST observations detected three of the remaining four X-ray
flares (\#1, \#4 and \#5) corresponding to flares 2A, 4A and 4B,
respectively.  Figure 21a,b show the X-ray light curves of these newly
detected NIR flares at 1.70$\mu$m.  Table 1 presents the flux and
spectral index of the NIR flare 4A. The remaining X-ray flare detected
in this campaign (\#3) was not observed contemperaneously in the IR.
The simultaneous monitoring of Sgr A* in X-rays and the IR have shown
that X-ray flares are always accompanied by flaring in the IR but that
the reverse is not necessarily true (Porquet et al.\ 2009).

The NIR flare emission from Sgr A* have been shown to be highly
polarized \citep{eckart06a,meyer06} and is therefore likely to be produced 
by  synchrotron
emission from GeV electrons in the $\sim 10$\,G magnetic field
strengths thought to be present in the vicinity of Sgr A*.  The 30
minute duration of the flares is broadly in line with the synchrotron
cooling time scale of these electrons.  Substructure in the NIR flare
lightcurves has been attributed to doppler beaming and lensing of an
orbiting hotspot \citep{meyer06}, although this remains to be
confirmed.

The X-ray flares are always seen in concert with flaring in the IR
(when the IR has been simultaneously observed).  Thus scenarios for
the X-ray emission are directly associated with the acceleration of
the GeV electrons responsibly for the IR synchrotron emission, either
through upscattering of submillimeter seed photons (Markoff et al.\ 2001,
 Yusef-Zadeh at al.\
2006, 2008), 
synchrotron self-Compton (Eckart et al. 2006a) 
simply as synchrotron emission from the high 
energy
tail of the accelerated electrons (Yuan et al.\ 2003, Dodds-Eden et al.\ (2009) 
or from the NIR emitting electrons \citep{eckart06a}.
Dodds-Eden et al.\ (2009) have recently shown that the ICS scenario
requires an uncomfortably small submillimeter source source size ($\la
R_s$) to match the observed X-ray and NIR fluxes for flare \#2.  In
addition, although the other flares can be modeled with source sizes
of a few Schwarzschild radii and field strengths in the 10\,G range
(e,g., Yusef-Zadeh et al.\ 2006), the energy density of the GeV
electrons in the NIR-emitting region exceeds the magnetic energy
density by more than an order of magnitude, and their acceleration and
confinement becomes problematic (this is also the case for flare \#2).
Finally, the duration of the X-ray flare is shorter than the IR flare
whereas one might expect them to be identical.  Dodds-Eden et al.\
2009 therefore strongly preferred the synchrotron scenario.  This
implies that the acceleration mechanism must continuously resupply the
100\,GeV electrons for the 30 minute duration of the observed flares
as the synchrotron loss time of the $\sim 100$\,GeV electrons
responsible for the synchrotron emission is $\sim 30$\,seconds.

Here we consider an alternative ICS scenario: the upscattering of NIR
seed photons emitted during the flare by the mildly relativistic $\sim
10\,$MeV electrons responsible for the quiescent radio-submillimeter
emission.  If the submillimeter emission region were optically thin
this would produce a similar X-ray luminosity as the upscattering of
submillimeter seed photons.  However, as the submillimeter source region is
optically thick below 1000\,GHz, the observed submillimeter flux is
produced by a fraction of the underlying electrons.  The emission
region is optically thin to NIR photons, and so all of these electrons
are available to upscatter NIR seed photons to X-ray energies.  As a
result, the ICS luminosity produced through this scenario will
dominate that produced by the original ICS scenario.

To estimate the resulting X-ray flux we characterize the electrons
responsible for the submillimeter emission by electron number density
$n_e$, a relativistic Maxwellian energy distribution at temperature
$T$, and a quasi-spherical region of size $R$.  These electrons
upscatter the NIR seed photons arising from synchrotron emission from
the relativistic electrons producing the NIR flare with observed flux
$S_\nu$ at the earth.  The ICS flux depends on the direction-averaged
intensity, $J_\nu$, of seed photons which in turn depends on the
location and size of the flare emission region; we estimate this to
order of magnitude by simply assuming that the flare region is of
similar size to the submillimeter emitting region, such that
$J_\nu = (d^2/\pi r^2) \, S_\nu\,,$
where $d=8$\,kpc is the distance to the Galactic center.
The energy of the upscattered photons is small compared to the
MeV-range of the electron energies, so we can use the Thomson
scattering cross-section.  To a good approximation, scattering by an
electron with energy $E\gg m_ec^2$ boosts the seed photon energies by
a factor $(E/m_ec^2)^2$ irrespective of the scattering angle.  The
differential ICS luminosity per unit energy interval is then simply
\begin{equation}
    L(E_\gamma)  = \frac{16\pi\sigma_T}{3h}\int N(E) J_\nu \,dE
    \label{eq:L_gamma}
\end{equation}
where $N(E)\,dE$ is the \emph{total} number of electrons in the energy
interval $[E,E+dE]$, and $J_\nu$ is the direction-averaged intensity
of the seed photons at frequency $\nu = E_\gamma/(E/m_ec^2)^2$.
For temperatures in excess of a few MeV, the vast majority of the
electrons have $v\approx c$, so we approximate the relativistic Maxwellian
distribution by
$f(E) \approx (2kT)^{-1}\, \left(E/kT\right)^2\,\exp\left(-E/kT\right)$
and then $N(E) = \frac{4}{3}\pi R^3 n_e f(E)$.
If the spectrum of the seed photons is a simple power-law, ie.\
$S_\nu = S_0 (\nu/\nu_0)^{-\beta}$,
the differential luminosity is
\begin{equation}
     L_\gamma(E_\gamma) = \frac{32\pi \sigma_T d^2}{9h}\,\Gamma(3+2\beta)
S_0
     \,  n_e  R
     \left(\frac{E_\gamma}{E_{\gamma0}}\right)^{-\beta}
\end{equation}
where $E_{\gamma0} = (kT/m_ec^2)^2 \, h\nu_0$.

By way of illustration we adopt reasonable choices $R=10\,R_s$, $n_e =
10^7$\,cm$^{-3}$, and compute the ratio of 2--10\,keV luminosity to
2.2\,$\mu$m flux as a function of spectral index $\beta$.  The results
for various adopted electron temperatures are plotted as solid curves in
Figure 22.  The X-ray to NIR flux ratio declines with increasing
spectral index $\beta$ (where $S_\nu \propto \nu^{-\beta}$ created by
upscattering of optical/NIR photons from an electron population with
radius $R=10\,R_s$, and uniform density $n_e = 10^7$\,cm$^{-3}$, and
temperatures of 3, 5, 7, and 10\,MeV).  Because X-rays in the
2-10\,keV band are produced by upscattering of photons that are
shortward of 2.2\,$\mu$m, and for fixed flux at 2.2\,$\mu$m there are
less of these as $\beta$ is increased.  Also shown for comparison are
the measured ratios and spectral indices of the 7 coincident IR and
X-ray flares seen to date (Yusef-Zadeh et al.\ 2006; Belanger et al.\
2005; Eckart et al.\ 2006; Hornstein et al.\ 2007; Marrone et al.\
2008; Porquet et al.\ 2008; Dodds-Eden et al.\ 2008; this paper). 
The high value of $\beta$ at high temperature correlates with 
a high ratio of X-ray to NIR flux, as shown in Figure 22.  
This correlation  is 
consistent  with the low flux value of NIR flare emission
for high value of the spectral index, as described in section 4.2. 
  We
conclude that the fluxes of the observed X-ray flares are broadly
consistent with this ICS scenario.

Our simple model predicts that the spectral indices of the X-ray and
IR flares should be identical, but this not need be the case for a
broken power-law electron energy distribution.  In the case of the
strongest X-ray flare (\#2) which coincides with a strong NIR flare,
the NIR spectral index and X-ray spectral index are different with
$\beta_{X-ray} = 1.3 \pm 0.3$ (90\% confidence) whereas $\beta_{NIR} <
1.0$ (3$\sigma$) \citep{dodds-eden09}.  This could be explained by a
broken power law of NIR emitting electrons with a steeper spectral
index shortward of 3.8$\mu$m, perhaps resulting from a shorter time
scale for synchrotron cooling of electrons at high energies.  In this
scenario, the flux in the 2-10\,keV band is produced predominantly by
upscattering of photons with wavelengths shortward of 1\,$\mu$m.  This
may explain why the width of the bright 2007 April 04 flare is less in
X-rays than in the infrared: the infrared flare may have decayed more
rapidly shortward of 1\,$\mu$m than at 3.8\,$\mu$m and so the X-ray
flux declines without a corresponding decrease at NIR wavelengths.

The extent of the submillimeter-emitting electron population may on
occasion give rise to significant time delay between infrared flaring
and their X-ray counterparts.  A sufficiently hard IR flare would lead
to X-ray production by inverse Compton scattering on the extended
$\sim1000\,R_s$ outer envelope of low-temperature electrons, producing
weak post-main-flare X-ray emission lasting for tens of minutes after
the main flare has subsided.  Theoretically, the electron temperature
is set by a balance between heating by Coulomb interactions with
protons and by plasma effects and synchrotron cooling.  kT for the
protons is a reasonable fraction of their virial energy because of
inefficient cooling in the accretion flow.  This implies that the
protons should be non-relativistic at $\sim1000\,R_s$, with MeV-range
energies.  The electrons are likely to have similar energies because
of formalization and are then mildly relativistic.  Empirically, Loeb
\& Waxman (2007) estimate from the radio/submillimeter spectrum that
the electron temperature is a few MeV all the way out to $1000\,R_s$.
Detection of these ``echoes'' would confirm the scenario proposed here
and help determine the size of the outer region which is rendered
inaccessible to direct by the effects of interstellar scattering in
radio wavelengths.

\subsection{Cross-Correlation of Light Curves}

As pointed out in the previous section the adiabatic expansion picture
of the flare emission from Sgr A* makes  the predictions that i) 
the NIR and X-ray emission are expected to 
be simultaneous and therefore optically thin whereas 
optical depth effects become  important at 
lower frequencies, thus 
a time delay is 
expected between their peak emission.  To examine these issues, 
 a great deal of  data have been obtained simultaneously in this 
campaign, which allows us to cross-correlate the multi-wavelength data for each
day of observation.  The light curves that we have presented thus far indicate  that the 
flux of Sgr A* is constantly changing 
as  there is low-level flare activity in almost all wavelength bands.
There are very few measurements that were taken simultaneously with the same 
time coverage with few exceptions, as described below. 

The  cross-correlation analyses  in this paper use the Z-transformed discrete correlation
function algorithm \citep{alexander97}; see also \citep{edelson88}.
This algorithm is particularly useful for analyzing sparse, unevenly
sampled light curves.  We identify the peak likelihood value, and a
1-$\sigma$ confidence interval around that value, using a maximum
likelihood calculation (Alexander 1997).

\subsubsection{NIR, X-ray and  Radio flare Emission}

Altogether there are four detected NIR flares that have shown X-ray counterparts.
There is no evidence that there is time delay between the peaks of any of the detected  
flares, thus supporting the fact that both NIR and X-ray  emission are  optically 
thin. The lack of time delay places a strong constraint on the ICS picture. 

The strong  flare and simultaneous  coverage in both X-ray and NIR wavelengths  
observed on 2007 April 4 is one example in which 
a cross correlation peak  with small error bars can be obtained. 
The cross-correlation of  the light curves at X-ray and NIR wavelengths
is shown in Figure 23. 
The  peak of the cross-correlation shows that X-ray emission  is  delayed by 
29 seconds with a one-sigma error bar  of  -6.5 and +7.0 minutes.
In the ICS picture, as  
described in section 4.4, the region from which NIR  photons are 
upscattered should be less than 0.8 AU.   
Dodds-Eden et al. (2009) presented first   the simultaneity of X-ray and L$'$bands to  within 
one sigma error bar of three minutes. 

The relationship between radio and NIR/X-ray flare emission has remained 
unexplored due the very limited simultaneous time coverage between radio, infrared 
and X-ray telescopes. The continued variations of the radio flux on hourly time 
scale also makes the identification of radio counterparts to infrared flares 
difficult. In spite of this, the strong flaring in NIR/X-ray wavelengths on 2007, 
April 4 has given us an opportunity to examine whether there is a correlation with 
variability at radio frequencies.  One of the key motivation of our observing 
campaign was to examine the adiabatic expansion picture of flaring activity of Sgr 
A*. One of the prediction of this model is a time delay between the peaks of 
optically thin NIR emission and optically thick radio emission. This implies a NIR 
flare with its short duration is expected to have a radio counterpart shifted in 
time with a longer duration. Given that there is zero time delay between NIR and 
X-ray light curves, as shown in Figure 23, we argue below for a radio counterpart 
to a strong X-ray flare by shifting and stretching the time axis of the X-ray 
light curve.

Figure 24a shows composite light curves of Sgr A* obtained with VLA, VLT, 
HST and XMM on 2007, April 4. 
 The flux increase at 43 GHz is 
$\sim$40\% which is higher than those from the first three days of VLA 
observations which is $\sim20$\%. We also know that there was no 
significant variation at 240 GHz during the period in which the strong 
NIR/X-ray flare took place. The IRAM-30m observation shows an average flux 
of 3.42$\pm$0.26 Jy between 5 and 6h UT when the powerful NIR flare took 
place. The flux is mainly arising from the quiescent component of Sgr A*.  
Comparing the light curves of the  43 and 240 GHz data, there is no 
evidence for a simultaneous radio counterpart to the NIR/X-ray flare with 
no time delays.

We now argue that the radio flare detected between 10h and 15h UT is a 
time-delayed counterpart to the NIR/X-ray flare for the following reasons: 
i) the highest percentage of the flux increase at 43 GHz on 2007, April 4 
compared to other three days of radio observations, ii) Similar morphology 
between radio and X-ray light curves as well the presence of three peaks 
in NIR and radio light curves and iii) the lack of significant flux 
variation above the quiescent flux of Sgr A* at 240 GHz during the 
NIR/X-ray flaring events. We suggest that flare emission at 43 GHz is time 
delayed with respect the NIR/X-ray flare emission. To explore this 
further, we have empirically shifted and stretched the time axis of the 
X-ray light curves by 5.25 hours and a factor of 3.5, respectively. The 
shift and stretch operation to the time axis is carried out by eye and 
then examined by cross correlating  the time-shifted and time-stretched 
X-ray data and the 43 GHz light curves. 
The top panel of Figure 24b presents the 
light curve of the 
time-shifted and time-stretched X-ray data. The middle plot shows a 
baseline subtracted radio light curve. The subtraction is used to remove 
the contribution by the quiescent flux. We find that the  best fit 
shows a peak in the cross correlation plot of 4.6$^{+9.4}_{-7.6}$ minutes 
which is consistent with zero. The 1-$\sigma$ error to the cross 
correlation peak of the shift is given in Table 2.

Given that NIR/X-ray and radio emission are expected to be optically thin and 
thick, respectively, the similarity in the substructures in radio and X-ray and 
NIR  light 
curves and the way that they trace each other, as shown in Figure 24, are 
remarkable. For example, two main peaks before and after 12h UT
are detected in both NIR, X-ray  and radio light curves. The 
dips near 13:30h UT 
and 13h UT between radio and NIR  are also seen in Figure 24a, 
 To make a stronger case that the X-ray/NIR  and radio flares are related to 
each other, the morphological agreements between radio and NIR could have been 
improved had we used a a varying time shift to different subflares 
or components in the NIR light curve.  
In fact, the
shift and stretch values measured here  is  not unique  as it is 
possible to decrease 
and increase the time-shifted and time-stretched  values, 
respectively,  
and yet 
obtain a reasonable zero time delay between radio and X-ray data.  
A detailed account of these light curves in 
the context 
of adiabatic cooling plasma model will be given elsewhere.

Given that there is continuous coverage for about ten hours between 5h-15h UT in 
X-ray, NIR and radio wavelengths using XMM, VLT, HST, IRAM and VLA with an 
exception of a two-hour gap between 7 and 9h UT in radio wavelengths and a 
2.5-hour gap between 10.5h and 13h UT in NIR wavelengths, we believe the lack of 
association between flaring activity in NIR and radio wavelengths is highly 
contrived. Obviously, we can not prove conclusively that radio flare seen on April 
4, 2007 is associated with the NIR/X-ray flare because of two gaps in our 
coverage. Nevertheless, the comparison of the 43 GHz light curve with the NIR data 
suggest that these  variations are tied closely with each other. 

\subsubsection{Other Cross Correlations}

There were no simultaneous observations that were taken with the same instrument 
except with the VLA and NMA but with a small frequency separation. In spite of the 
small separation between the observed frequencies at 134 GHz (2.23 mm) and 146 GHz 
(2.05 mm), the data are taken simultaneously on 2007, April 4 with the NMA.  The 
cross correlation of the light curves at these frequencies, as shown in Figure 
25, peaks with 6$^{+6.6}_{-4.8}$ minutes time delay.

Due to the limited UT coverage with individual telescopes as well as the lack of 
strong flaring event in this campaign (with the exception of the strong NIR/X-ray 
flare on 2007, April 4), the cross correlation of the light curves had difficulty 
following accurately the time evolution of a flare as a function of frequency. In 
spite of these difficulties, we have obtained cross-correlation plots of low-level 
fluctuations evident in four light curves. Although most of the individual cross 
correlation peaks have low signal-to-noise, the peaks of optically thick emission 
all show a tendency to lag rather lead, thus, consistent with the adiabatic 
expansion picture of flare emission. We have selected the best light curves to 
show the time lag but in fact almost all light curves systematically showed a time 
lag rather than a lead in their cross correlation peaks, though low signal to 
noise ratios.  We believe the data presented here supports the plausibility of the 
time delay, as has also been shown in earlier cross-correlation measurements. We 
give four examples that indicate higher probability that the peak flare emission 
at high frequencies leads those at low frequencies.


Figure 26a presents
the cross-correlation plot of the light curves taken at 450$\mu$m using the CSO and
230 GHz using the combined data taken from the SMA and SMT on 2007 April 3. 
The
cross-correlation plot at the
bottom of the panel  shows a maximum likelihood time delay at
1.32$^{+1.66}_{-0.69}$ hours.
Another example shows  the evidence for a time delay between 1.2mm (230 GHz)  and 
450$\mu$m wavelength bands on 2007, April 1 and the cross correlation plot 
is displayed in Figure 26b.
The cross-correlation peak between these wavelength bands 
is 0.24$^{+1.48}_{-0.04}$ hours time delay.

The cross-correlation peak between 1.70$\mu$m and 1.3mm wavelengths on 
2007, April 5 is shown in Figure 26c and gives a time delay 2.64 hours 
with a a 2$\sigma$ uncertainty range of -1.66 to 3.3 hours. The NIR data 
for this plot combined the 1.70$\mu$m data of HST (flare 5A of Fig. 3) and 
the 3.8$\mu$m VLT data (flare 6 in Dodds-Eden et al. 2009). Because the 
spectral index of the HST data is determined, we assumed that the 
preceding flare detected by the VLT at 3.8$\mu$m has the same spectral 
index and its duration is continuous with the brightest HST flare emission 
seen in this campaign. Lastly, Figure 26d shows the light curves taken 
with CARMA and VLA on 2007 April 02 at frequencies of 94\,GHz and 43\,GHz 
respectively. The cross correlation plot shows the strongest peak with a 
time delay of 1.02$^{+0.16}_{-0.31}$ hours. 
Given that there are three 
peaks shown in the cross correlation plot, there is ambiguity in the 
determination of the true time delay from the comparison of the peaks 
alone. However, when the cross correlation is considered in the context of 
many other cross correlations plots showing similar delays of peak 
emission at long wavelengths following those at short wavelengths, it is 
clear that the primary peak signifies most likely the real time delay.

As discussed before, the strongest NIR/X-ray flare  was detected on 
2007, April 4 showing a peak at 5.9h UT with a full duration of about two hours in NIR 
wavelengths.
The 2.1mm (140 GHz) light curve taken with IRAM during this period of  
flaring activity  placed a 
constraint by showing a lack of flux variation with a one-sigma error of 0.26 Jy 
at millimetre 
wavelength during 
a strong flaring activity.

\subsection{Adiabatic Expansion of Hot Plasma vs. Hot Spot Model}

We discuss two models that attempt to explain the nature of the flare
emission.  One is an expanding hot plasma model in which the peak
frequency of emission (e.g., the initial optically thin NIR flare)
shifts toward lower frequencies (submillimeter, millimeter and then
radio) as a self-absorbed synchrotron source cools adiabatically away
from the acceleration site \citep{shklovskii60,vanderlaan66,zadeh06b,zadeh08a}. 
 A variation of this model is a jet model in which the expansion speed of the
plasma is relativistic and is collimated in the form of outflow.  In
the expanding blob model, polarized flare emission does not follow
classical Faraday rotation and a frequency-dependent rotation measure
(RM) is predicted (Yusef-Zadeh et al.  2007).  The expanding blob
picture considers hot plasma being launched from the disk.  The
cooling plasma is dominated by the magnetic pressure as the plasma
escapes  or remains bound to the system.  The second model assumes that
flares are hotspots that are orbiting within few R$_s$ of the black
hole \citep{broderick06a} where Doppler boosting and GR effects become
important.  The hot spot picture requires the hot plasma be embedded
within the disk where the emission is dominated by the gas pressure in
the disk before the hot spot plunges into the hole.  Quasi-periodic
flaring events are expected under the assumption that hot spots
survive longer than the period of the last stable orbit.  There are
several issues that the hot spot model appears to be inconsistent with
observations.  One is the time delay between the peaks of flare
emission which is not expected in this picture.  It is possible that
the optically thin and thick flare emission is not related to each
other and that the hot spot model is applicable only to the NIR flare
emission.  However, recent measurements as well as measurements presented in 
this campaign indicate time delay 
between
the peaks of NIR/X-ray and submillimeter flare emission
\citep{zadeh08a,marrone08b,eckart09}.

Another difficulty with the hot spot model is the lack of evidence for
power on the quasi-periodicity in  NIR light curve (Do et al.
2008, Meyer et al.  2008).  Claims of quasi-periodic variations in the
NIR lightcurves were the motivating observation for the hotspot model,
but the most thorough analysis of NIR lightcurve variability have not
shown evidence for significant quasi-periodic power
\citep{do08,meyer08}.  Furthermore, MHD simulations of accreting gas
indicate that hot spots can last less than an orbital time scale
before they disperse \citep{hawley02}. 

A  major motivation for the observational study of Sgr A* in the 2007 
observing campaign was  to test predictions of the plasmon model of
flare emission such as the time delay between the peak emission at
different wavelengths. 
 Previous observing campaigns to monitor Sgr A* have found evidence
for time delay between the peak emission at 43 and 22 GHz.  
(Yusef-Zadeh et al.\  2006b; Yusef-Zadeh et al.\ 2007a).
These measurements were  consistent with the predicted time delays in the plasmon model.
As described above, we  carried  out the cross correlation of the
peak of radio flare emission and the emission at other wavelength
bands using simultaneous space- and ground-based observatories.  These
measurements would  allow us to determine if radio flares are correlated
with flaring in the near-IR, X-ray and sub-millimeter.
The data presented here show an
increasing chance of high frequency flare emission leading the low
frequency emission when simultaneous data between 43 GHz, 94 GHz, 230GHz,
0.45mm and 1.70$\mu$m are examined.  It is only the collection of the
cross correlation plots that make the time delay between the peaks of
flare emission compelling.

Another intriguing example that could be used against the orbiting hot spot model 
is the relationship between radio and NIR/X-ray flare emission. 
The light curve 
at 43 GHz using the VLA began covering 
the flare 4.5
hours after the start of the NIR flare on 2007 April 4. 
The morphology of the peaks
in the two light curves appears to show a rise of flux followed by
flattening of the emission.  Figure 24b shows the comparison of the
X-ray and radio  light curves by shifting the X-ray light curve by about 5.25  hours and
stretching by a factor of 3.5.  We suggested  that shifting and
stretching of the light curves serve as the time delay and duration of
flare as it evolves in time.  
The stretching of the NIR light
curve by a factor of 3.5 can be  viewed in the context of the
expanding blob model of an initial 
 flare or a compact blob observed in NIR followed by
the expansion of a blob of hot plasma emitting in radio wavelengths.
Individual NIR and radio flares show typical durations of $\sim$20 min 
and 1-2 hours, respectively. 
The ratio of the observed durations is similar to
the stretching factor that was applied to the time axis of the NIR light curve 
of  flare emission.  A more detailed account of this analysis in the context of the 
plasmon model will be given 
elsewhere. 

Recently, Marrone et al. (2008) criticized the expanding blob model on
the grounds that their measured ratio of 1.3\,mm to 850\,$\mu$m flux
during a flare on 2006 July 17 was higher than expected from the blob
model, both in the optically thick precursors (where one would expect
a spectral index of 5/2 rather than the measured value
$\beta=0.1\pm0.5$), and in the ratio of the amplitudes of the flares
at the two wavelengths.  However, the continual variability at radio
and millimetre means that there is a large uncertainty in determining
the underlying background flux level for a particular flare, and we
have determined that there are reasonable choices of the levels that
renders flare profiles that are consistent with the plasmon framework.
Fig.\ 27  demonstrates simultaneous fit to both the 1.3\,mm 
and  850\,$\mu$m flux of the 2006 July 17 flares (see Figure 3 of Marrone 
et al.  2008) using the plasmon model.    The parameters of the successful 
fits  to the  first  flare peaking near
5.7\,hr UT are  $p=1$, $R_0 = 0.52\,R_s$, $v=0.011c$, and $B=73\,G$ 
whereas the parameters of the strong flare peaking near 
7.5\,hr UT are  $p=0.5$, $R_0 = 0.42\,R_s$, $v=0.003c$, and $B=75\,G$. 
These fits show clear evidence that a simple picture of plasmon model 
can easily  be applied to previously published light curves in 
submillimeter wavelengths (see additional fits to light curves in 
Yusef-Zadeh et al. 2008). 

The other criticism of \citet{marrone08b} is that the model requires
adiabatic expansion of the blobs at speeds $\sim0.03$c much below than
the canonical sound speed $c/\sqrt{3}$ approaching the black hole that
would be expected were the blobs filled with plasma and sitting in
vacuum.  Given that the blob diameters are in the range of several
$R_s$ they may well be located at 10 or more $Rs$, where the local
sound speed would be $\sim 0.1\,c$.  In any case, they may be embedded
in the outer layers of an accretion flow where they would be only
mildly overpressured with respect to their surroundings and the
expansion time scale would be comparable to the buoyancy or orbital
time scale.

\subsection{Distribution of Electrons in the Flares}
Previously we have modeled the time delays at submillimeter to radio
frequencies in the expanding hot plasma  model assuming that the accelerated
particles have a power-law energy distribution.  This is motivated by
the long time scale of the flares compared to the synchrotron loss
time for the expected magnetic field strengths of 10--30\,G. These
models assume a homogeneous sphere threaded by a uniform magnetic
field.  As the region expands, the relativistic particles cool by
adiabatic expansion with $E\propto 1/R$ and the magnetic field is
diluted as $B\propto R^{-2}$ because of flux freezing.  Assuming that
the relativistic electron energies run between 1\,MeV and 100\,MeV and
that they are in equipartition with the magnetic field, the models are
characterized by the particle spectral index, $p$ (with $n(E)\propto
E^{-p}$), the expansion speed $v$ (assumed constant), and the timing
and amplitude of the flaring at a single frequency.  From this we can
infer a magnetic field strength $B$ and the size of the emitting region
at $t_0$, $R_0$.  Figure  28a shows an approximate fit to the CARMA
and VLA data obtained on 2007 April  02 using two flares.  The derived
parameters of the first flare at 10.9\,h UT are $p=0.5$, $R_0 =
3.6\,R_s$, $v=0.070c$, and $B=15\,G$, while the 
later flare at
14.7\,hr UT has $p=1.5$, $R_0 = 9.8\,R_s$, $v=0.065c$, and $B=13\,G$.
These numbers should be regarded as illustrative given the rough
fitting, the simplicity of the model, and the freedom in choosing the
baselines at each frequency.

While a power-law electron spectrum is plausible, the inferred spectra
are significantly harder than the $E^{-2}$ expected on the basis of
the simplest version of diffusive shock acceleration.  This suggests
that the derived power law may instead be the effective power-law of
the particle energy spectrum over the small (5\%) range of initial
energies relevant to our observing frequencies at 96\,GHz to 43\,GHz.
Other spectral forms are easily introduced within the context of this
model.  
By way of example, in Fig.\ 28b we show the ``best''
relativistic Maxwellian model, with the particle spectrum
characterized by the electron temperature $T_e$ at time $t0$ instead
of $p$.  The parameters in this case ($kT_e=0.25$\,keV, $R_0 =
4.4\,R_s$, $v=0.077c$, and $B=10\,G$; $kT_e=4.1$\,MeV, $R_0 =
10.4\,R_s$, $v=0.091c$, $B=12$\,G) yields similar emission region
characteristics but is  worse in matching the 43\,GHz
data.  
This is because the synchrotron spectrum is more strongly
peaked than for a power-law electron population.  As a result, the
flare amplitude declines more rapidly at successively lower
frequencies than is the case for power-law models (except for large
choice of $p$).   Future simultaneous light curves with better 
time coverage are needed to confirm this results.

\section{Conclusions}

The main results of extensive observing campaign that took place in 2007 
can be summarized as follows: 

\begin{description}
\item[]
Simultaneous VLA and VLBA observations indicate that flare emission from Sgr A* at 43 GHz
arises from  within the scattering size  of Sgr A* 
which is $\sim0.3\times0.7$mas \citep{bower04} or within the inner 30$\times$70 R$_{s}$ of 
Sgr A*.

\item[]  We show the evidence of  varying spectral index values when 
weak and bright NIR flares are compared. In addition, 
the NIR flare statistics indicate that the probability of flare emission 
is proportional to the inverse of the flux density. 
Simulations of the histogram of such flares 
assuming   uniform distribution of  peak flare emission is  consistent
with observations.  The significance of the probability of flare 
emission is inversely proportional to the flux of flare is not understood. 

\item[] 
In addition to a powerful X-ray flare with a NIR counterpart and 11.8$\mu$m  upper limit 
on 2007 April 4 that had been 
reported  
earlier by Porquet et al. (2008), Dodds-Edden et al. (2009), and  Trap et al.\ (2009),  
we show evidence of  three  new X-ray flares with NIR counterparts. 
The origin of X-ray production  is explained in the context of ICS employing the structure details 
of the Sgr A* emitting region inferred from   intrinsic size measurements.  
In this picture,   the seed 
photons  associated with  flares
 in NIR wavelengths are upscattered by the sea of  electrons that are responsible for 
the quiescent emission  of the Sgr A* in radio and submillimeter wavelengths. 
A prediction of this model is a 
time delay between the peaks of X-ray and NIR flare emission.

\item[] The comparison of the  light curves at  multiple wavelengths indicated 
time delays implying optically thick emission.  We also argue a tantalizing 
radio flare three hours after the strongest NIR and X-ray flare detected 
on 2007 April 4. 
These measurements are consistent 
with an adiabatic  expansion of  hot plasma. Although these measurements weaken 
the hot spot model of flare  emission, we can not distinguish whether there is 
jet activity associated with the observed time delays or the expansion 
 of hot plasma that is bound 
to Sgr A*.

\end{description}

Acknowledgments:

This work is partially supported by the grant  AST-0807400 from the National Science Foundation.
Some of the data presented here 
were
obtained from Mauna Kea observatories. We are grateful to the
Hawai'ian people for permitting us to study the universe from
this sacred summit.  Research at the Caltech Submillimeter Observatory
is supported by grant AST-0540882 from the National Science Foundation. 
Research grants are also given by Australian Research Council
(DPO986386)  and Macquarie University.  
The SMT is operated by the Arizona Radio
Observatory (ARO), Steward Observatory, University of Arizona.
The XMM-Newton project is an ESA Science Mission with instruments
and contributions directly funded by ESA Member State and the USA (NASA).

\bibliography{sgra}
\include{sgra}


\begin{deluxetable}{lrrr}
\tablecaption{Spectral Index Distribution Using NICMOS}
\tabletypesize{\scriptsize}
\tablewidth{0pt}
\tablehead{\colhead{Event} & \colhead{F(1.45$\mu\rm m\pm\sigma$)} & \colhead{F(1.70$\mu\rm m\pm\sigma)$} 
&
\colhead{$\beta\pm\sigma$}}
\startdata
 5A  & 8.55$\pm0.4$ & 9.61$\pm0.4$ &   0.73$\pm$0.39\\
 2A  & 6.77$\pm0.6$ & 7.92$\pm0.5$ &   0.97$\pm$0.68\\
 2C  & 4.54$\pm0.1$ & 6.54$\pm0.7$ &   2.29$\pm$0.27\\
 4A  & 4.77$\pm0.1$ & 6.14$\pm0.3$ &   1.59$\pm$0.33\\
 5B  & 2.31$\pm0.5$ & 3.62$\pm0.4$ &   2.82$\pm$1.51\\
 7A  & 2.85$\pm0.3$ & 3.63$\pm0.2$ &   1.52$\pm$0.74\\

\enddata
\end{deluxetable}

\begin{deluxetable}{lcccccccccc}
\tabletypesize{\scriptsize}
\setlength{\tabcolsep}{0.02in}
\tablewidth{0pt}
\tablecaption{Measured Time Lags with 1-$\sigma$ Errors}
\tablehead{
\colhead{Date} &
\multicolumn{4}{c}{Time Delay}&\\
\colhead{2007} &
\colhead{84-43} &
\colhead{134-146} &
\colhead{450-230} &
\colhead{1.70-230} &
\colhead{X-ray-43} &
\colhead{3.8-X-rays} & \\
\colhead{April} &
\colhead{(GHz-GHz)} &
\colhead{(GHz-GHz)} &
\colhead{(\micron-GHz)} &
\colhead{(\micron-GHz)} &
\colhead{(2-10keV-GHz)}\tablenotemark{a} &
\colhead{(\micron-2-10keV)} &\\
\colhead{} &
\colhead{(hr)} &
\colhead{(min)} &
\colhead{(hr)} &
\colhead{(hr)} &
\colhead{(min)} &
\colhead{(min)} &\\
}
\startdata
1  & --- & ---  &0.24(-0.29, +1.12) & --- & ---  & ---  \\
2  & 1.02(-0.31, +0.16)&  & --- & ---    & --- \\
3  & --- &  &1.32(-0.63, +0.33) & & --- & --- \\
4  & ---  & 3(-8.0, +3.4) &---& --- & 4.6(-7.6,+9.4) & 0.5(-6.5,+7)   \\
5  & & --- & --- &  2.64 (-0.67, +0.5) & ---   & --- \\
\enddata
\tablenotetext{a}{The X-ray data is time-shifted by 5.25 hours and time-stretched by 3.5} 
\end{deluxetable}

\vfill\eject

\begin{figure}
\centering
\includegraphics[scale=0.7, angle=0]{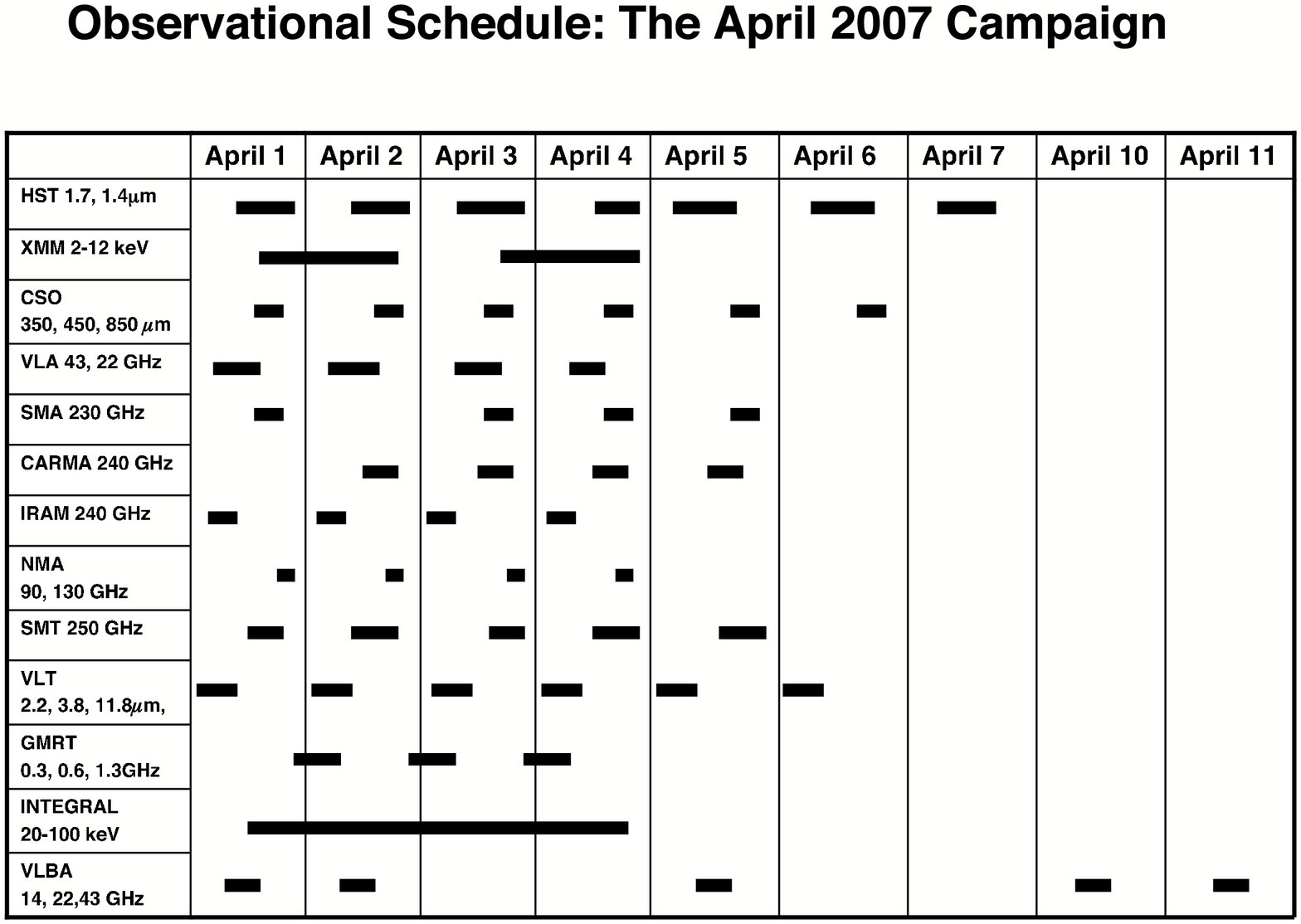}
\caption{A schematic diagram showing 13  telescopes 
joining the 2007 April campaign. The width of individual observing
period is not scaled. }
\end{figure}

\begin{figure}
\includegraphics[scale=0.5, angle=0]{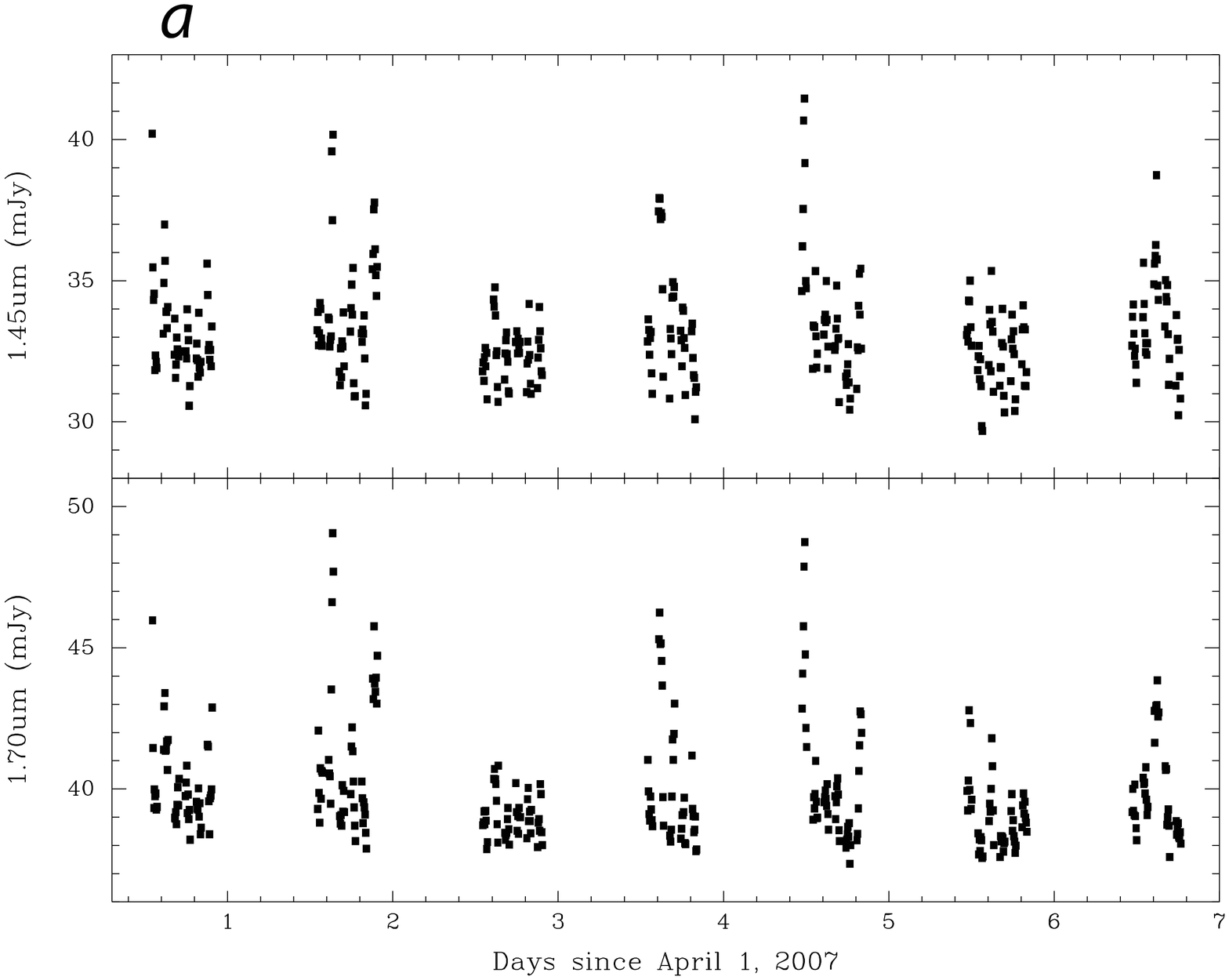}\\
\includegraphics[scale=0.5, angle=0]{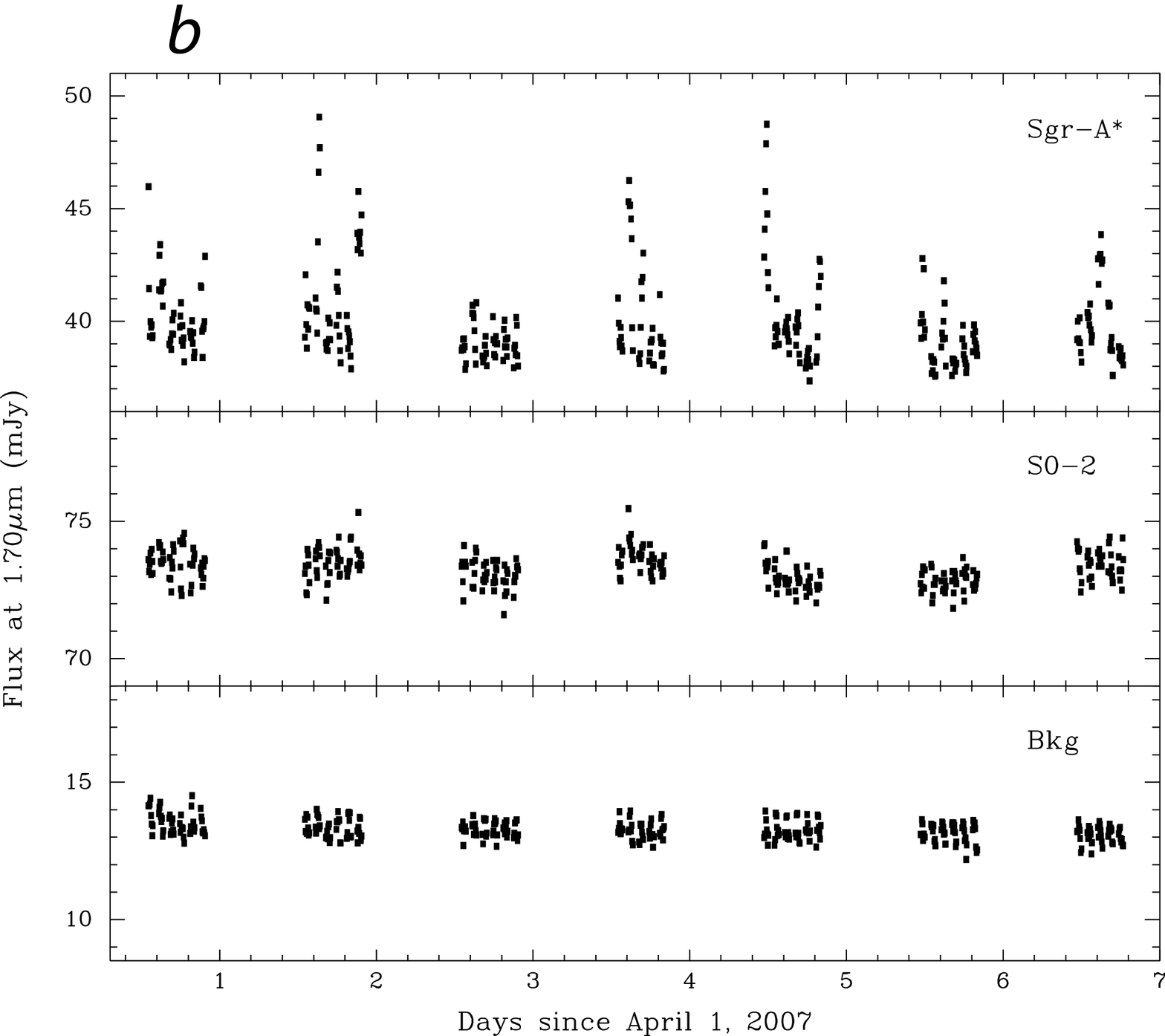}
\caption{
(\textit{a}) The light curves of \sgra\ at 1.45$\mu$m and 1.70$\mu$m 
for the seven windows of HST observations on 2007 April 1--7. No background 
flux is removed from these plots. 
(\textit{b})  The HST 1.70\mic\ light curves of \sgra, the star S0-2,
and a region of background emission.
The constancy of the S0-2 and background light curves strongly suggest
that the variability of \sgra\ emission is intrinsic to \sgra . 
}\end{figure}

\begin{figure}
\centering
\includegraphics[scale=0.3, angle=0]{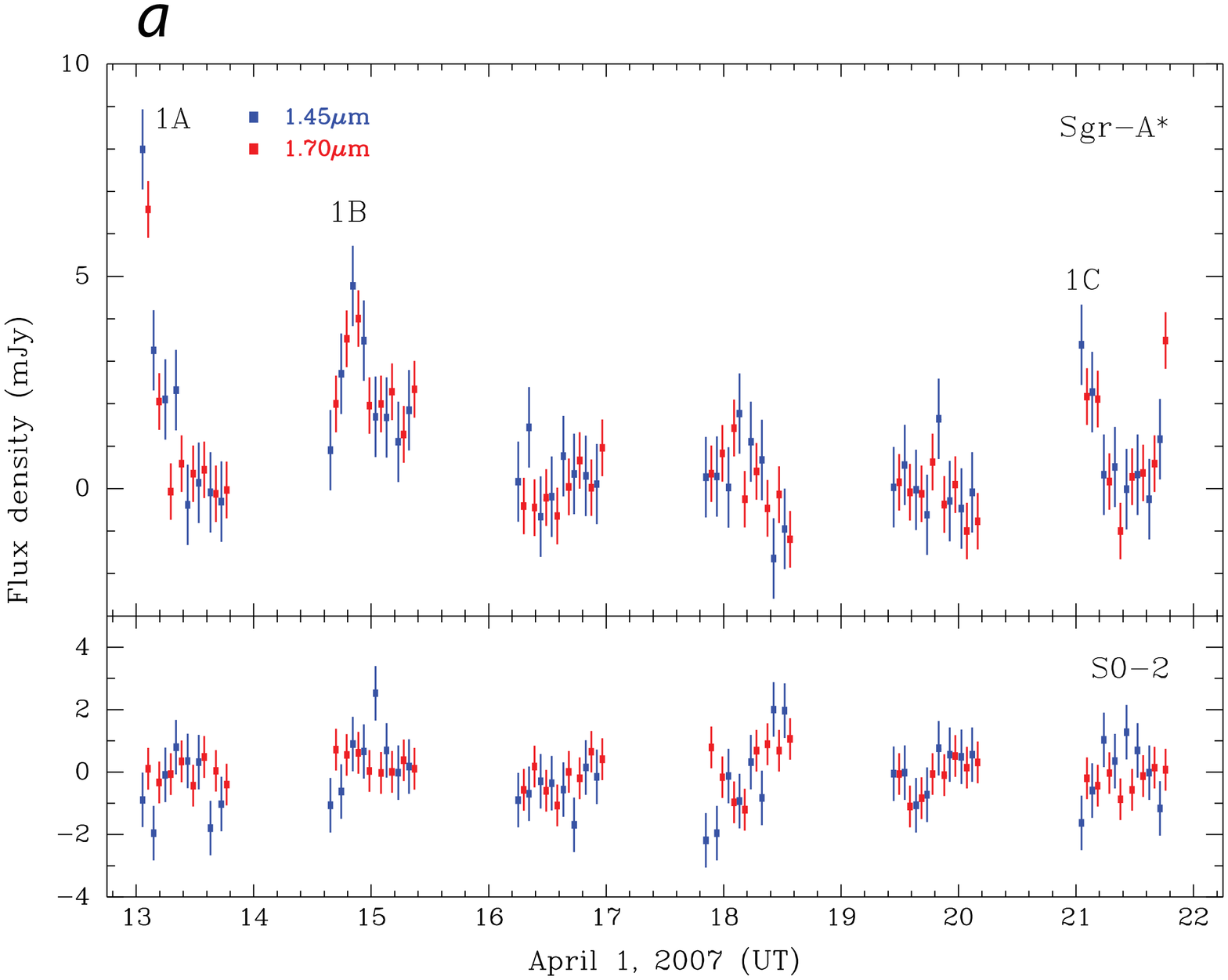}
\includegraphics[scale=0.3, angle=0]{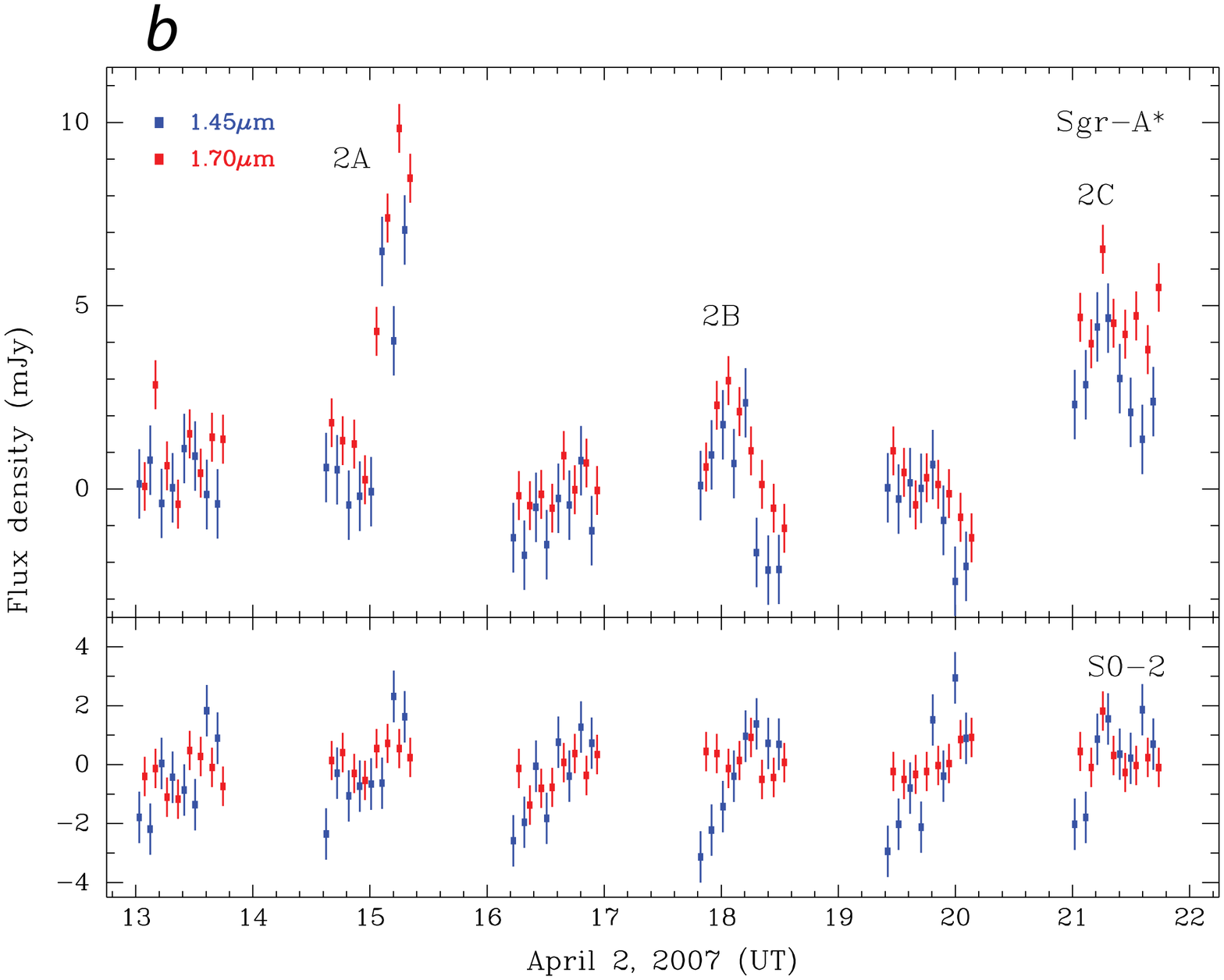}\\
\includegraphics[scale=0.3, angle=0]{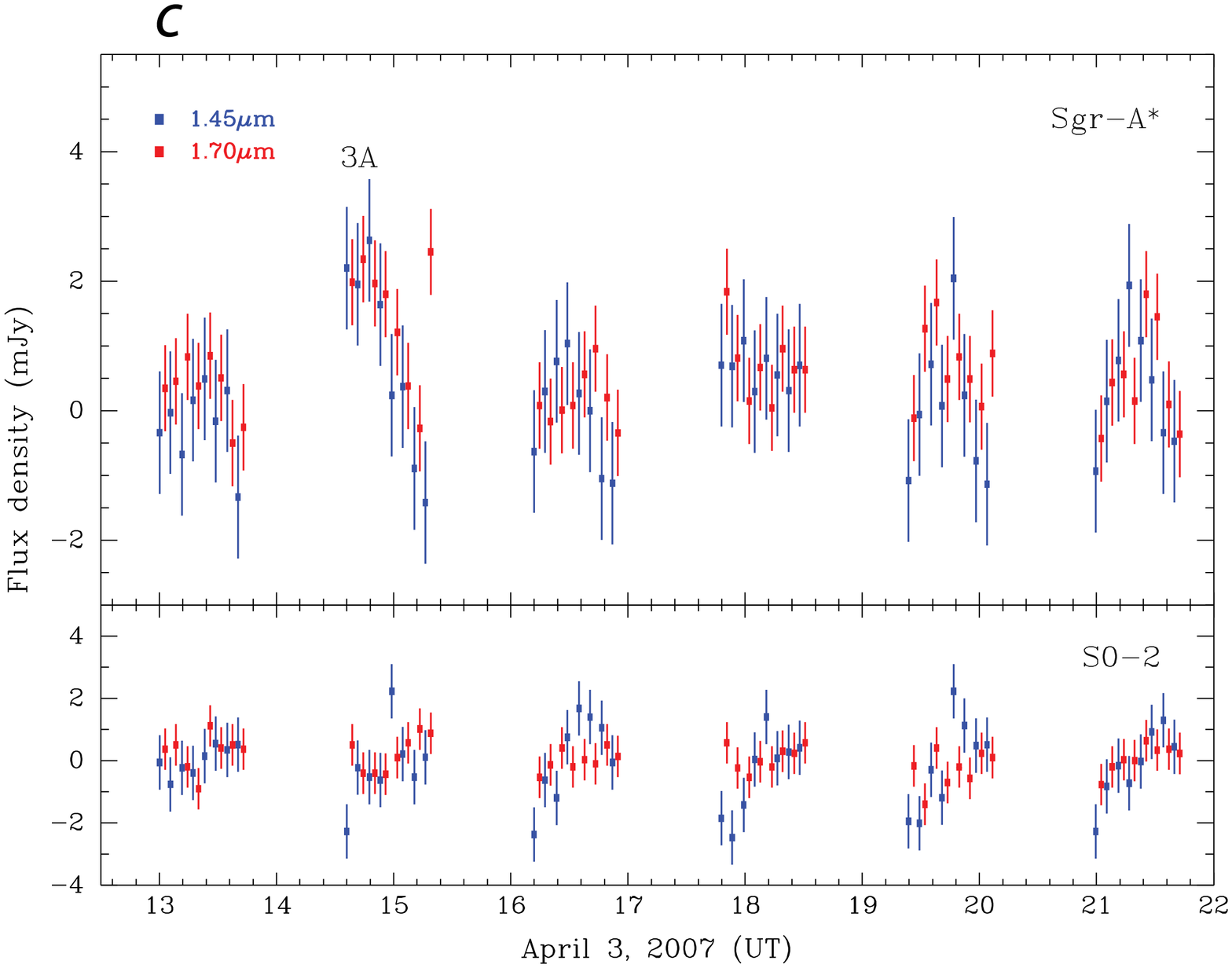}
\includegraphics[scale=0.3, angle=0]{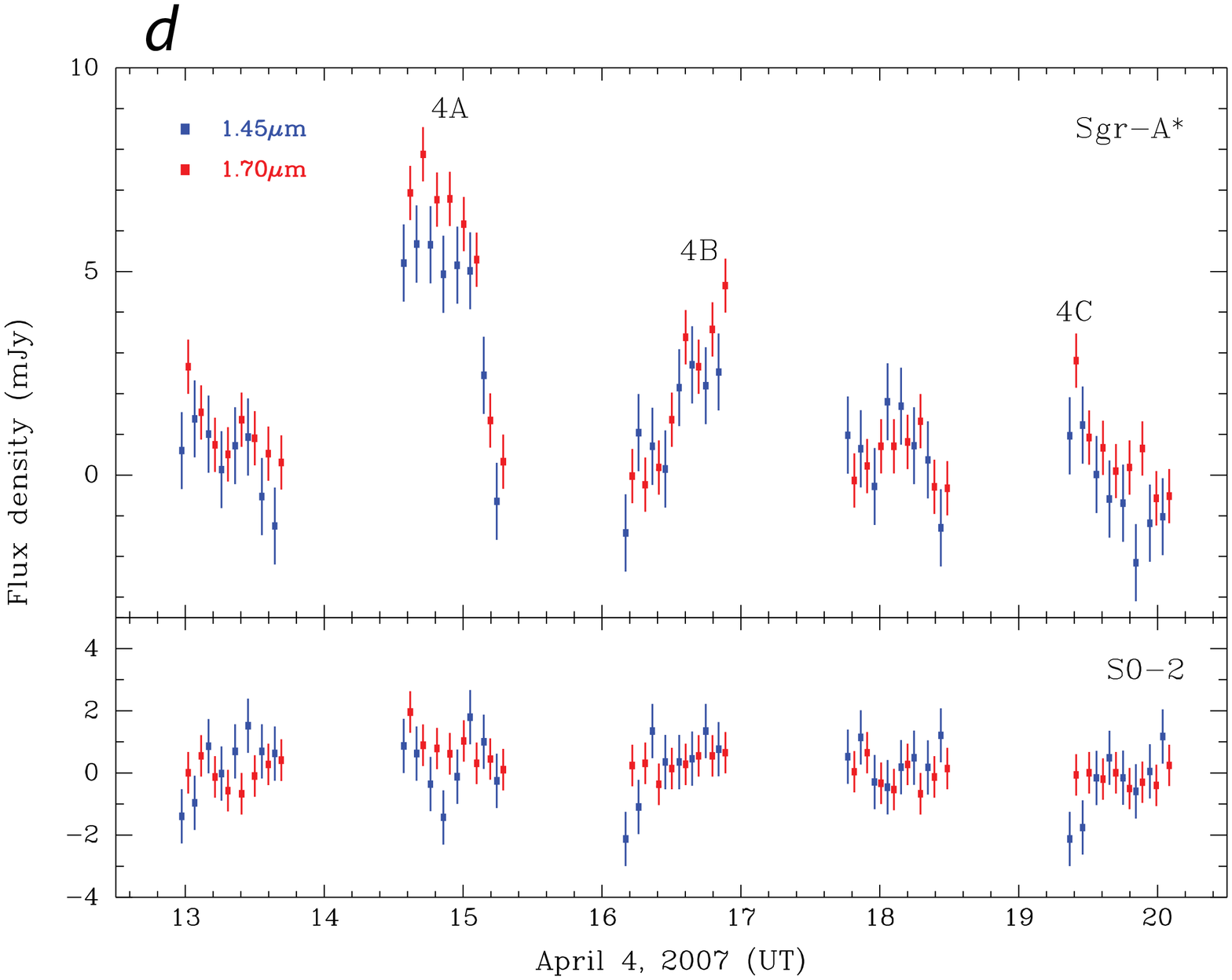}\\
\includegraphics[scale=0.3, angle=0]{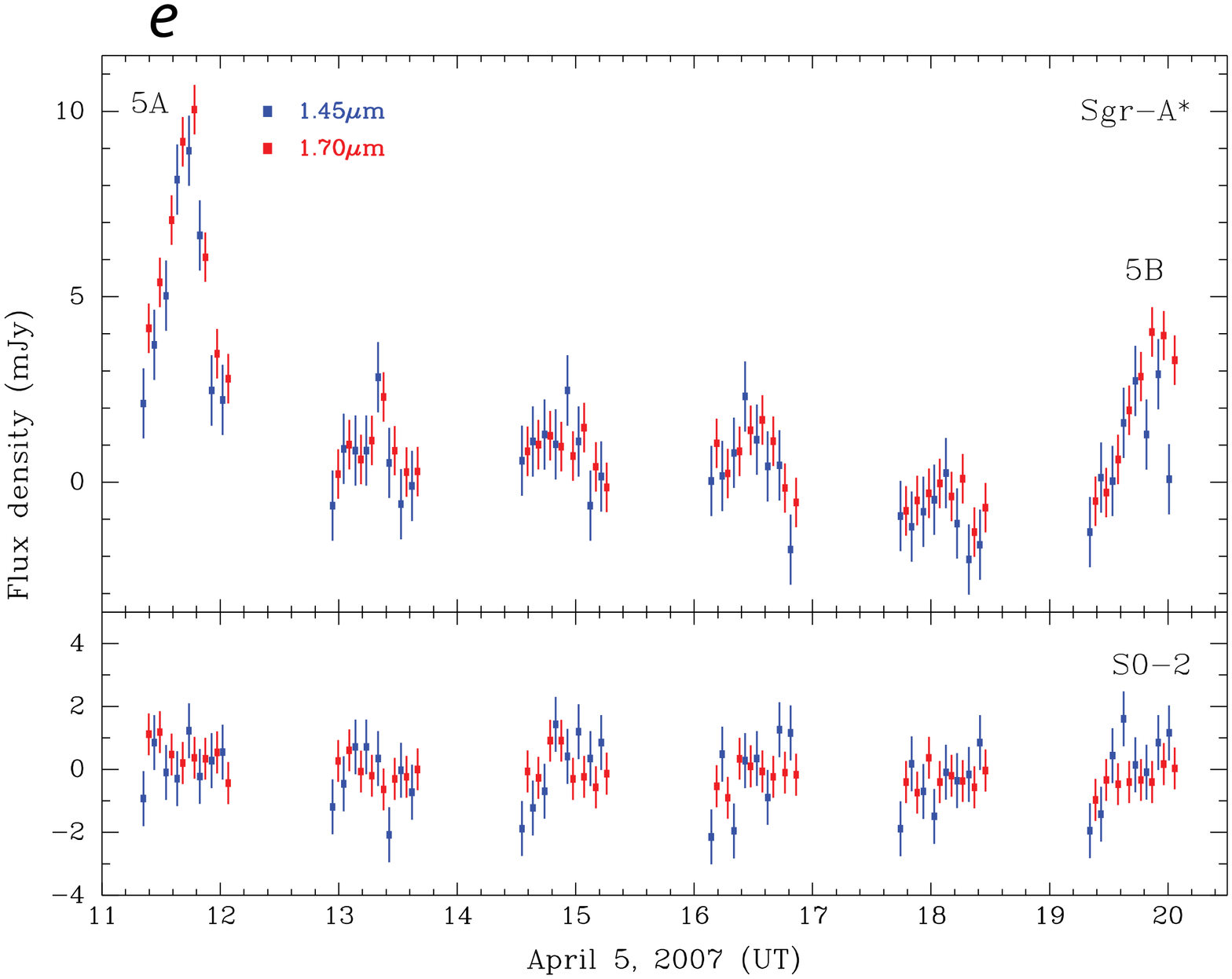}
\includegraphics[scale=0.3, angle=0]{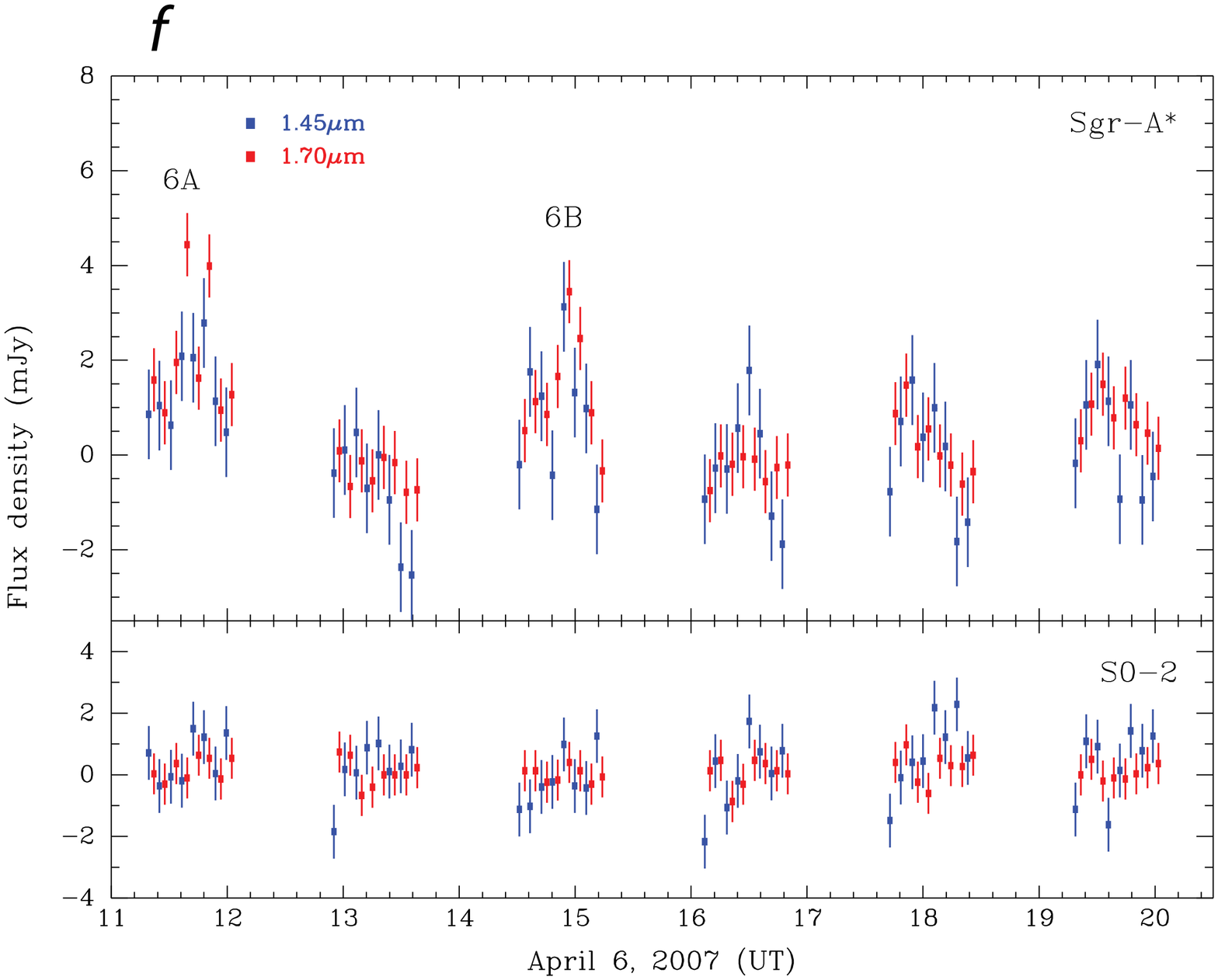}
\includegraphics[scale=0.3, angle=0]{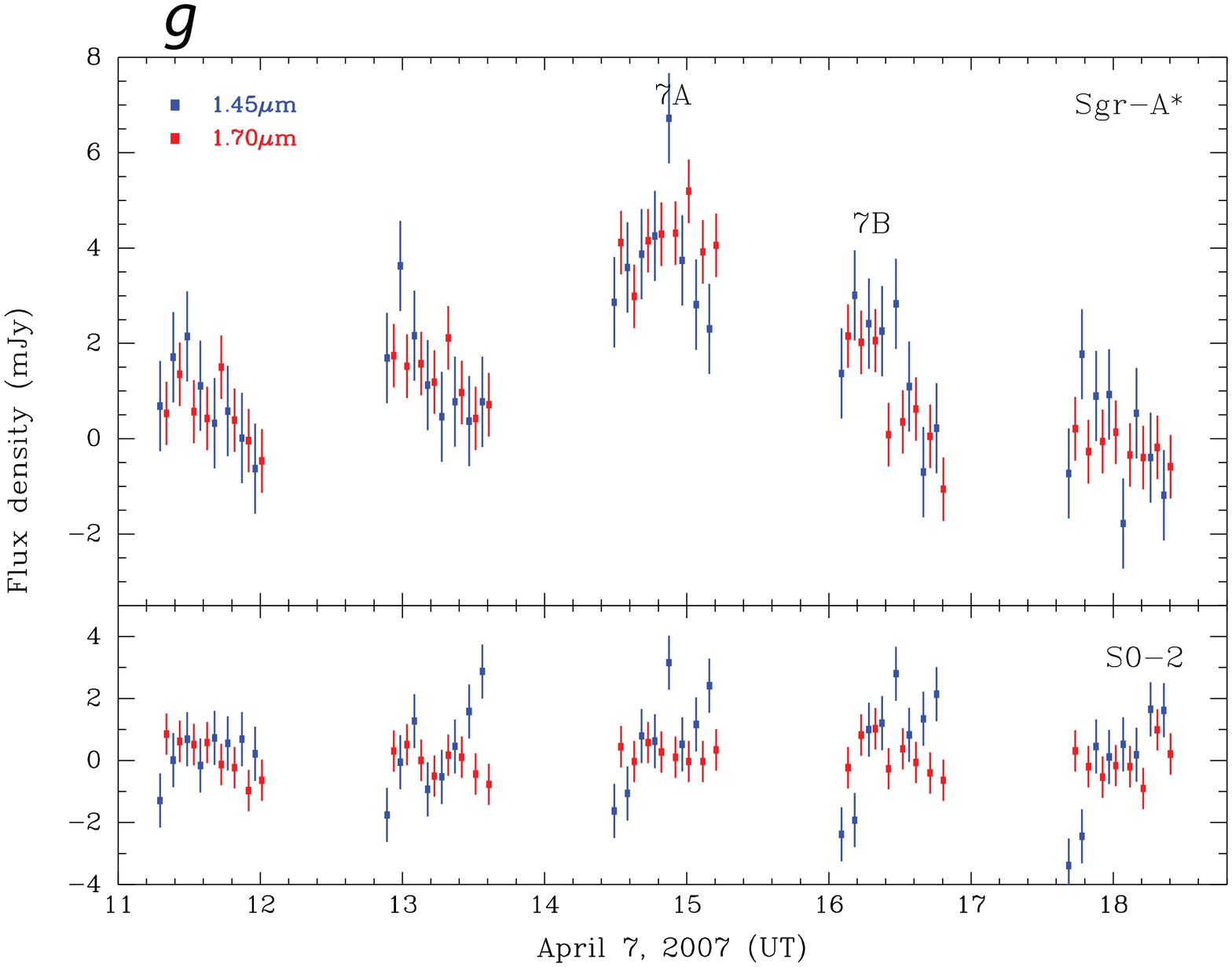}
\caption{
The background subtracted light curves of \sgra\ and S0-2 for each of the seven 
HST observing
windows, with flare events labeled. The data points at 1.70 and 1.45 
$\micron$ are sampled at 144
sec intervals. 
}\end{figure}

\begin{figure}
\centering

\includegraphics[scale=0.6, angle=0]{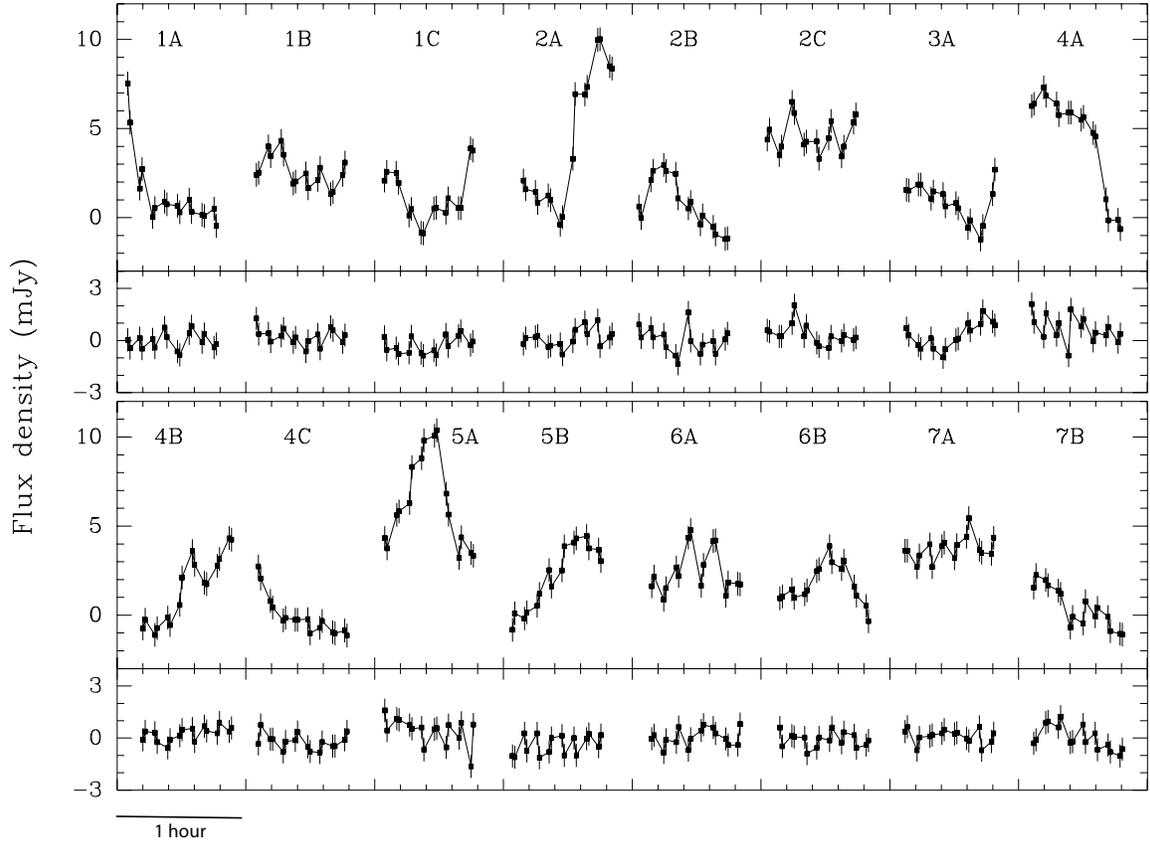}
\caption{ This plot shows all the 
64-sec sampled data for  the 16 periods  that flares are identified. 
The 16 periods are stacked on the top and bottom panels. 
The x-axis is the elapsed time - each flare episode occupies a
45-minute slice within the 8-hour axis. 
Each flare event is labeled, as defined in Figure 3, 
The Sgr A* and S0-2 light curves are on the upper and lower portion of the panel. 
The data show  only the 1.70\mic\ photometry
sampled at intervals of 64 sec but  the light curves  for each flare are laid 
side by side. 
}\end{figure}

\begin{figure}
\centering
\includegraphics[scale=0.4, angle=0]{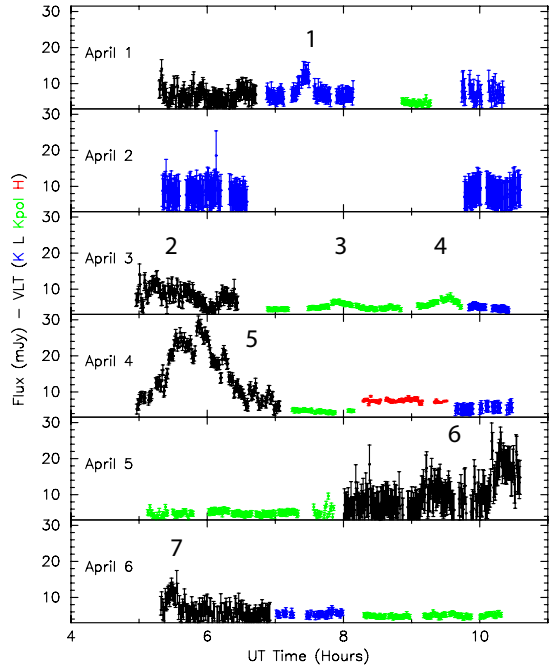}
\caption{Light curves of SgrA* 
for April 1-6 at 
H (1.66 $\mu$m), in red,  K$_s$ (2.12$\mu$m), in blue, K$_s$ in polarimetric mode in 
blue and L' (3.8$\mu$m), in black,  bands. 
are taken from  Dodds-Eden et al. (2009).  There are a total of 
seven periods of flaring activity reported in these observations. 
The brightest flares occurred on 2007, April 4 and April 5. 
}\end{figure}

\begin{figure}
\centering
\includegraphics[scale=0.4, angle=0]{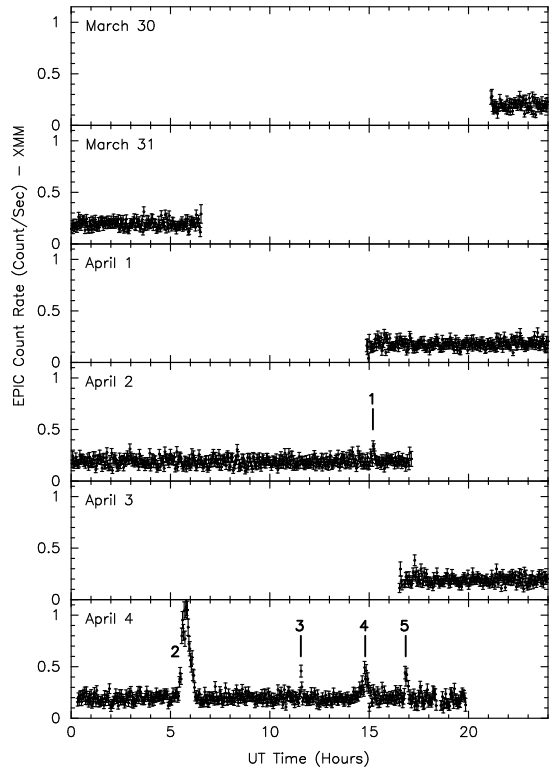}
\caption{Light curves of all the X-ray data taken with the 
XMM-Newton  during the 2007 April observing campaign (Porquet et al. 2008).
The data are  averaged  over  a 144sec sampling.  Five X-ray flares 
are  detected, four of which had simultaneous coverage with the VLT and HST and showed 
NIR  counterparts. 
}\end{figure}

\begin{figure}
\centering
\includegraphics[scale=0.35, angle=0]{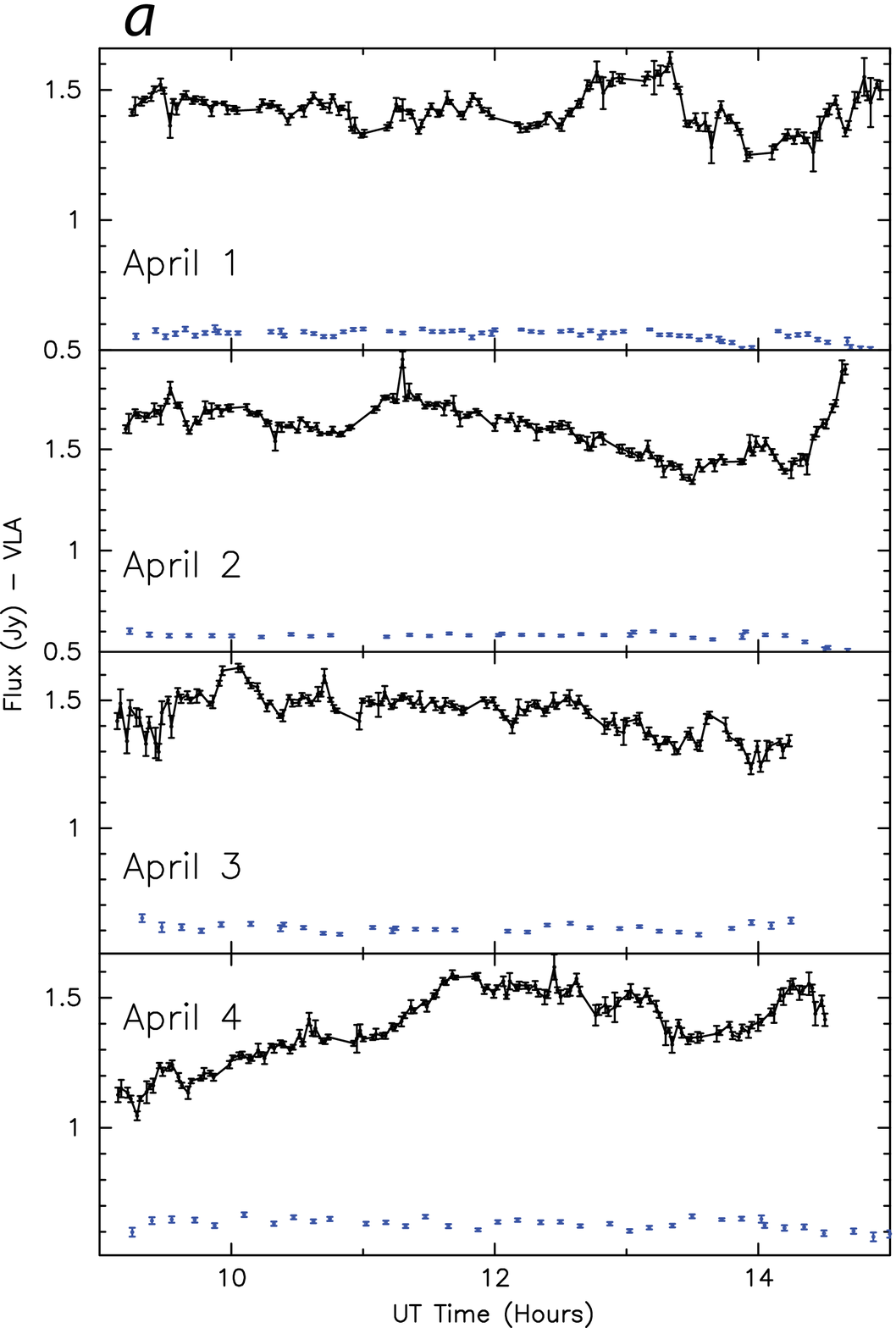}
\includegraphics[scale=0.35, angle=0]{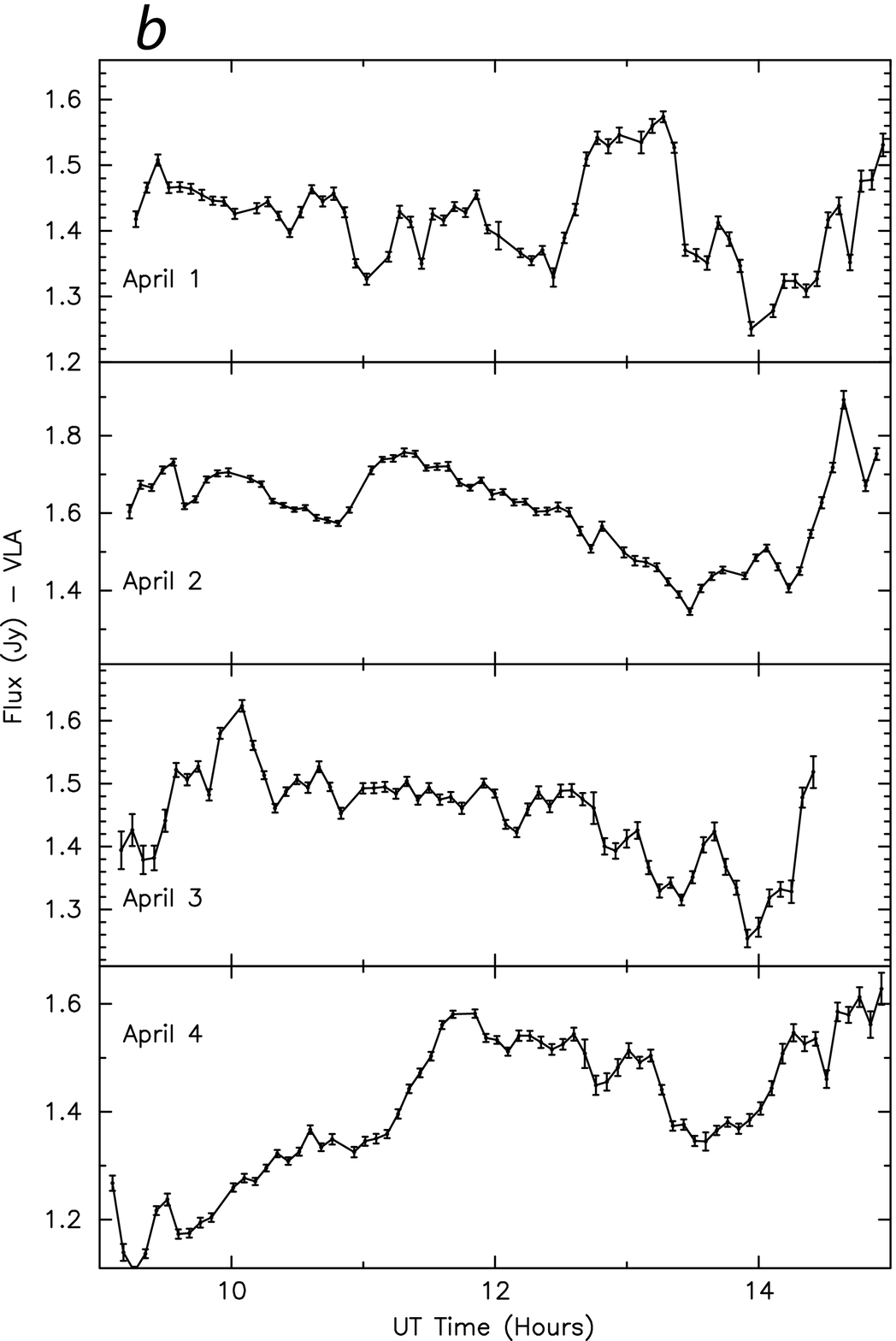}
\caption{
(\textit{a - Left}) 
 Light curves  of  Sgr A* and the calibrator 17444-31166 at 
43 GHz data 
obtained with  VLA observations taken during 2007, April 1-4. 
The sampling time is 87s 
for Sgr A*
and 90 sec for the calibrator at the bottom of each 
panel. The  Sgr A* {\it uv} data is restricted to $>$100k$\lambda$ in order to 
suppress the contribution of extended emission. 
(\textit{b - Right})  Similar to (a) 
except that only  the light curves of Sgr A* are  shown 
with a sampling time of 300 sec.
}\end{figure}

\begin{figure}
\centering
\ContinuedFloat
\includegraphics[scale=0.35, angle=0]{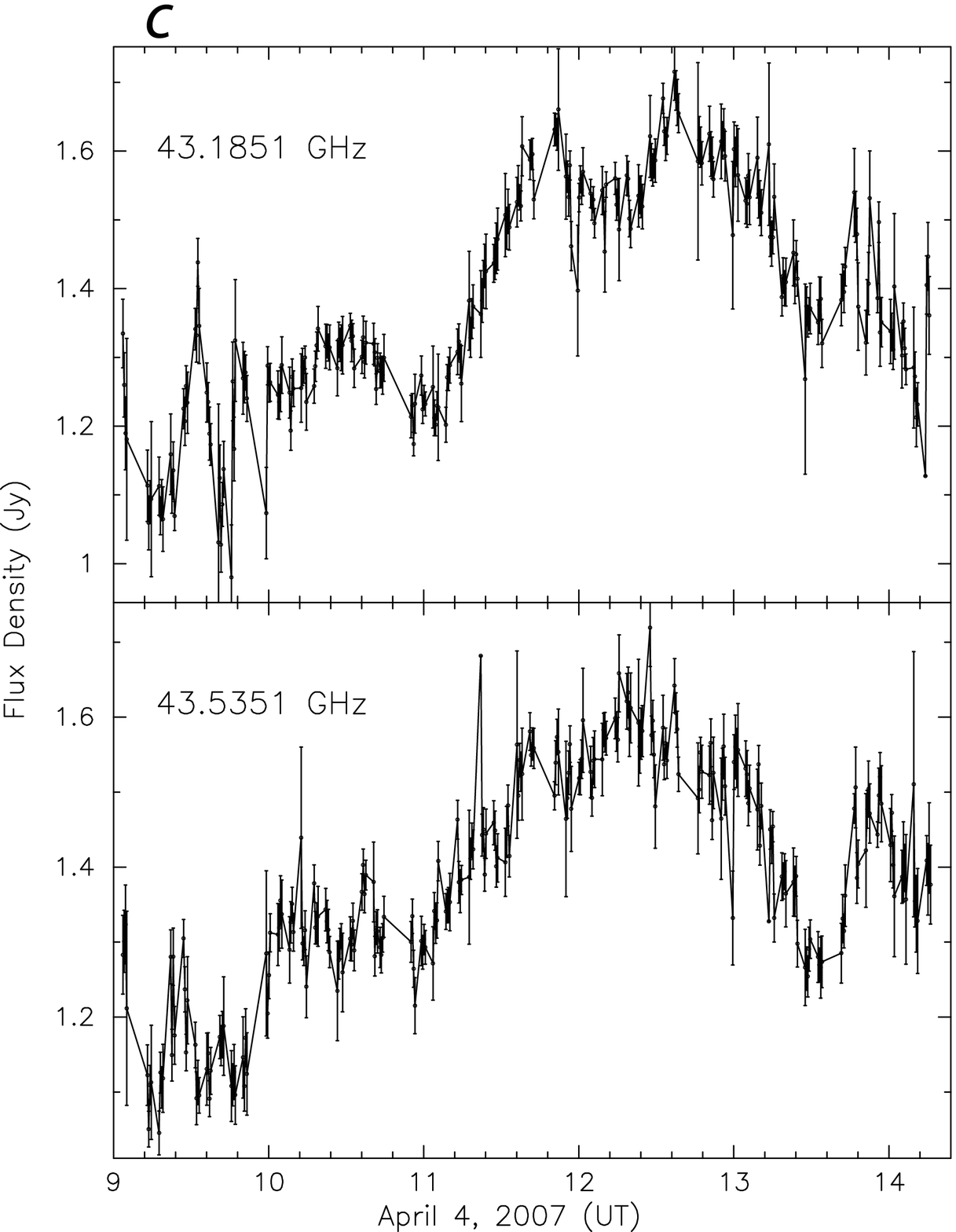}
\caption{
(\textit{c}) 
The light curves of Sgr A* on 2007, April 4 
are shown in the top and bottom panels
at 43.185 GHz and 43.535 GHz, 
respectively.  The light curves are shown with a 30s sampling 
time.The data points with large error bars   correspond to 
a small number of data points in a given sample.  
The data corresponding to minimum and maximum  {\it uv} baselines   are selected between    110 and  125 
k$\lambda$.
}\end{figure}

\begin{figure}
\centering
\includegraphics[scale=0.5, angle=0]{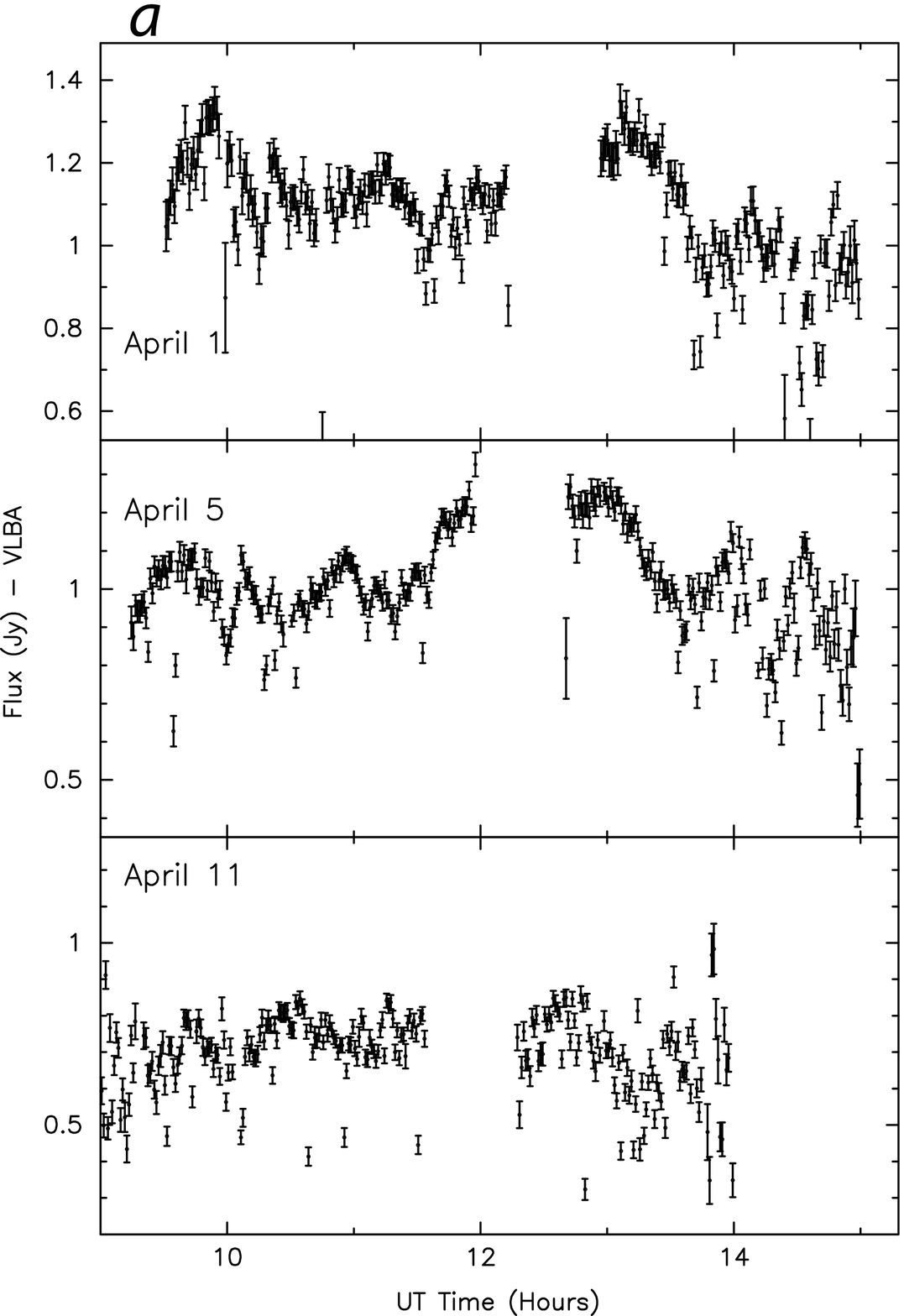}\\
\includegraphics[scale=0.37, angle=0]{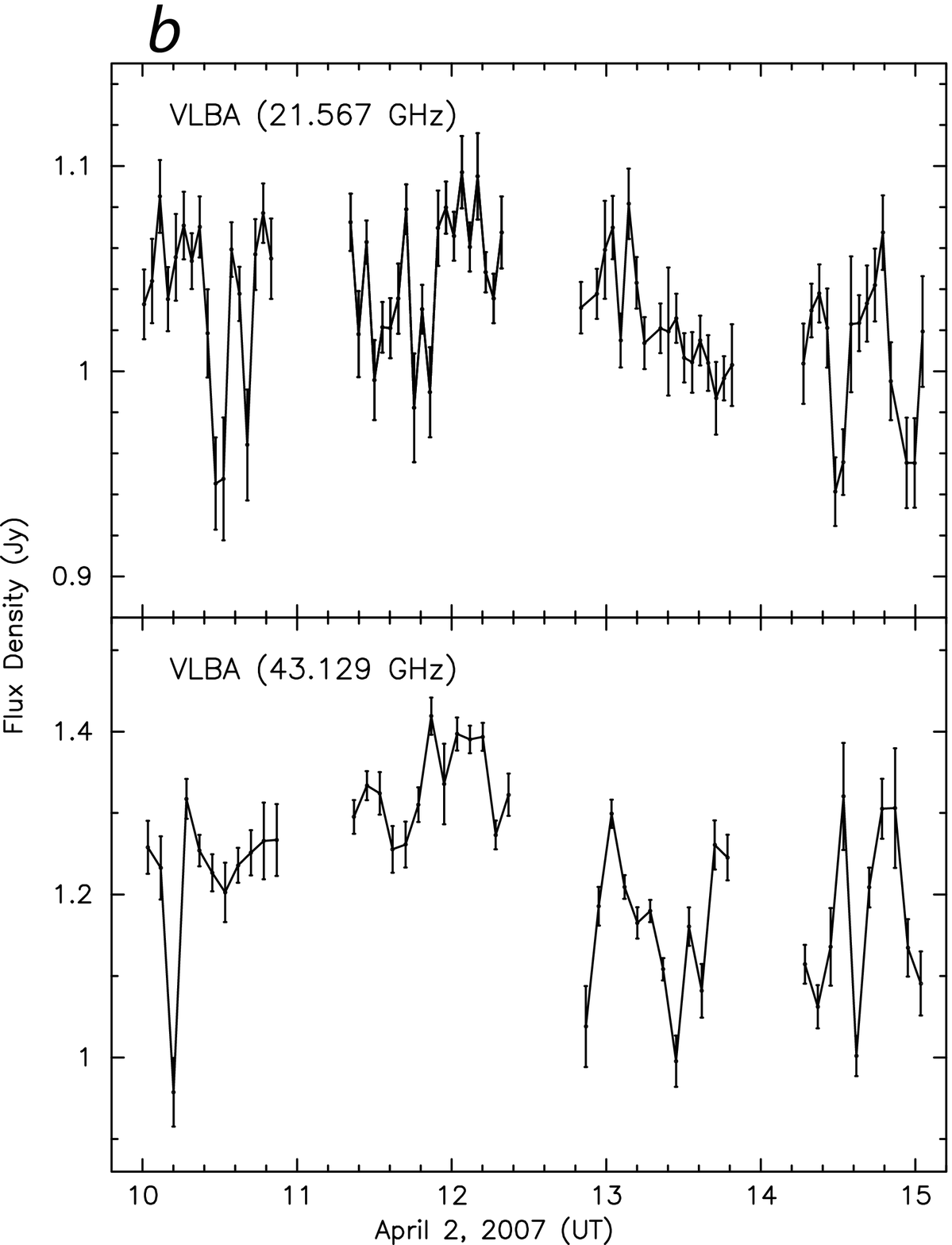}
\includegraphics[scale=0.37, angle=0]{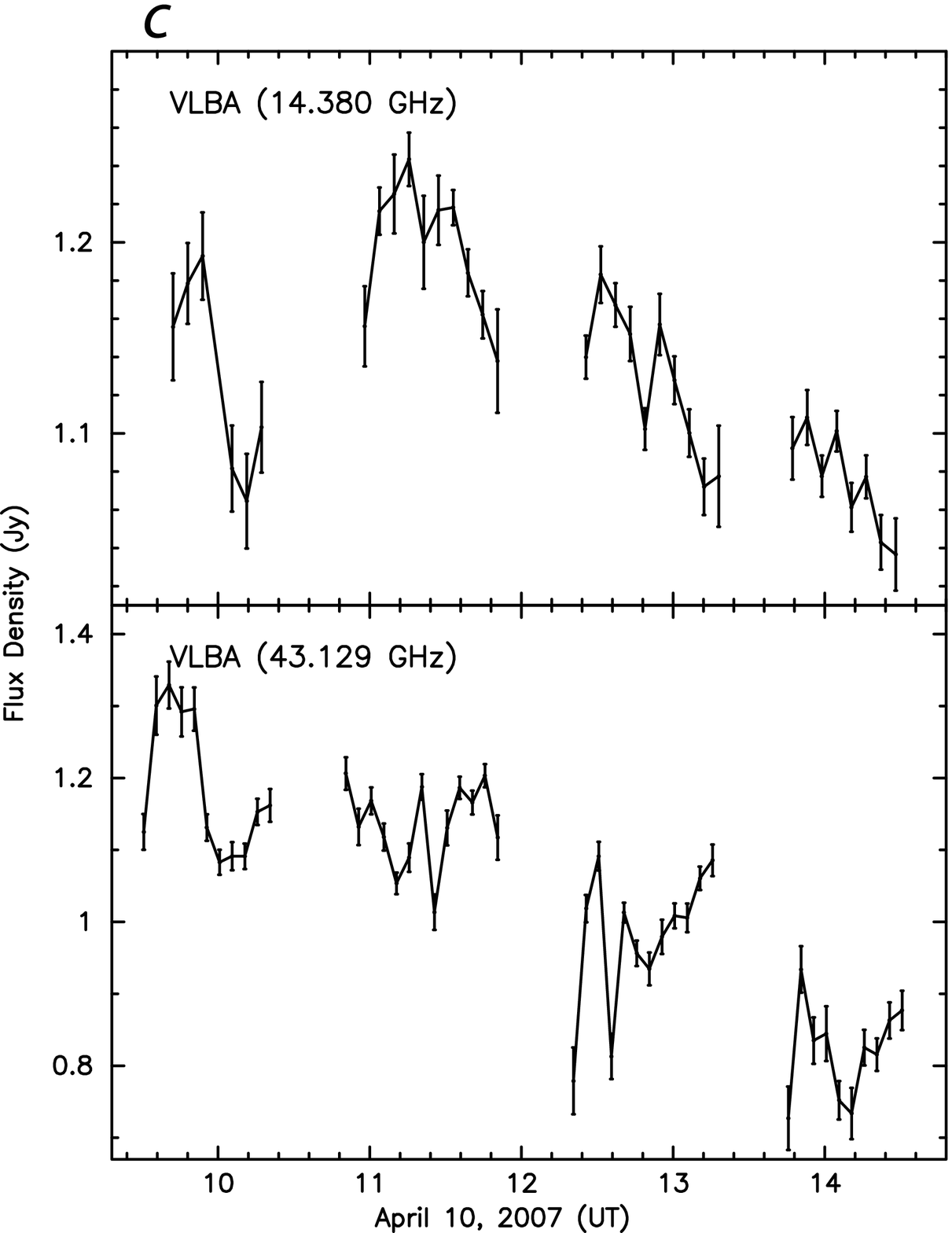}
\caption{
(\textit{a - Top }) 
 Light curves of Sgr A* on 2007, April 1, 5 and 11 using VLBA at 43.22 GHz. The sampling 
time is 60 sec.  
(\textit{b - Bottom Left}) 
 The light curve of Sgr A* observed with VLBA on April 2, 2007 at 
22 and 43 GHz. The sampling time is 300 sec.  
(\textit{c - Bottom Right }) 
 The light curve of Sgr A* observed with VLBA on April 10, 2007 at 
14 and 43 GHz. The sampling time is 300 sec.  
}\end{figure}

\begin{figure}
\centering
\includegraphics[scale=0.4, angle=0]{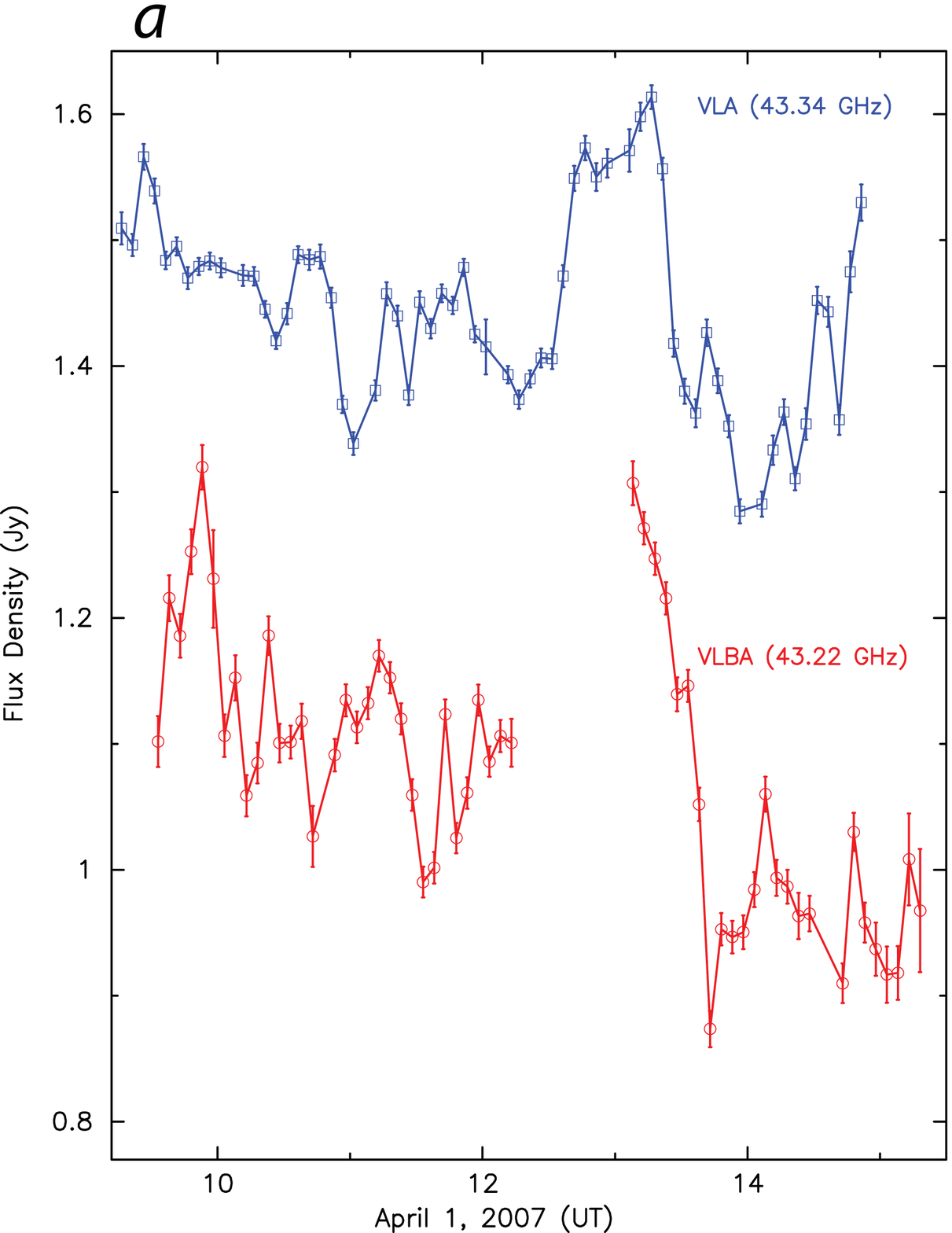}
\includegraphics[scale=0.4, angle=0]{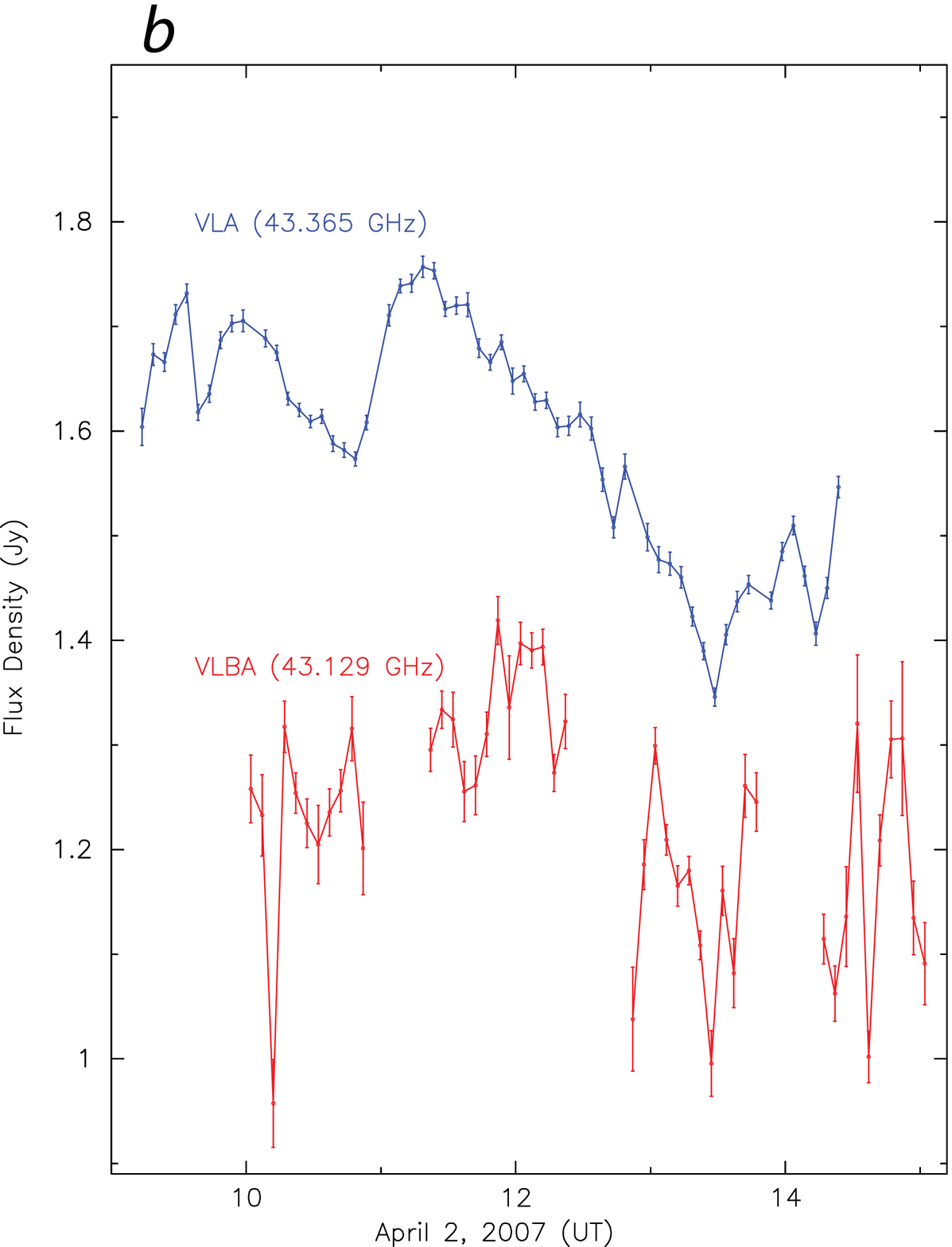}
\caption{
(\textit{a -  Left}) 
 The light curve of Sgr A* observed with VLA and VLBA on April 1, 2007. 
The sampling time is 300 sec.  The center frequencies  of the VLA and VLBA light curves 
correspond to  43.34 GHz 
43.22 GHz, respectively. The selected VLA {\it uv} data is $>$100k$\lambda$. 
(\textit{b - Right}) Similar to (a) except that the observations are
carried out on April 2 at 43 GHz. VLBA observations on April 1 (bottom panel) are sampled 
continuously  unlike those made on April 2. 
VLBA plots  are shown in red (bottom) whereas VLA plots 
are shown in blue (top).  
}\end{figure}

\begin{figure}
\centering
\includegraphics[scale=0.5, angle=0]{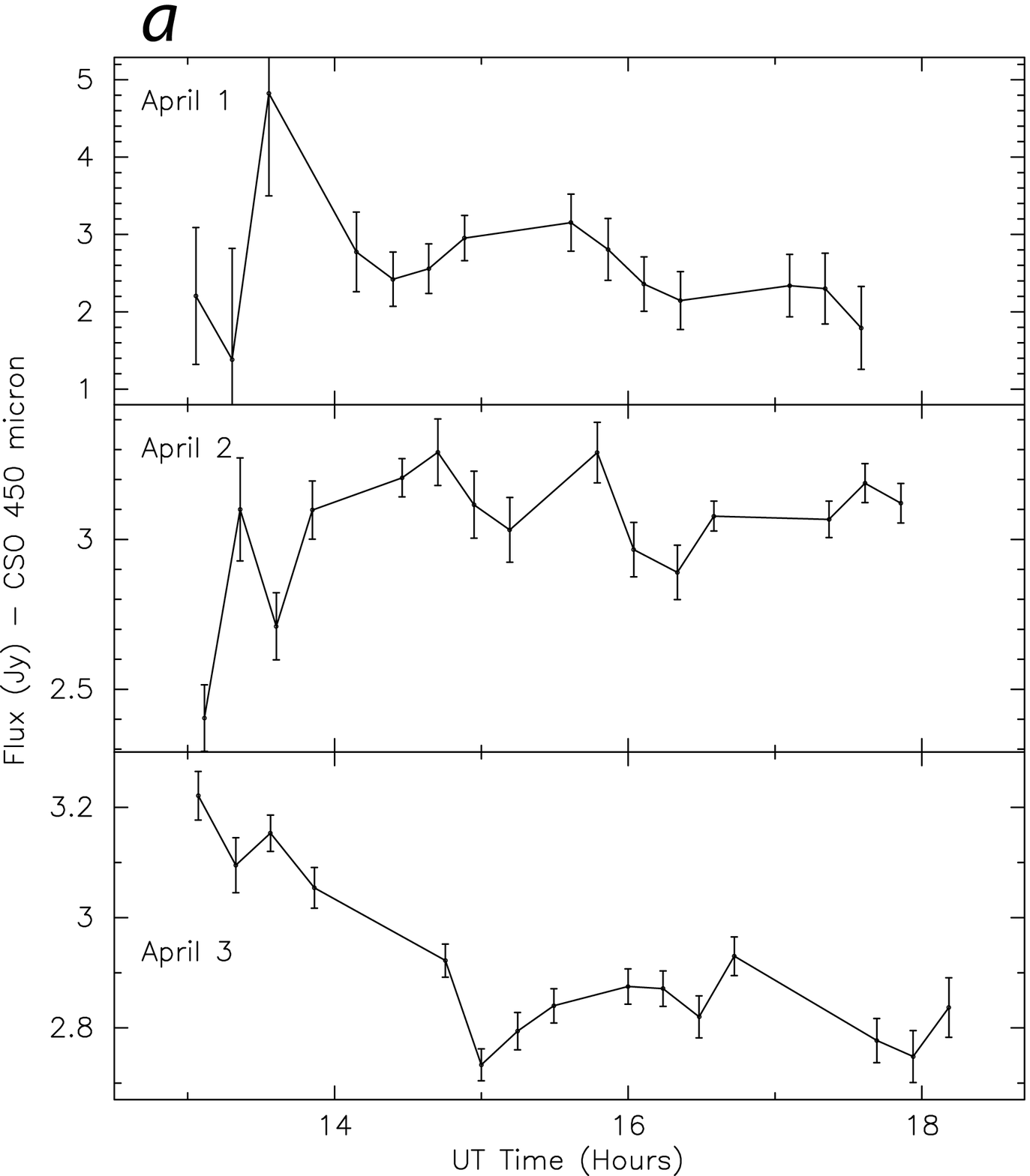}
\includegraphics[scale=0.5, angle=0]{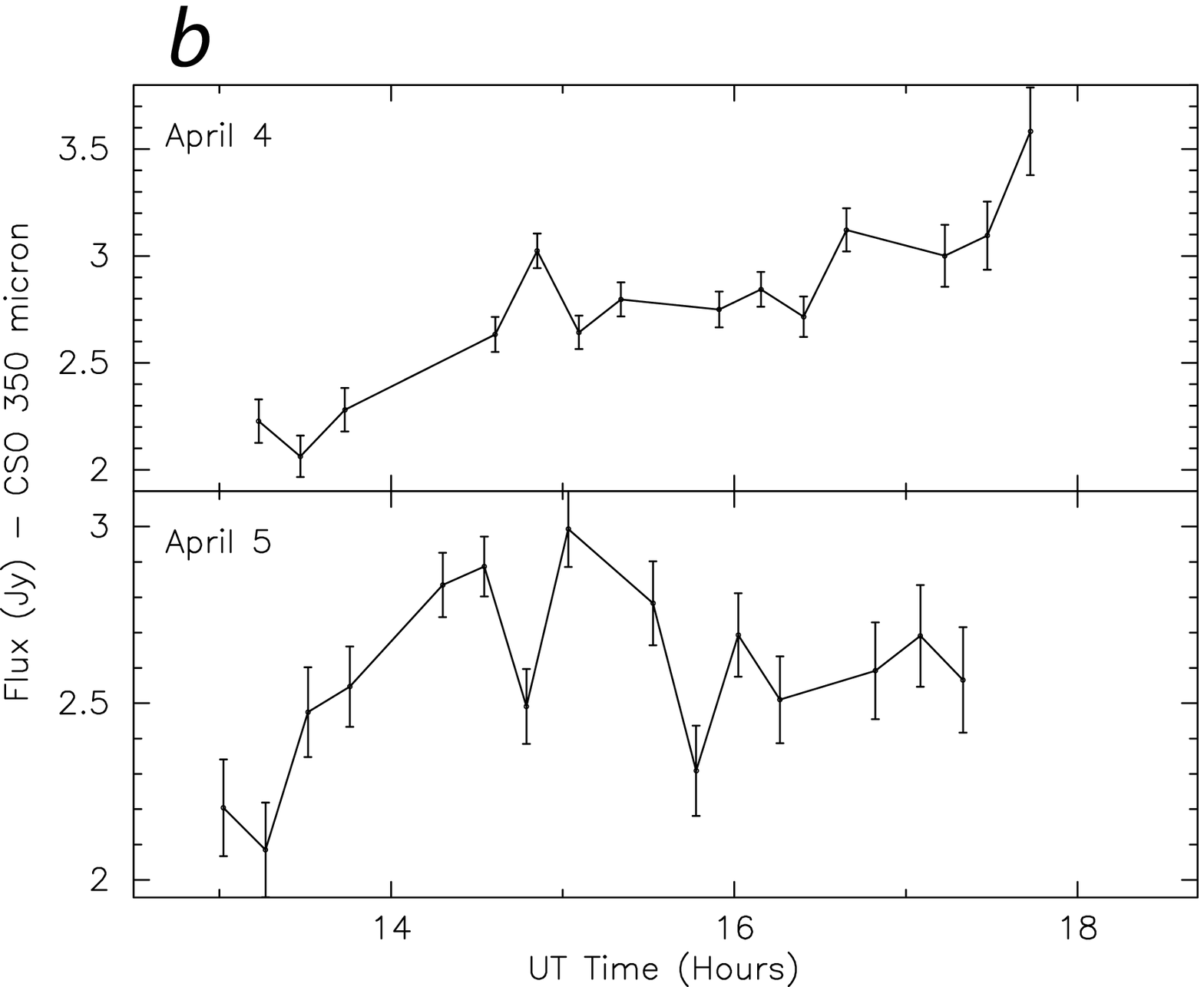}\\
\caption{
(\textit{a})  Light curves of Sgr A* at 450$\mu$m based on CSO observations on 2007, 
April 1-3 with a sampling time of 15 minutes. 
(\textit{b}) 
 Similar to (a) except at 350$\mu$m on 2007, April 4-5 with  a sampling 
time of 15 minutes. 
}\end{figure}

\begin{figure}
\centering
\ContinuedFloat
\includegraphics[scale=0.7, angle=0]{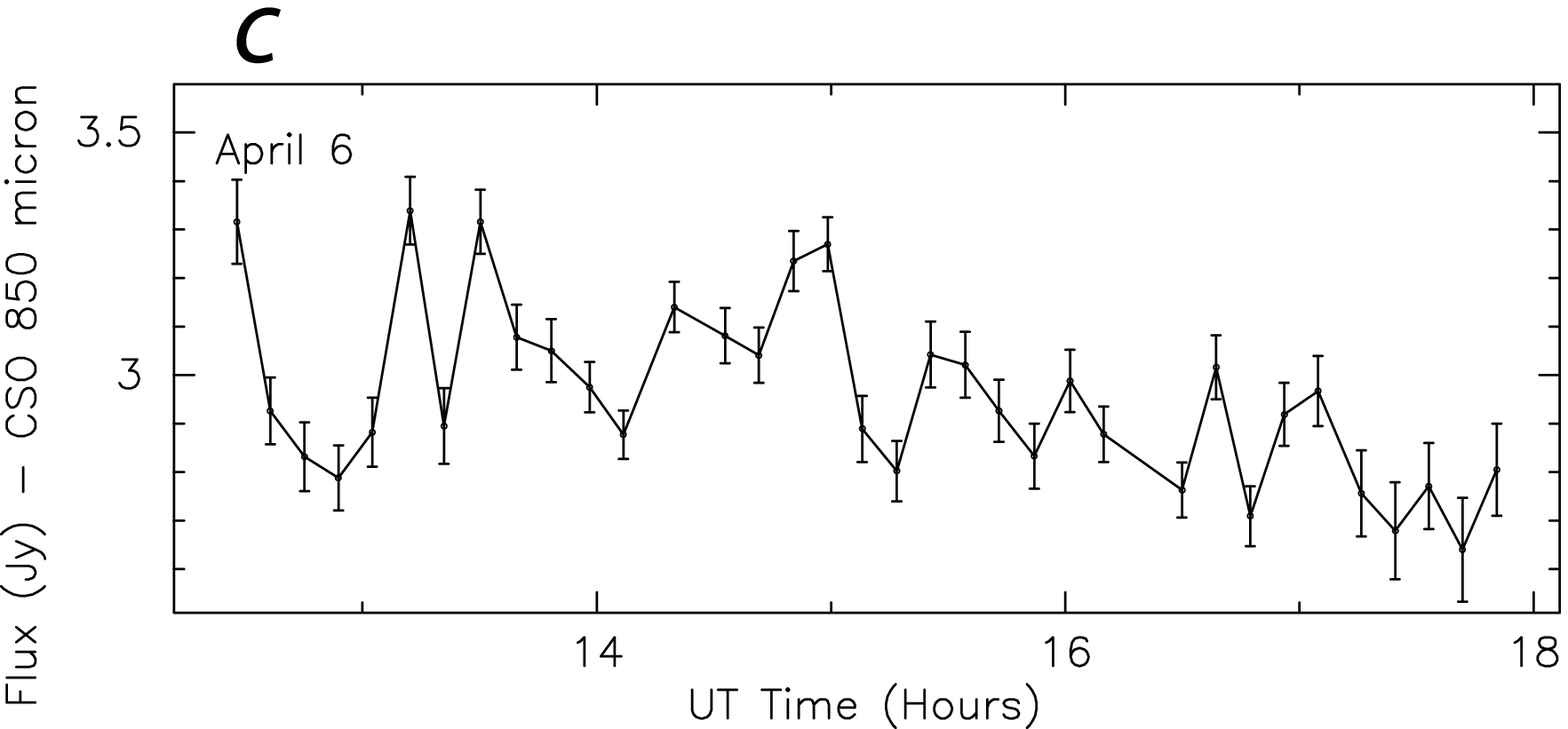}
\caption{
(\textit{c}) Similar to (a) except at 850$\mu$m on 2007, April 6 using a 
sampling time of 10 minutes.
}\end{figure}

\begin{figure}
\centering
\includegraphics[scale=0.4, angle=0]{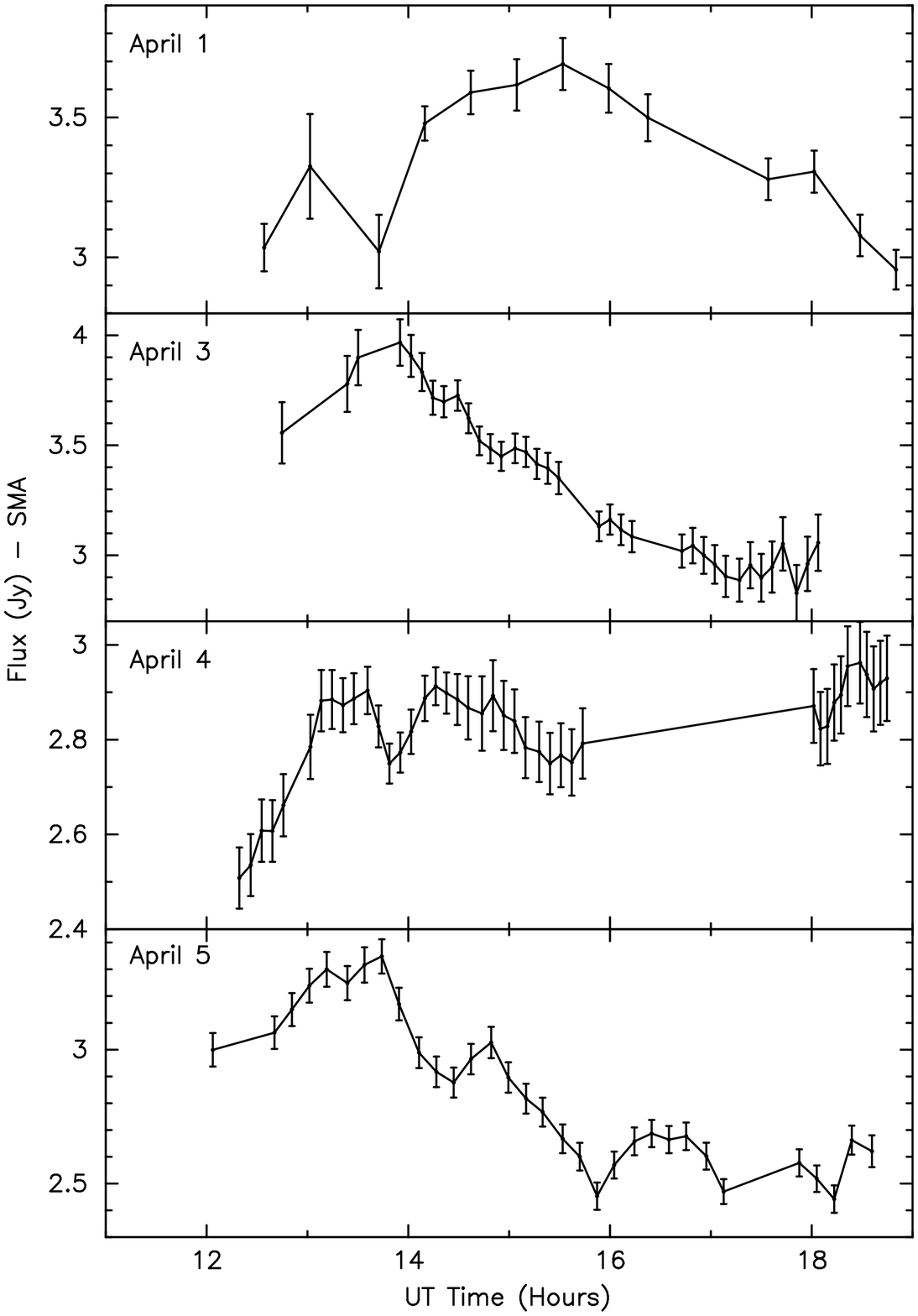}
\caption{Light curves of Sgr A* at 230 GHz using SMA on 2007 April 1, and 3-5. 
The sampling times are   27 min, 6.5 min, 8 min and 10 min for the April 1, 3, 4 and 5 light 
curves, respectively. 
}\end{figure}

\begin{figure}
\centering
\includegraphics[scale=0.4, angle=0]{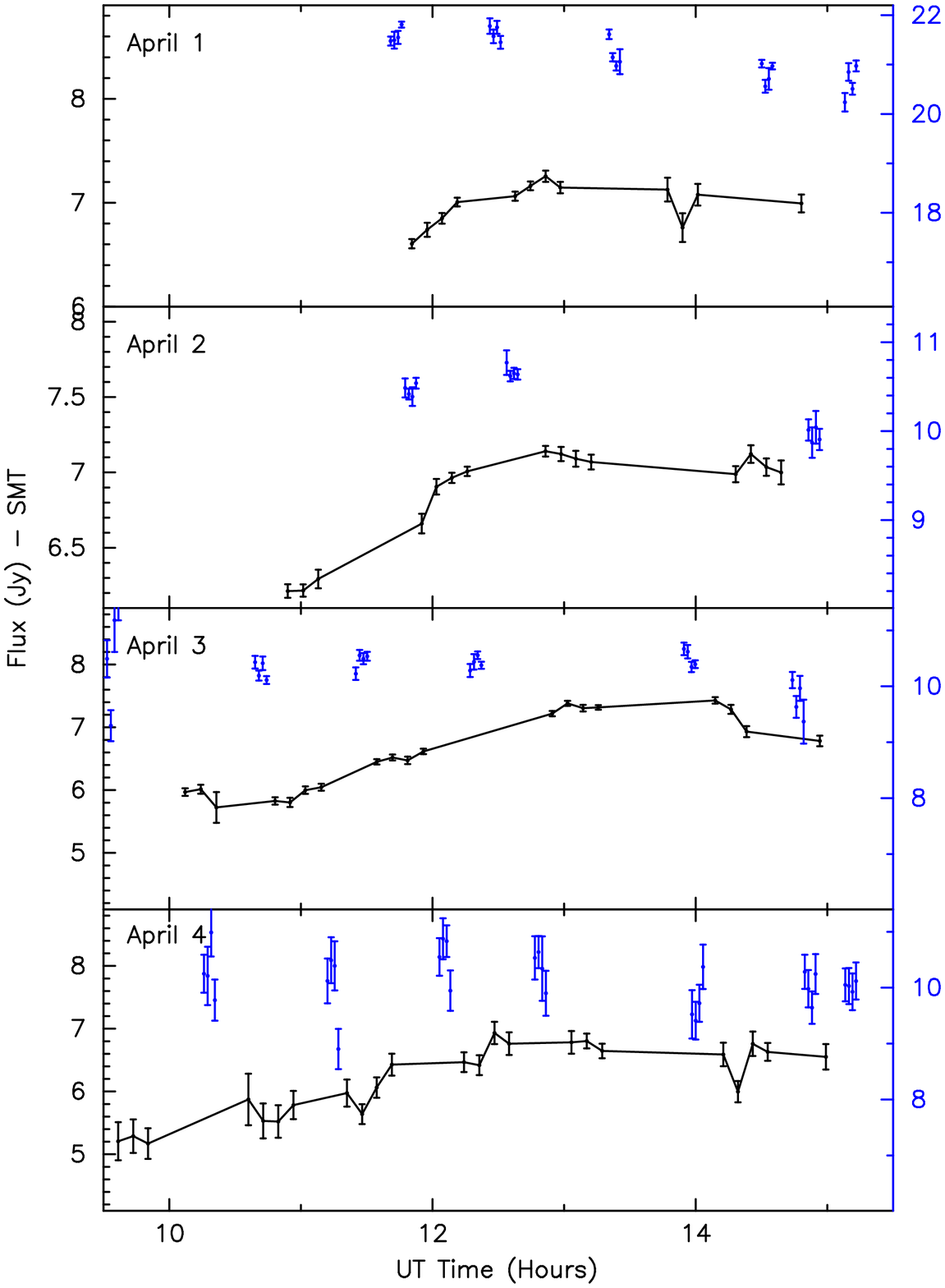}
\caption{Light curves of Sgr A* and the calibrator G34.3 and 1757-240 (blue) at 230 GHz using 
SMT on 2007 April 1 and April 2-4, respectively, with a sampling time of $\sim$7 minutes. 
}\end{figure}

\clearpage 

\begin{figure}
\centering
\includegraphics[scale=0.4, angle=0]{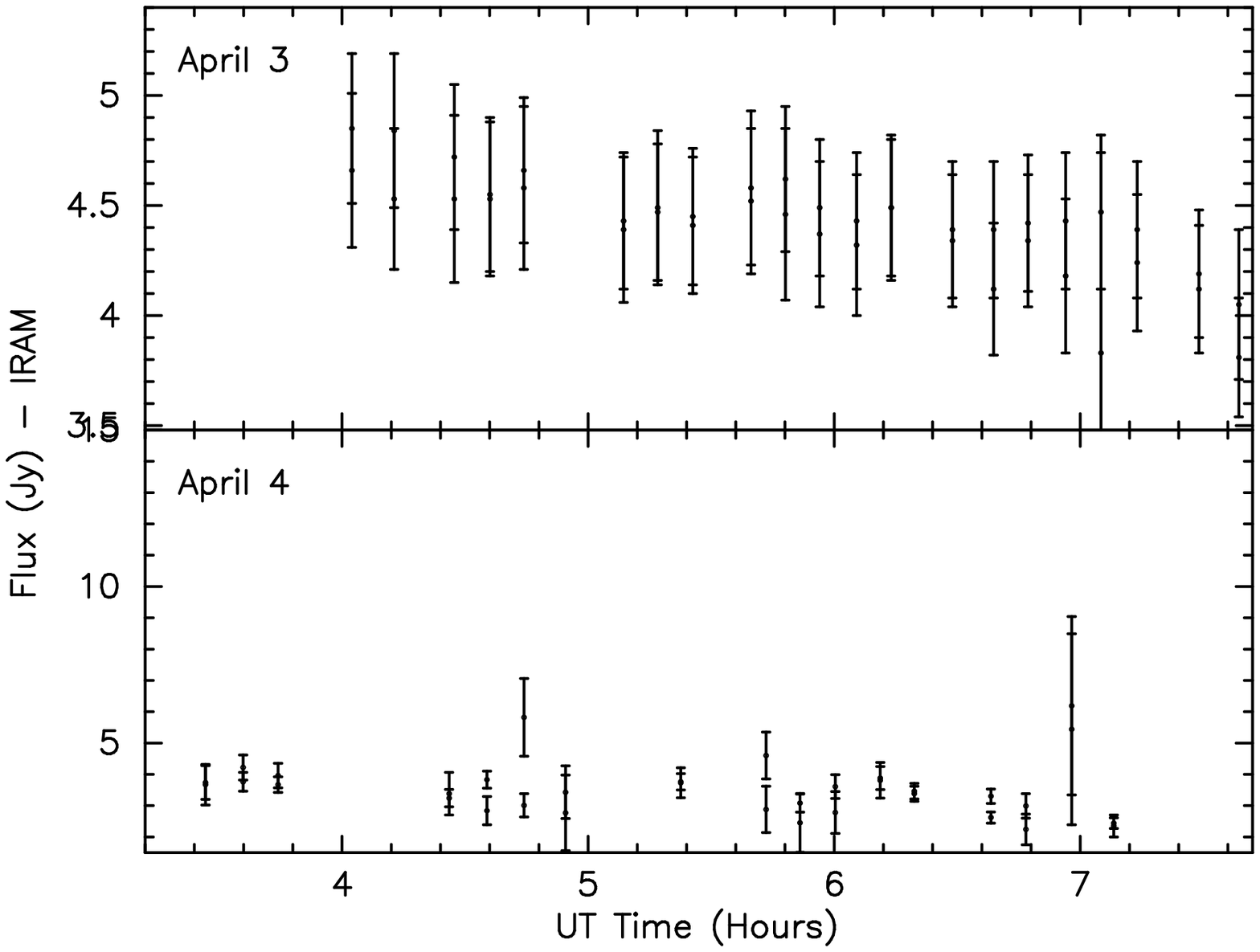}
\caption{Light curves of Sgr A* at 240 GHz (1.25mm) taken with IRAM on 2007, 
April  3-4 with sampling time of $\sim$10 minutes. 
For each time sample, there are two data points estimating the flux
of SgrA* from repeated pointing measurements.
}\end{figure}

\begin{figure}
\centering
\includegraphics[scale=0.4, angle=0]{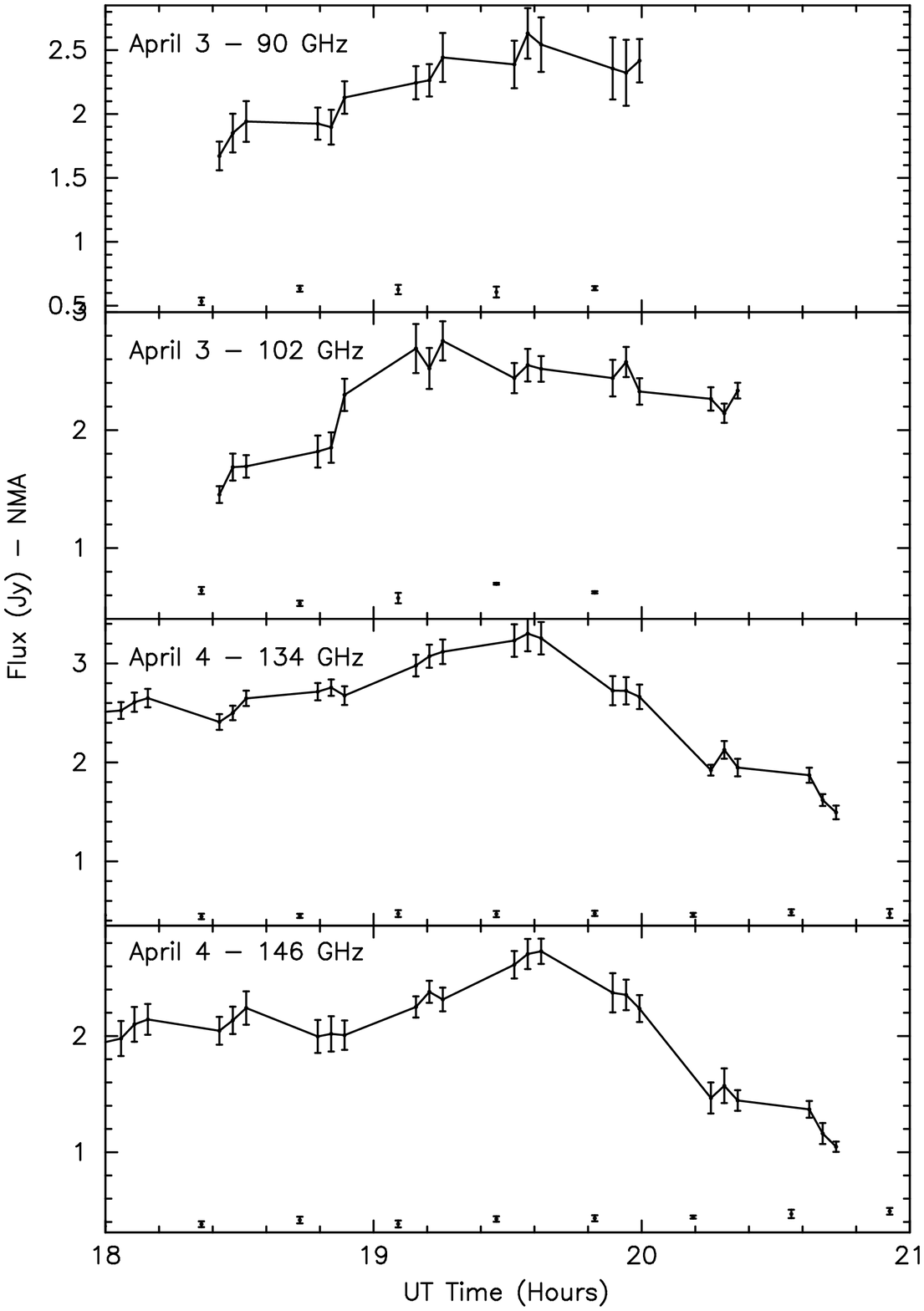}
\caption{Light curves of Sgr A* taken with the NMA.  The 
top panels show the light curves of Sgr A*
 at 90 GHz and   102 GHz on April 3
whereas the bottom two panels show simultaneous light curves at 2.23mm 
(134 GHz) 
and 146 GHz on 2007, April 4. The flux of the calibrator 1744-312 
is shown at the bottom of each panel. The sampling time is 3 minutes. 
}\end{figure}

\begin{figure}
\centering
\includegraphics[scale=0.4, angle=0]{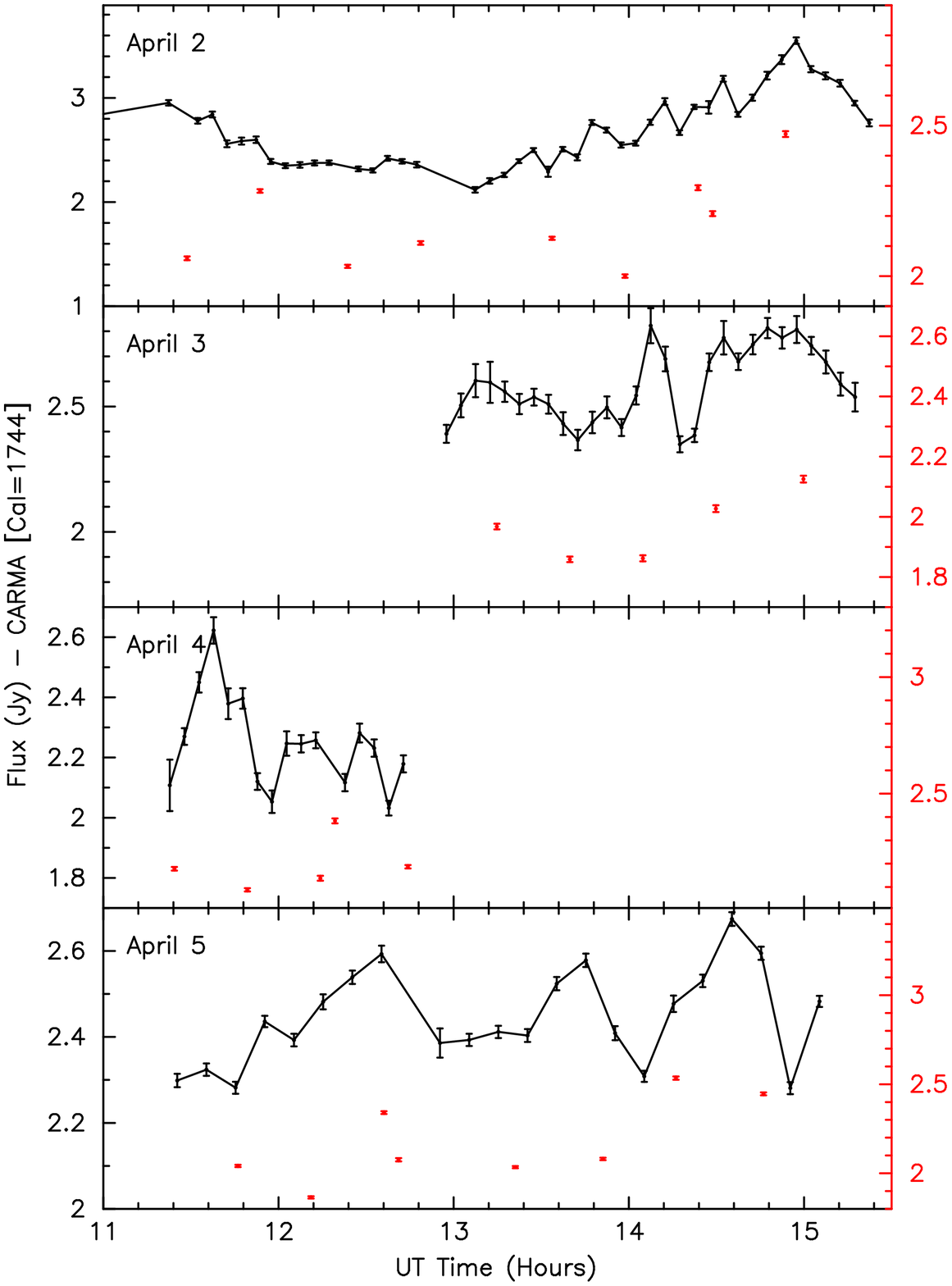}
\caption{Light curves of Sgr A* and the calibrator 1730-130 obtained with CARMA at 
90 GHz with a sampling time of 
300 sec on 2007, April 2-5. The {\it uv} data $> 20 \rm k\lambda$ are used to make 
the Sgr A* light curves. 
}\end{figure}

\begin{figure}
\centering
\includegraphics[scale=0.5, angle=-90]{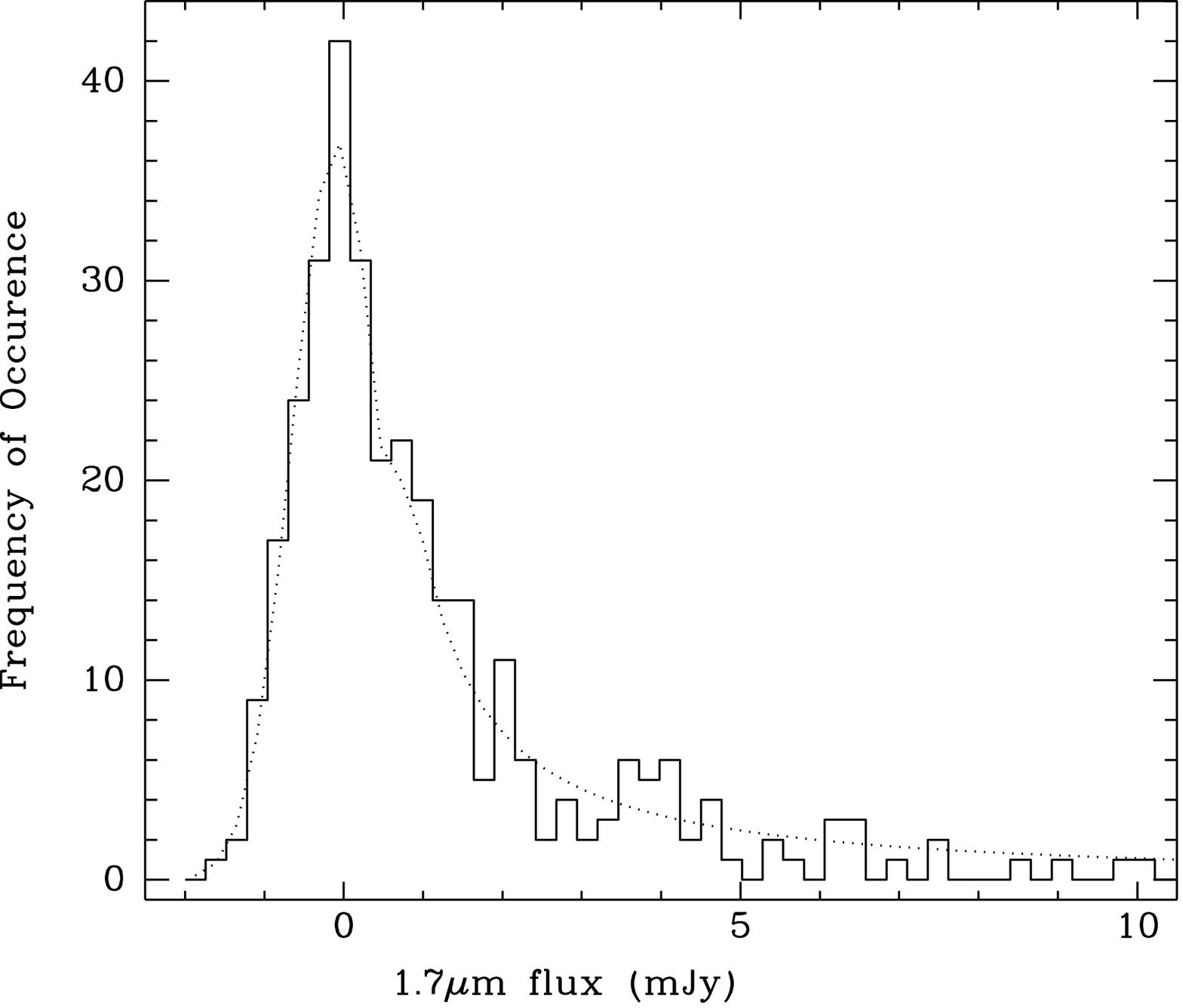}\\
\caption{
A histogram plot of the detected signals and the noise at 1.70$\mu$m 
as well as the simultaneous single Gaussian fit and power law fits to both the
noise and the flares. The dotted  lines show the 
Gaussian and power law fits.
}\end{figure}

\begin{figure}
\centering
\includegraphics[scale=0.5, angle=-90]{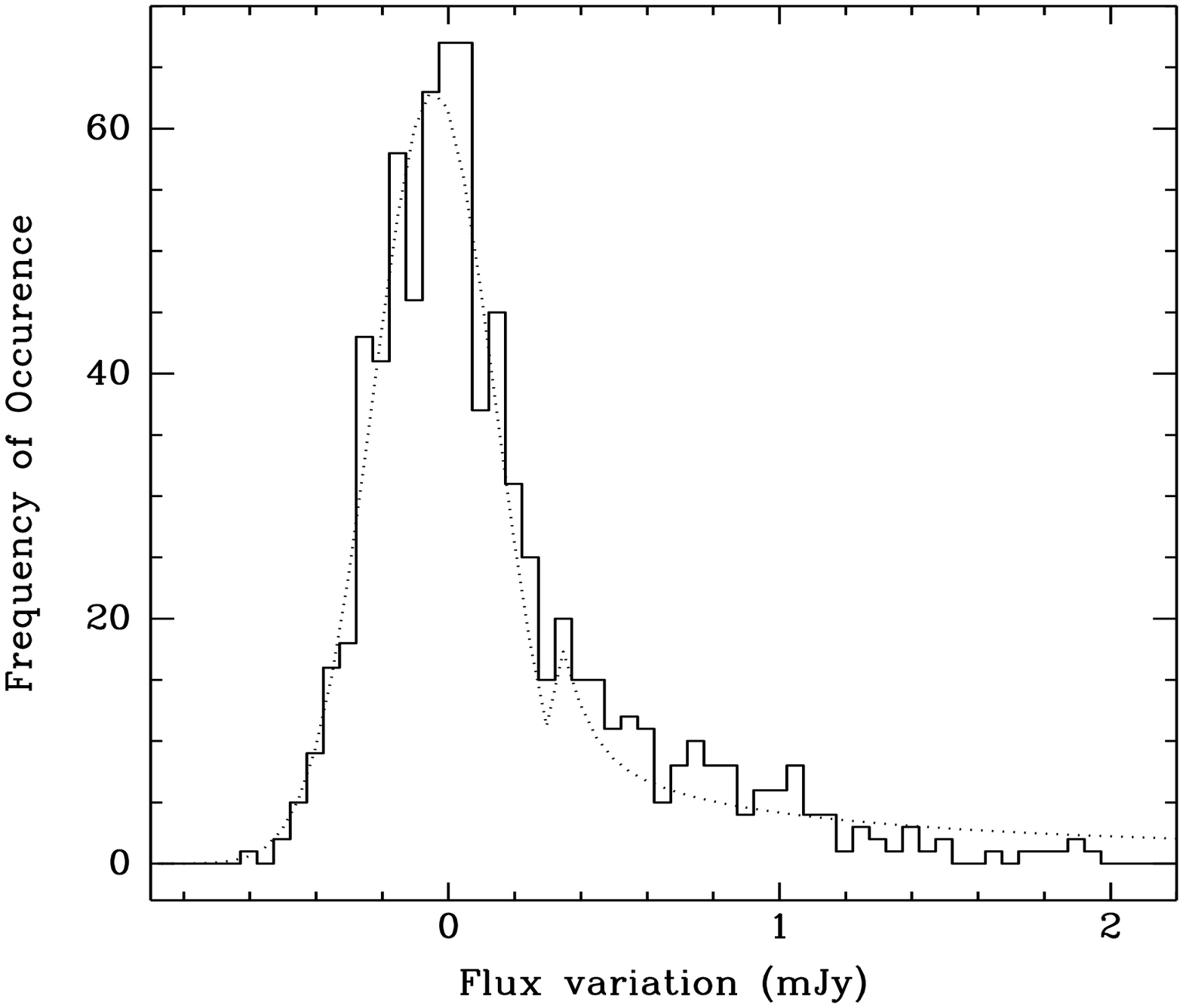}
\caption{
Similar to Figure 16 expect that the 2004 histogram of flare activity 
(Yusef-Zadeh et 
al. 2006) is plotted at 1.60$\mu$m.
}\end{figure}

\begin{figure}
\centering
\includegraphics[scale=1, angle=0]{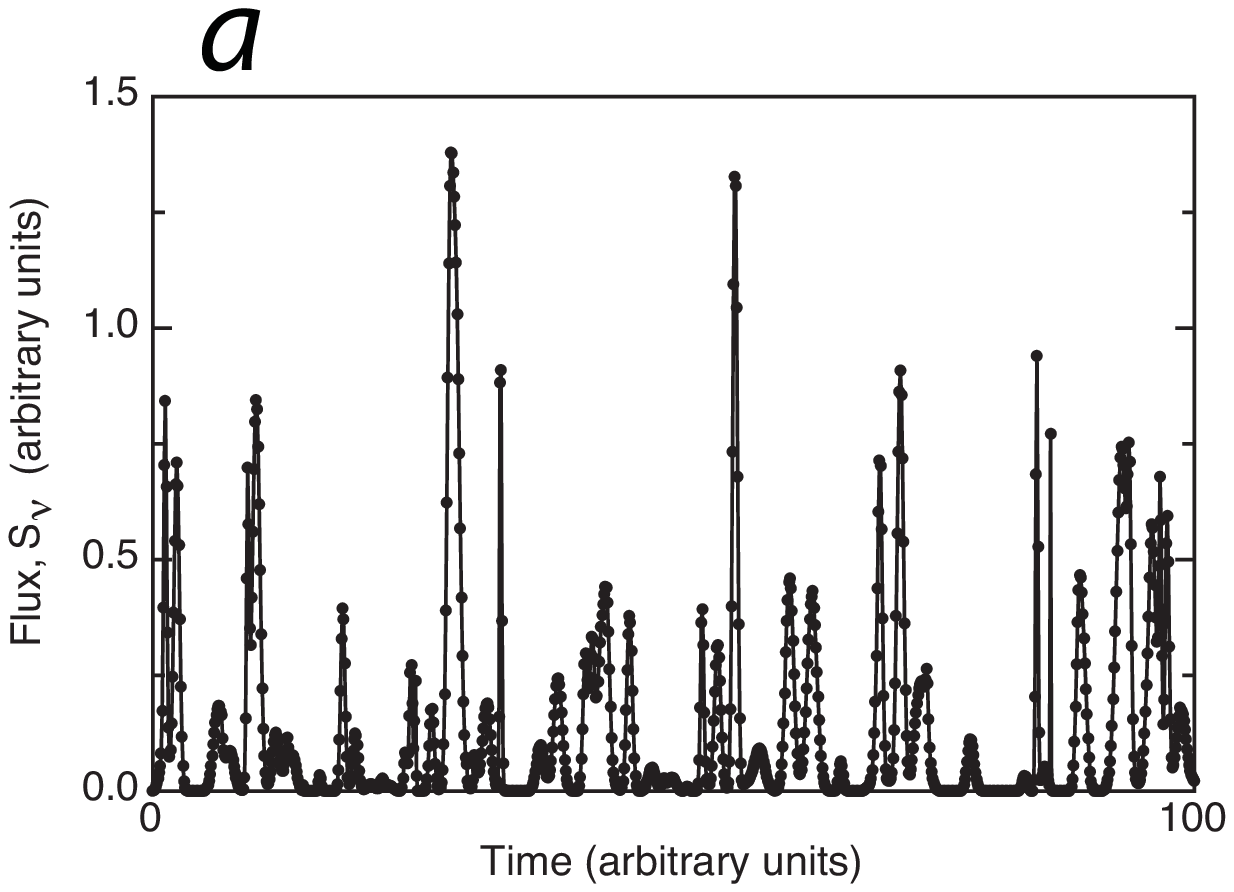}
\includegraphics[scale=1, angle=0]{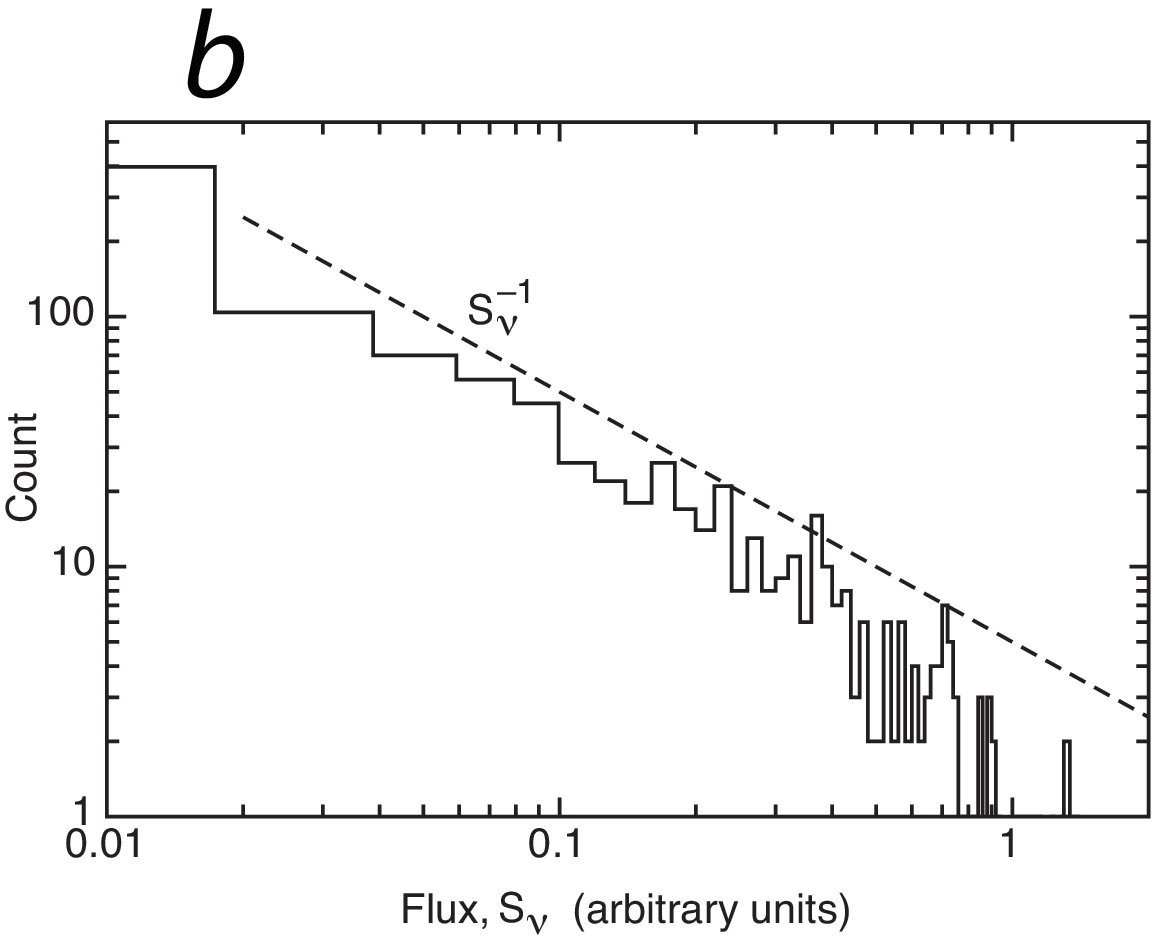}
\caption{
(\textit{a})
\emph{(Left)} A synthetic light curve constructed from the sum of 100 Gaussian profiles with peak
positions and, standard deviations drawn uniformly
between -1.5 to 101.5 and 0 to 0.5 time units respectively; the probability distribution of the peak
fluxes are distributed as 1/(peak flux) between 0.01 and 1 flux units.
(\textit{b}) 
\emph{(Right)}  Distribution of uniformly sampled flux values in the simulated flares. The dashed line
indicates a slope of 1/S$_{\nu}$.
}\end{figure}

\begin{figure}
\centering
\includegraphics[scale=0.8, angle=0]{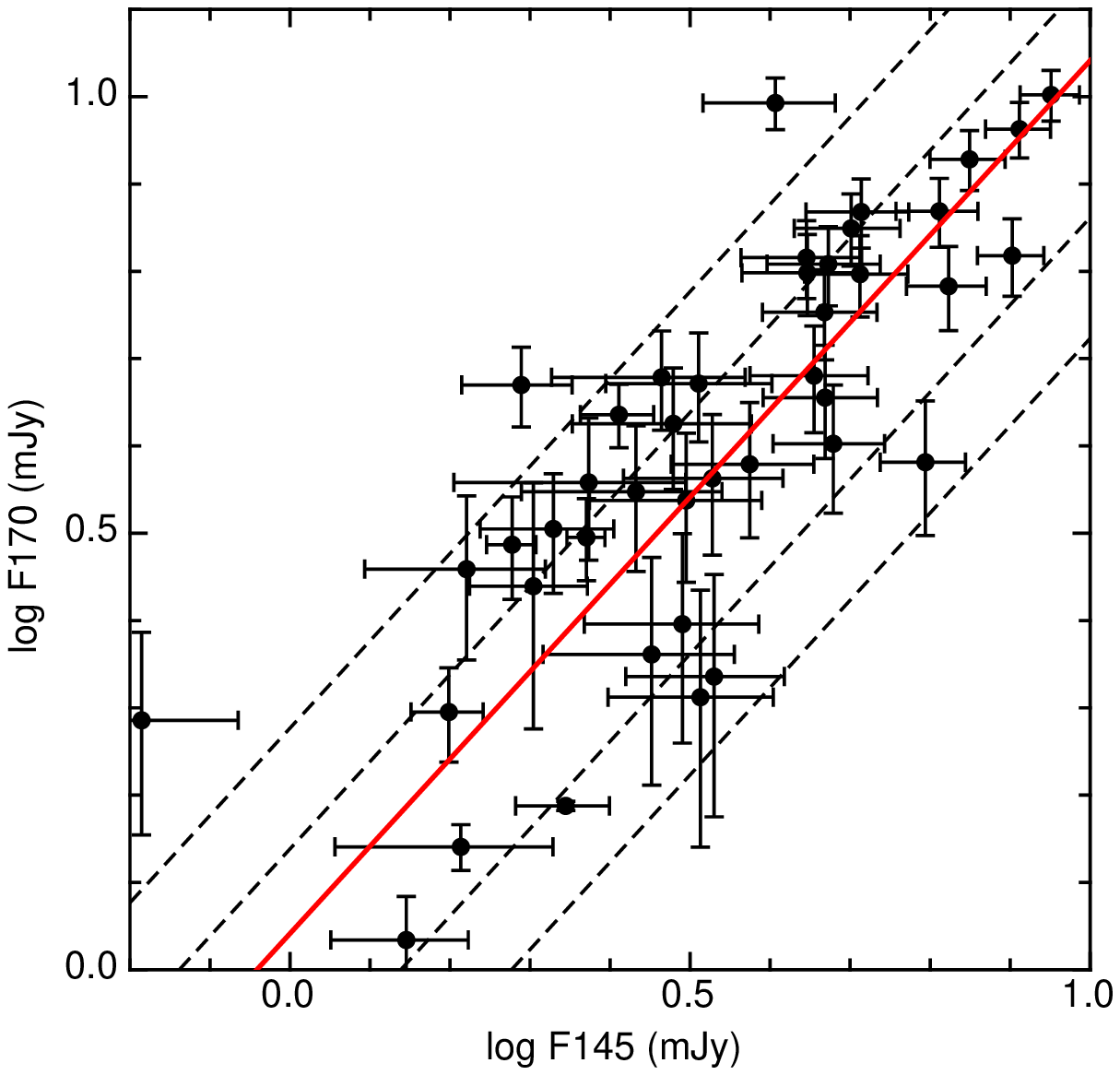}
\caption{
A log-log plot of NIR fluxes in  the F170 and F145 filters of 
NICMOS at 1.70$\mu$m and 1.45$\mu$m, respectively. The thick line in 
red shows the
spectral index $\beta$=0.6. The thin dotted lines to the right and left
 of the $\beta=0.6$ line correspond to $\beta=-2,  -4$ and 
$\beta=+2, +4$, respectively. 
}\end{figure}

\begin{figure}
\centering
\includegraphics[scale=0.4, angle=0]{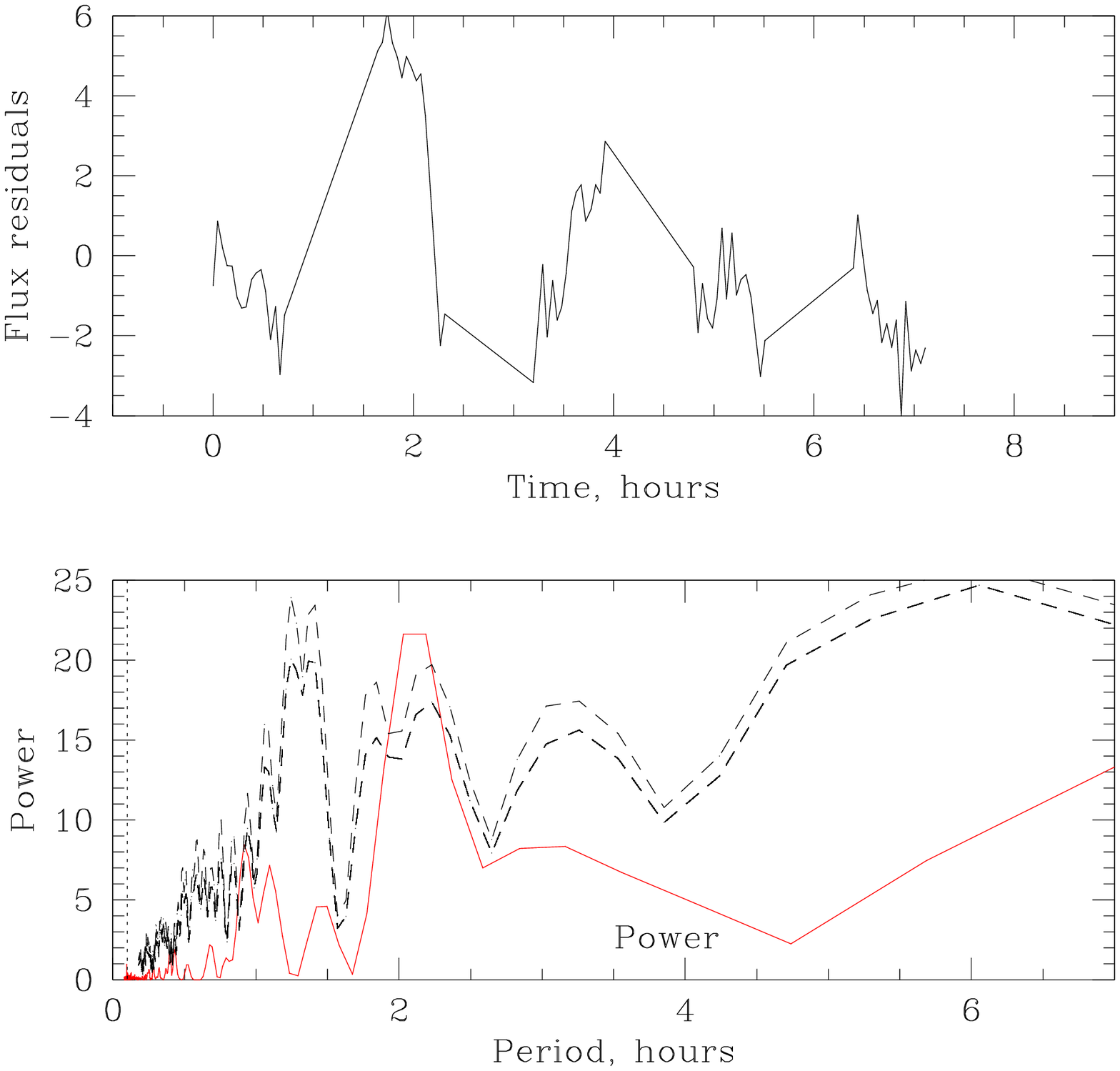}
\clearpage
\caption{The top and bottom boxes
 show the light curve of 2007,  April 4 based on HST observations
and the corresponding power spectrum  of the residual flux of Sgr
A*, respectively.  The dashed lines
show the significance of the power spectrum at   99\% and 99.9\%
confidence levels. We explain the significance of the peak in the text. 
}\end{figure}

\begin{figure}
\centering
\includegraphics[scale=0.4, angle=0]{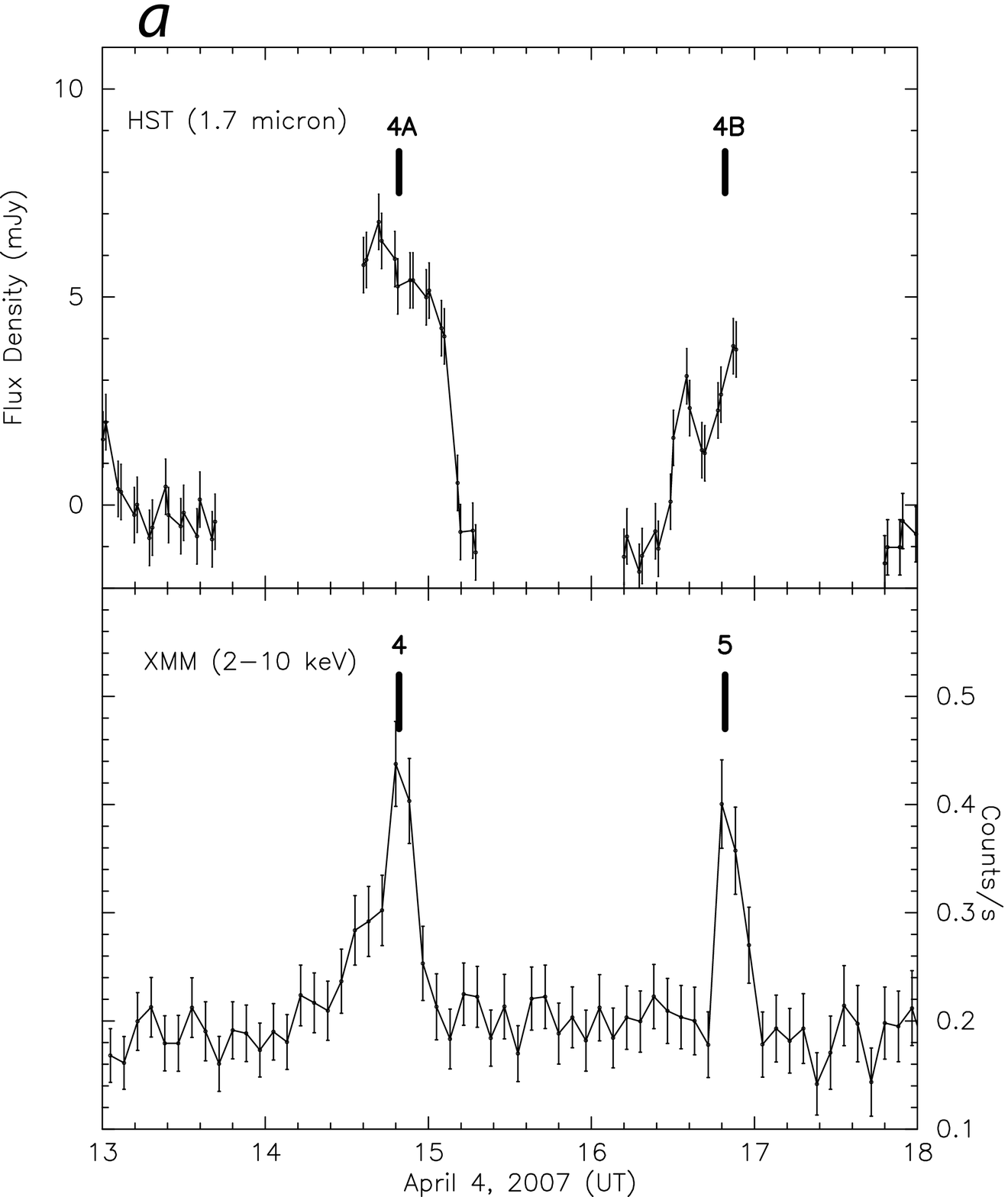}
\includegraphics[scale=0.4, angle=0]{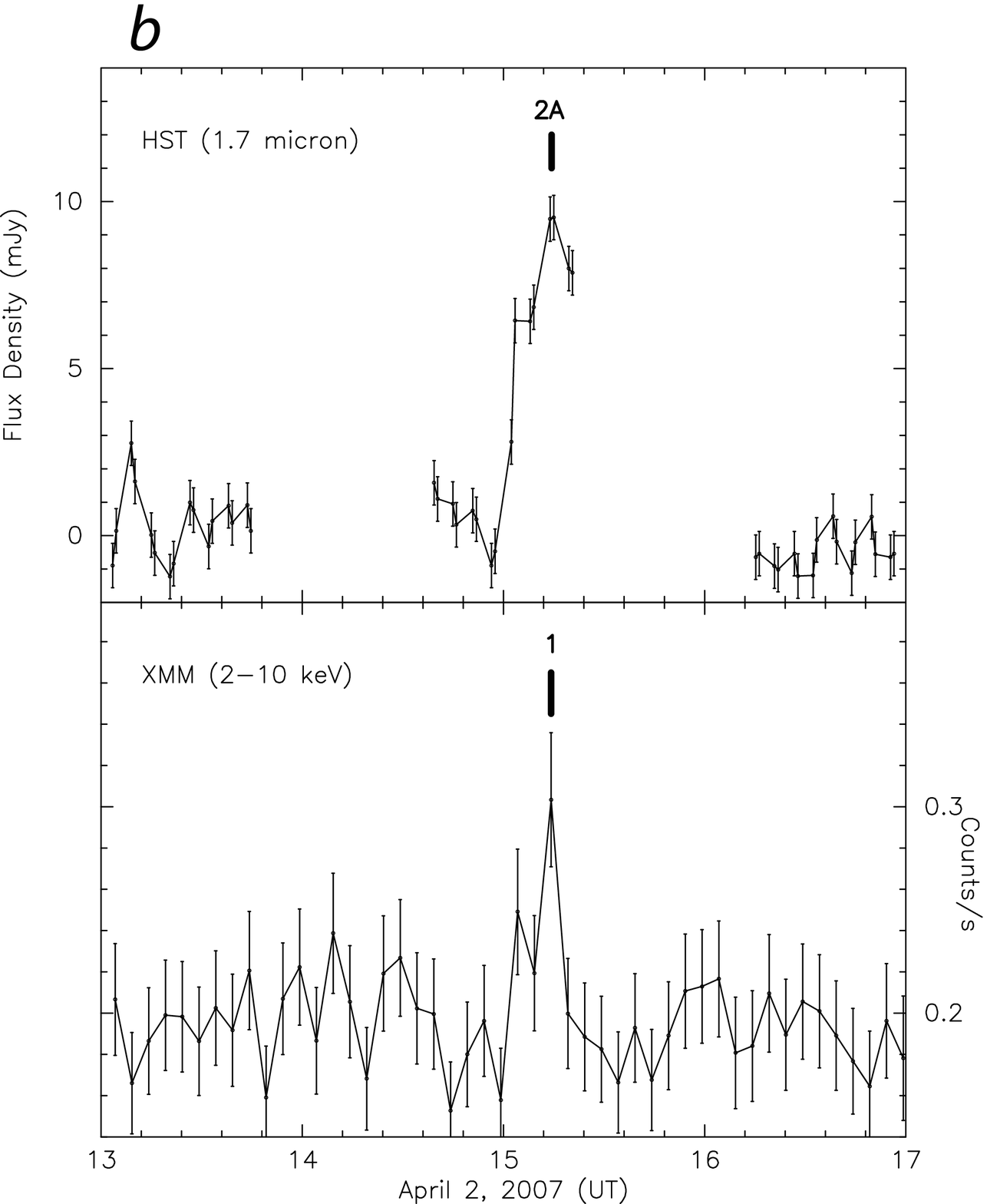}
\caption{
(\textit{a - Left})
The top panel shows the light curves of two near-IR flares 
identified as 4A ad 4B measured with NICMOS on 2007, April 4 with a sampling time of 64s at 
1.70$\mu$m.
The bottom panel shows the X-ray counterpart to these flares 
with a sampling time of 300 sec. These X-ray flares  are identified as flare \#4 and \#5 by 
Porquet et al. (2008).  
(\textit{b - Right}) Similar to (a) except for a  NIR and X-ray flare that 
occurred on April 2, 2007. 
}\end{figure}

\begin{figure}
\centering
\includegraphics[scale=1.1, angle=0]{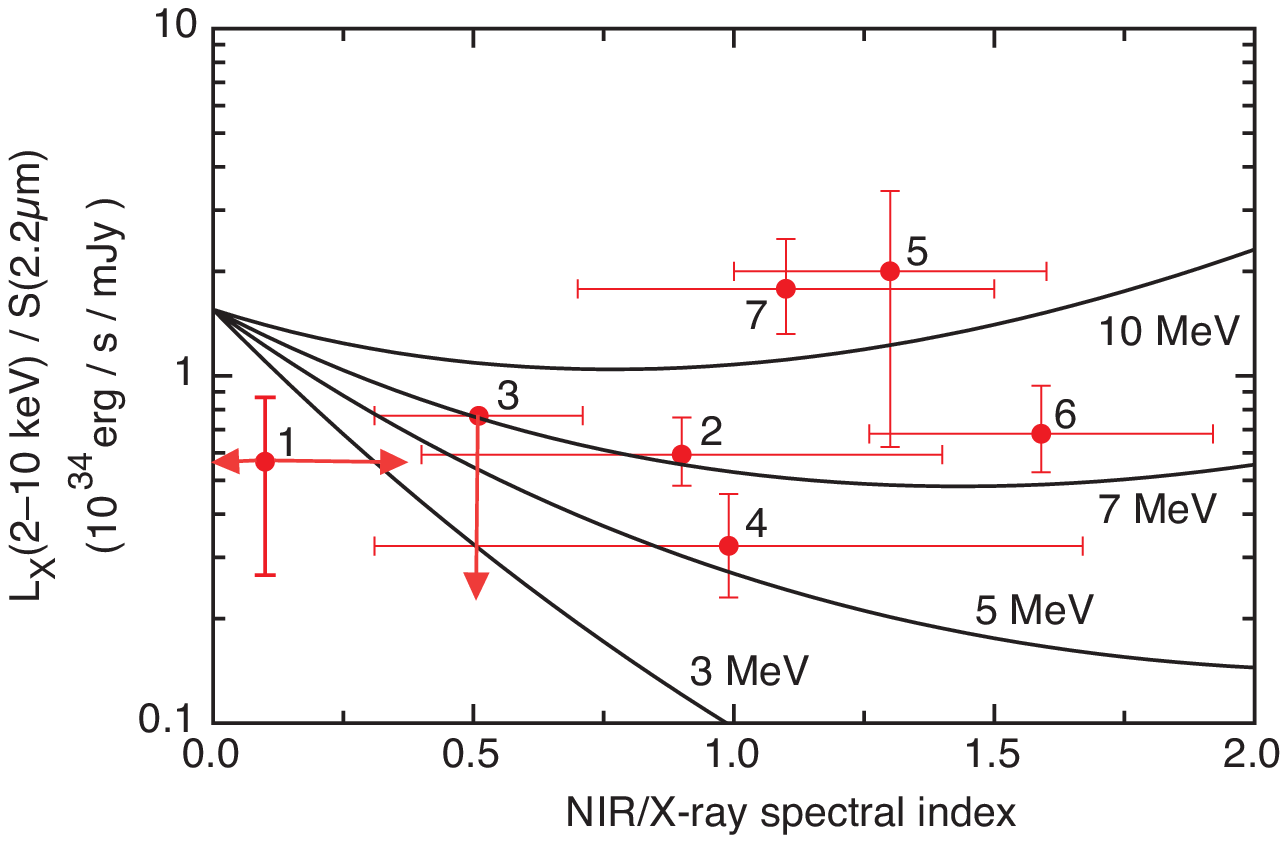}
\caption{
Solid curves show the ratio of inverse Compton X-ray luminosity (2--10\, 
keV) to NIR synchrotron flux (at 2.2$\mu\,$m) as a function of NIR or X-ray spectral index $\beta$ 
Points labeled 1 to 7 indicate the measured ratios and IR spectral 
indices of the seven X-ray flares with known NIR counterparts. \emph{(1)} 2004 July 07, 
Eckart et 
al.\ (2006, X-ray: flare $\phi_3$, IR: flare III); \emph{(2)} 2004 Aug 31, X-ray: Belanger et 
al.\ (2005), IR: Yusef-Zadeh et al.\ (2006); \emph{(3)} 2006 July 17, X-ray: Marrone et 
al.\ 
(2008), IR: lower limit from Hornstein et al.\ (2007); \emph{(4)} 2007 April 02, X-ray: 
Porquet et 
al.\ (2008, flare 1), IR: this paper (flare 2A); \emph{(5)} 2007 April 04 05:25, X-ray: 
Porquet et 
al.\ (2008, flare 2), IR Dodds-Eden et al.\ (2009); \emph{(6)} 2007 April 04 14:37, 
X-ray: Porquet 
et al.\ (2008, flare 4), IR: this paper (flare 4A); and \emph{(7)} 2007 April 04 16:45, 
X-ray: 
Porquet et al.\ (2008, flare 5), IR: upper limit from this paper (flare 4B).
Where the NIR spectral index is not known, the measured X-ray spectral
index is used (points 2, 5  and 7); the NIR and X-ray spectral indices are
both unknown for point 1.  NIR measurements for points 2--7 are scaled
to 2.2$\mu\,$m using either the NIR or X-ray spectral index.  The measured 2--8 keV
X-ray luminosities for points 1 and 3 have been rescaled to 2--10 keV
by multiplication by 4/3.
}\end{figure}

\begin{figure}
\centering
\includegraphics[scale=0.6, angle=0]{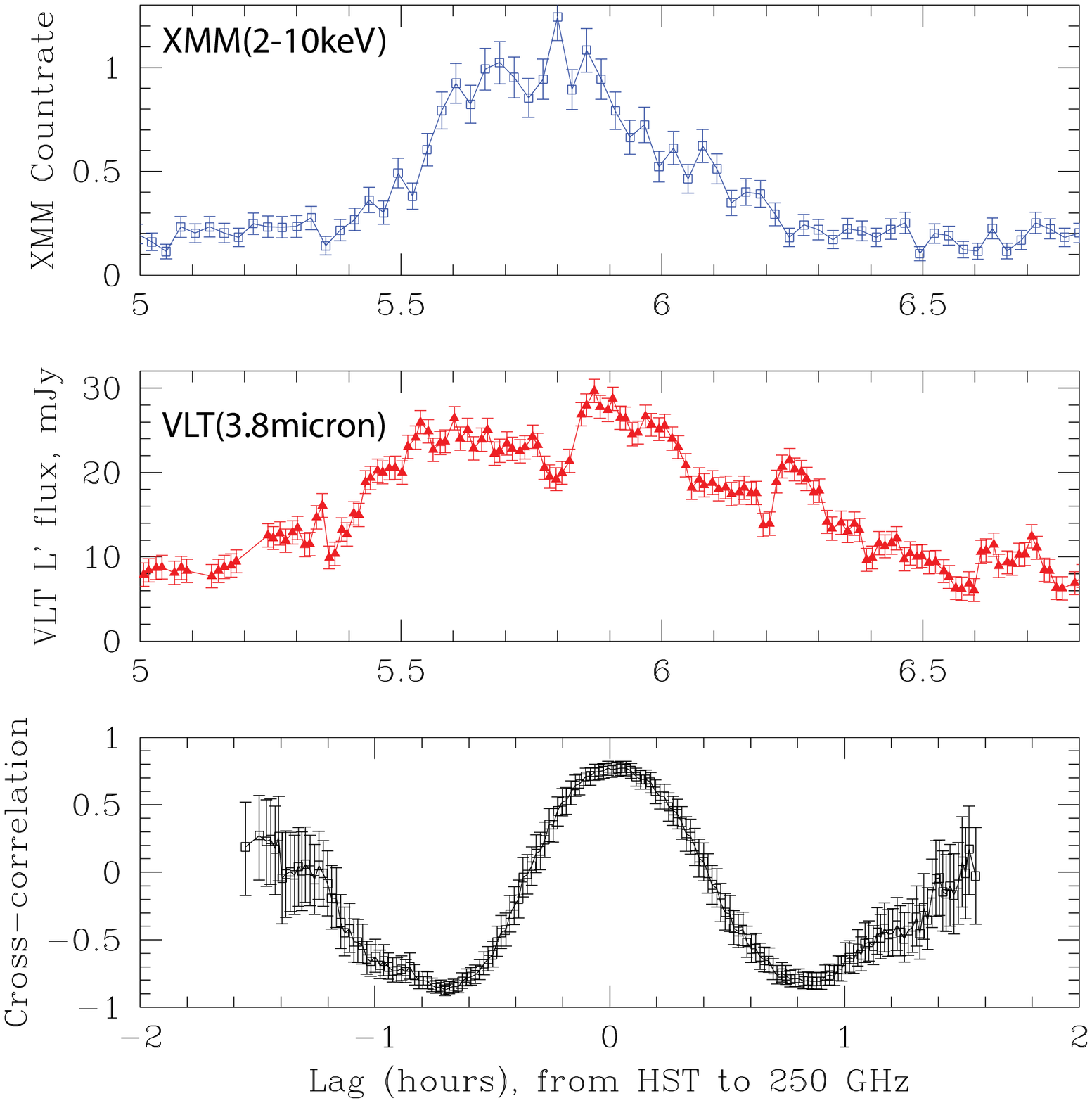}
\caption{
The top and middle plots show the X-ray and NIR light curves taken on 
2007, April 4. The cross correlation plot in the bottom panel indicates
a peak at -0.5$^{+7.0}_{-6.5}$ minutes time delay which is consistent with zero 
}
\end{figure}

\begin{figure}
\centering
\includegraphics[scale=0.4, angle=0]{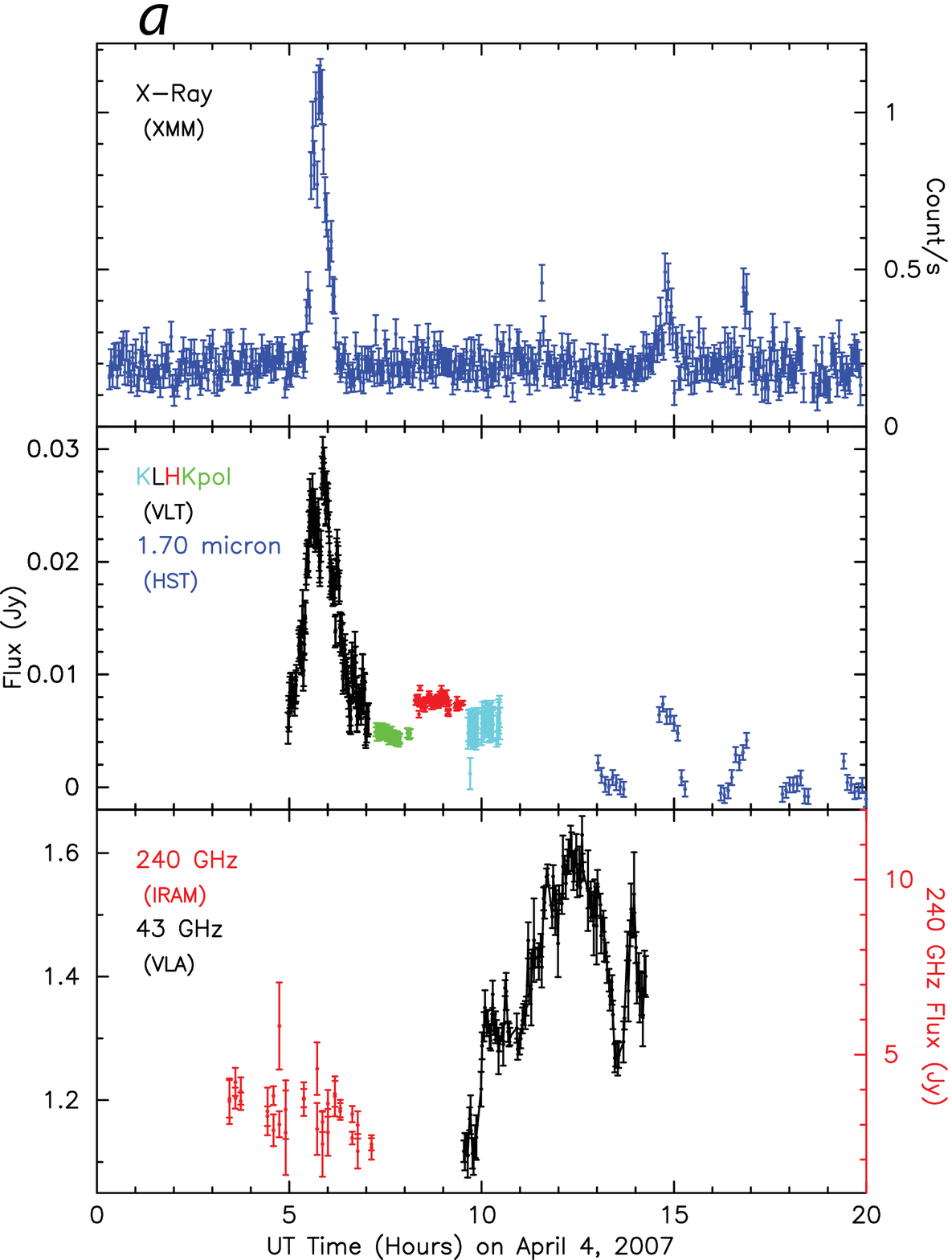}
\includegraphics[scale=0.4, angle=0]{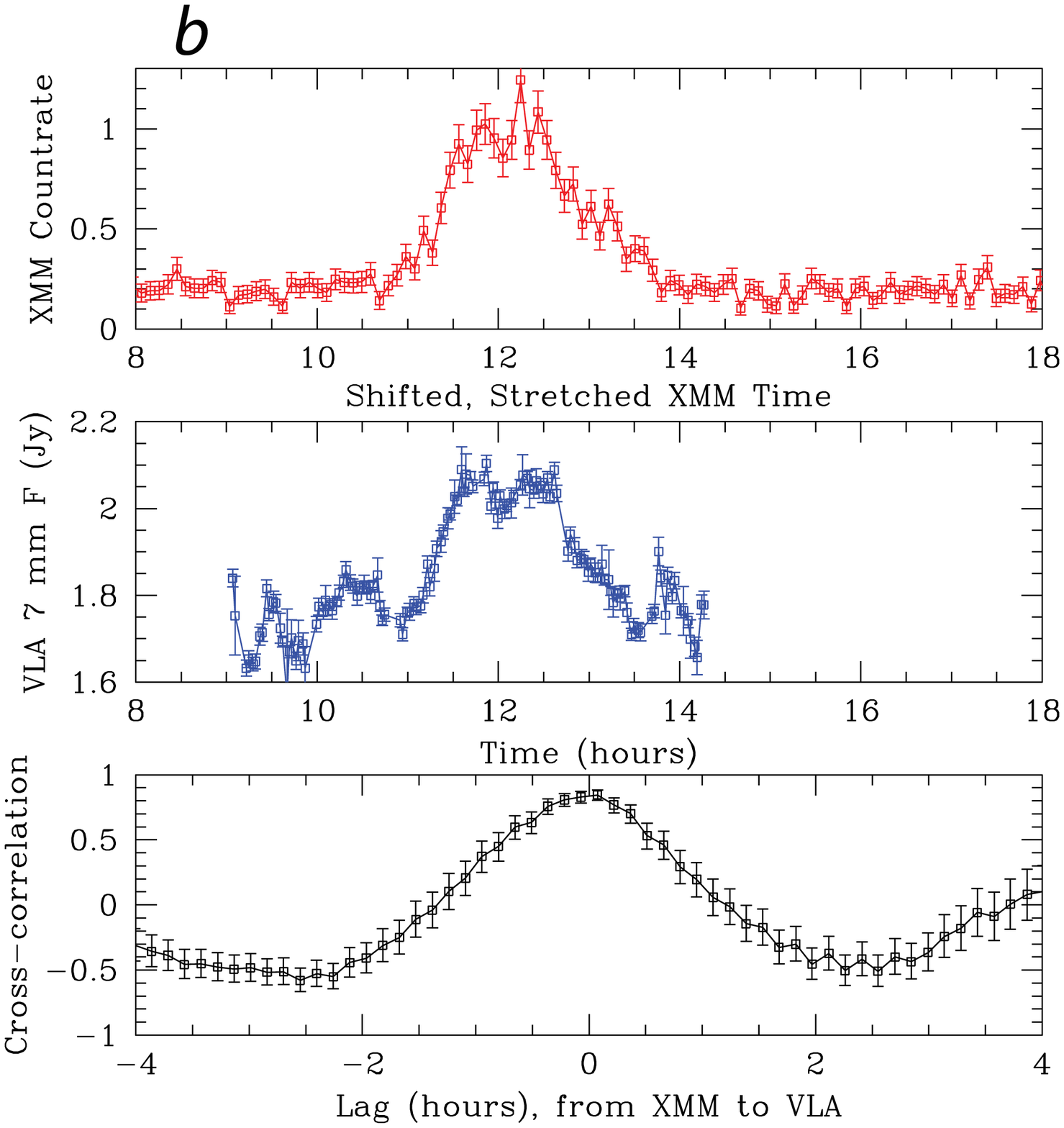}
\caption{
(\textit{a - Left}) 
The light curves of 
Sgr A* on 2007, April 4 obtained with  XMM in X-rays (top), VLT and HST in NIR 
(middle) 
and IRAM-30m  and  VLA at 240 GHz and 43 GHz, respectively (bottom). 
The NIR light curves in the middle panel are represented as H (1.66 $\mu$m) in red,  
K$_s$ and K$_s$-polarization mode 
(2.12$\mu$m) in green and light blue, respectively,  L' (3.8$\mu$m) in black
(Dodds-Eden et al. 2009), and 
NICMOS of HST in blue at 1.70$\mu$m. In the bottom panel, red and black colors 
represent the 240 and 43 GHz  light curves, respectively. 
(\textit{b - Right}) The top  panel shows the 
light curve of Sgr A* obtained with the XMM, the middle 
panel shows the 
light curve taken with the VLA at 43 GHz. The X-ray light curve 
is  shifted by 5.25 hours and  stretched by a factor of 3.5. 
The 43 GHz  light curve is baseline 
subtracted in order to remove the slope due to the quiescent component of 
Sgr A*.  The bottom panel shows the cross correlation plot of the shifted 
and stretched X-ray light curve with the radio data showing
the  maximum  likelihood delay of  4.6 (-7.6, +9.4) minutes. 
}\end{figure}

\begin{figure}
\centering
\includegraphics[scale=0.4, angle=0]{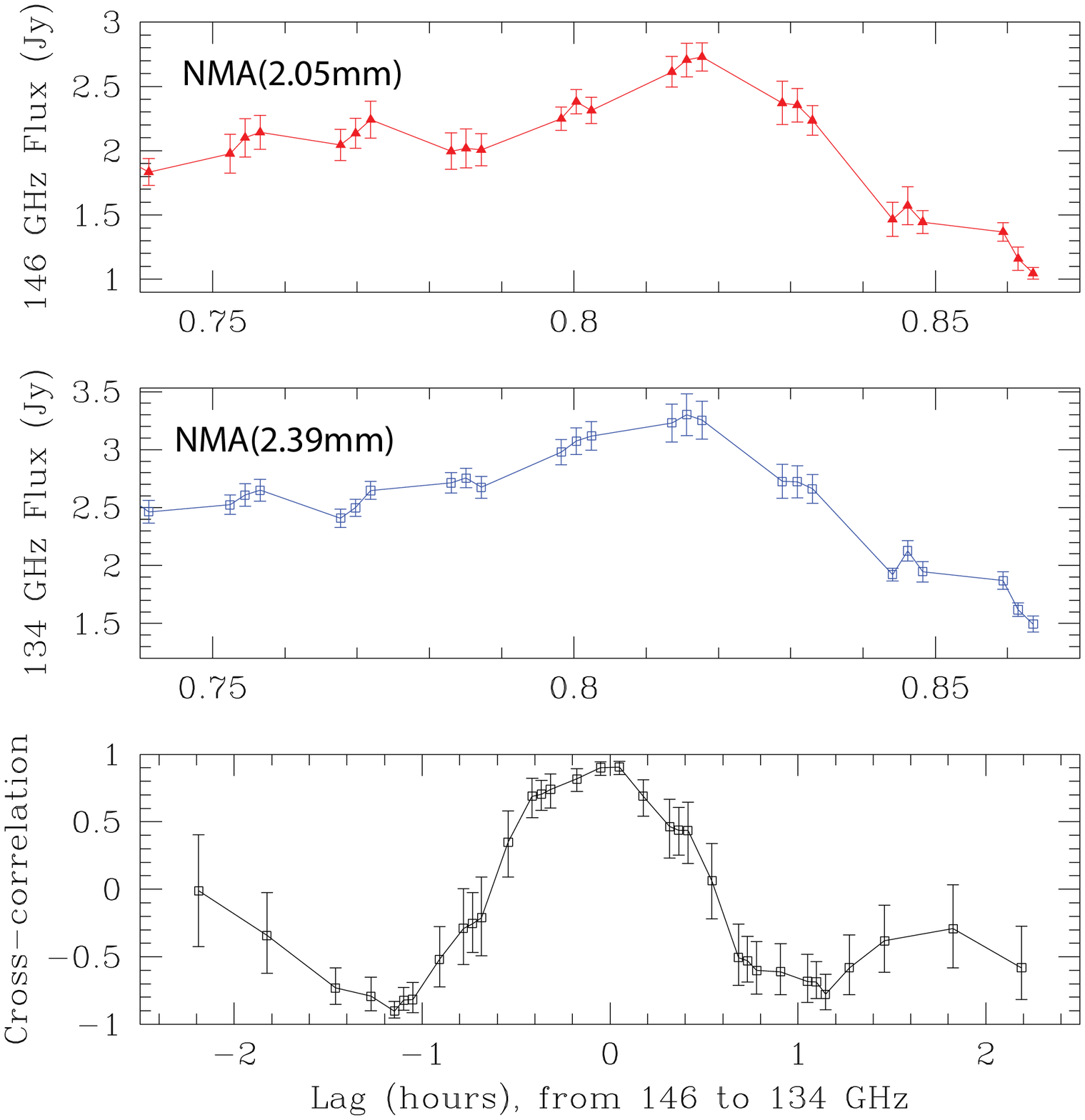}
\caption{
The light curves shown in the top and center  panels are measured 
simultaneously with the NMA at 
146 GHz and 134 GHz on 2007, April 4, respectively. The cross correlation plot
shows  a peak with  3$^{+3.4}_{-8.0}$ minutes  time delay.  The time delays in both
 plots are  consistent with zero. 
}
\end{figure}

\begin{figure}
\centering
\includegraphics[scale=0.4, angle=0]{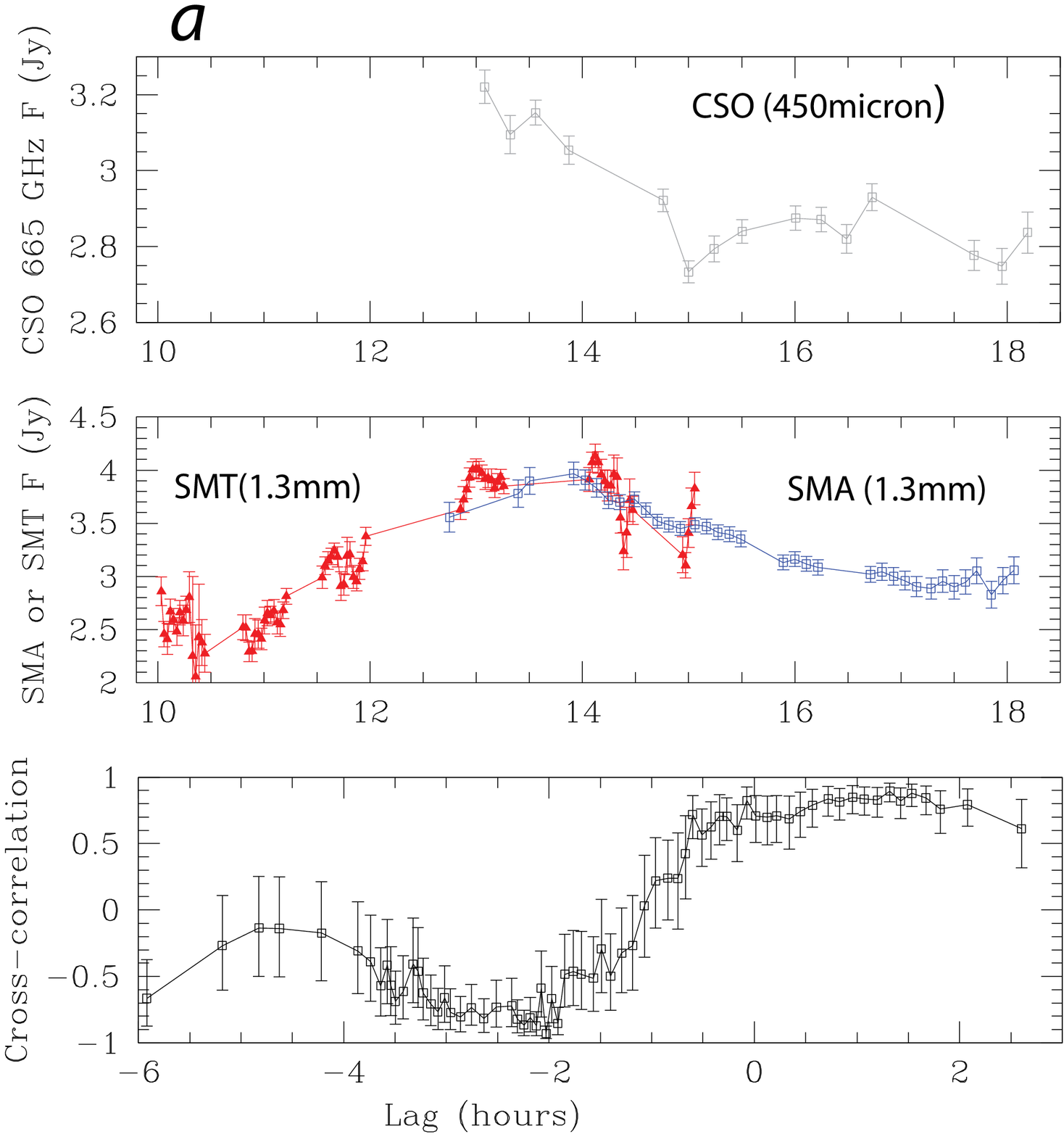}
\includegraphics[scale=0.4, angle=0]{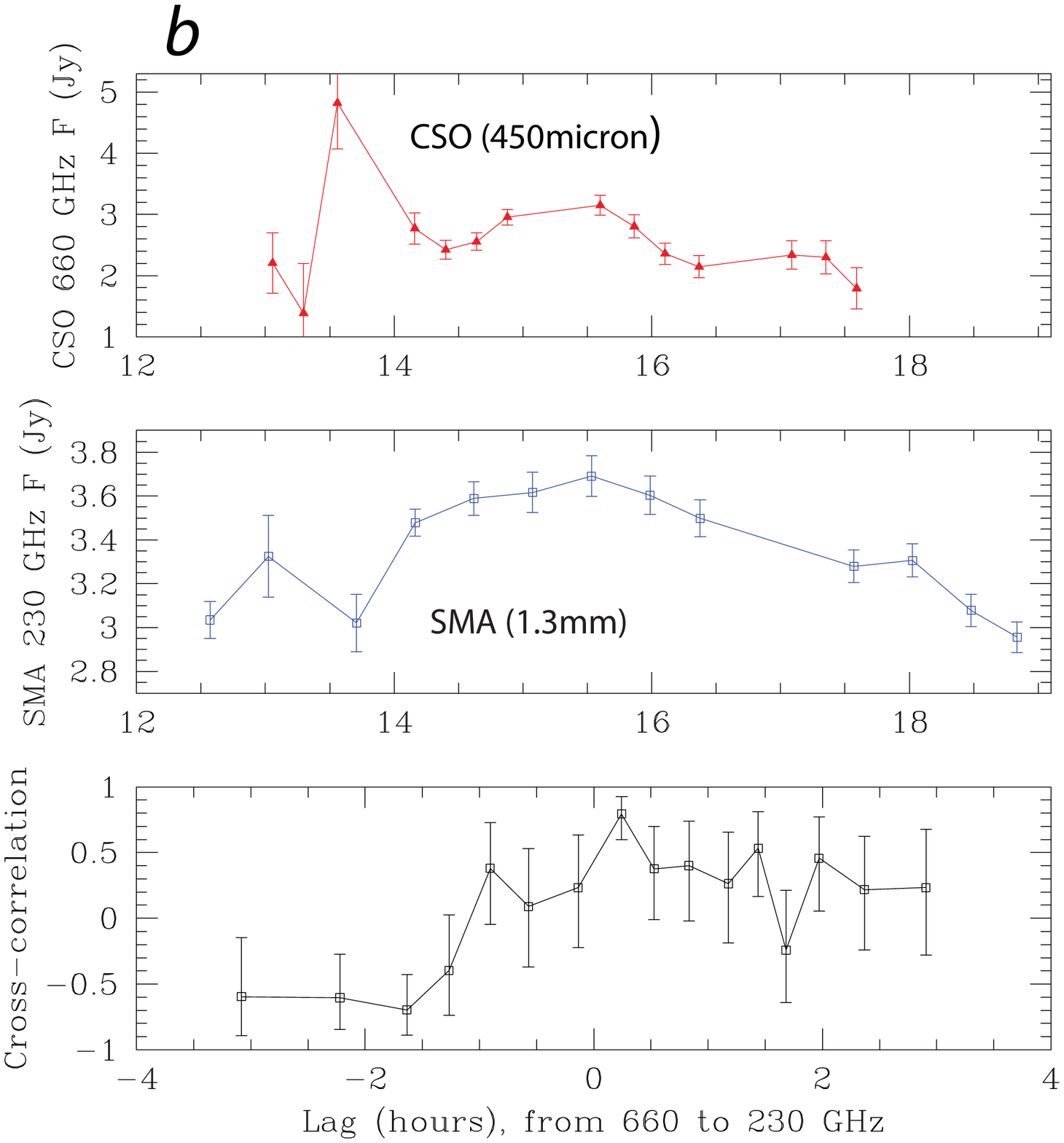}\\
\caption{ 
(\textit{a -  Left}) 
The  light curve of Sgr A* at 450$\mu$m (665 GHz)
is shown in the top panel whereas the
230 GHz (1.3mm)  light curve in the middle panel is  based on 
combining SMT
and SMA observations on 2007, April 3 with UT ranges 10:07h - 14:56h and 12:44h - 18:3.8h.
A constant offset due to the contamination
of a steady background emission is subtracted from the SMT data.
 The  bottom panel represents  the cross
correlation plot showing a peak with 1.32$^{+0.33}_{-0.63}$ hours  time delay.
(\textit{b -  Right}) 
Similar to (a) except that the light curves are taken 
on 2007, April 1 at 450$\mu$m and 230 GHz using the CSO and SMA, respectively. The peak of the cross 
correlation shows a 
 0.24$^{+1.12}_{-0.29}$ hours time delay. 
One $\sigma$ error bars are given for all the time lags. 
}\end{figure}

\begin{figure}
\centering
\ContinuedFloat
\includegraphics[scale=0.4, angle=0]{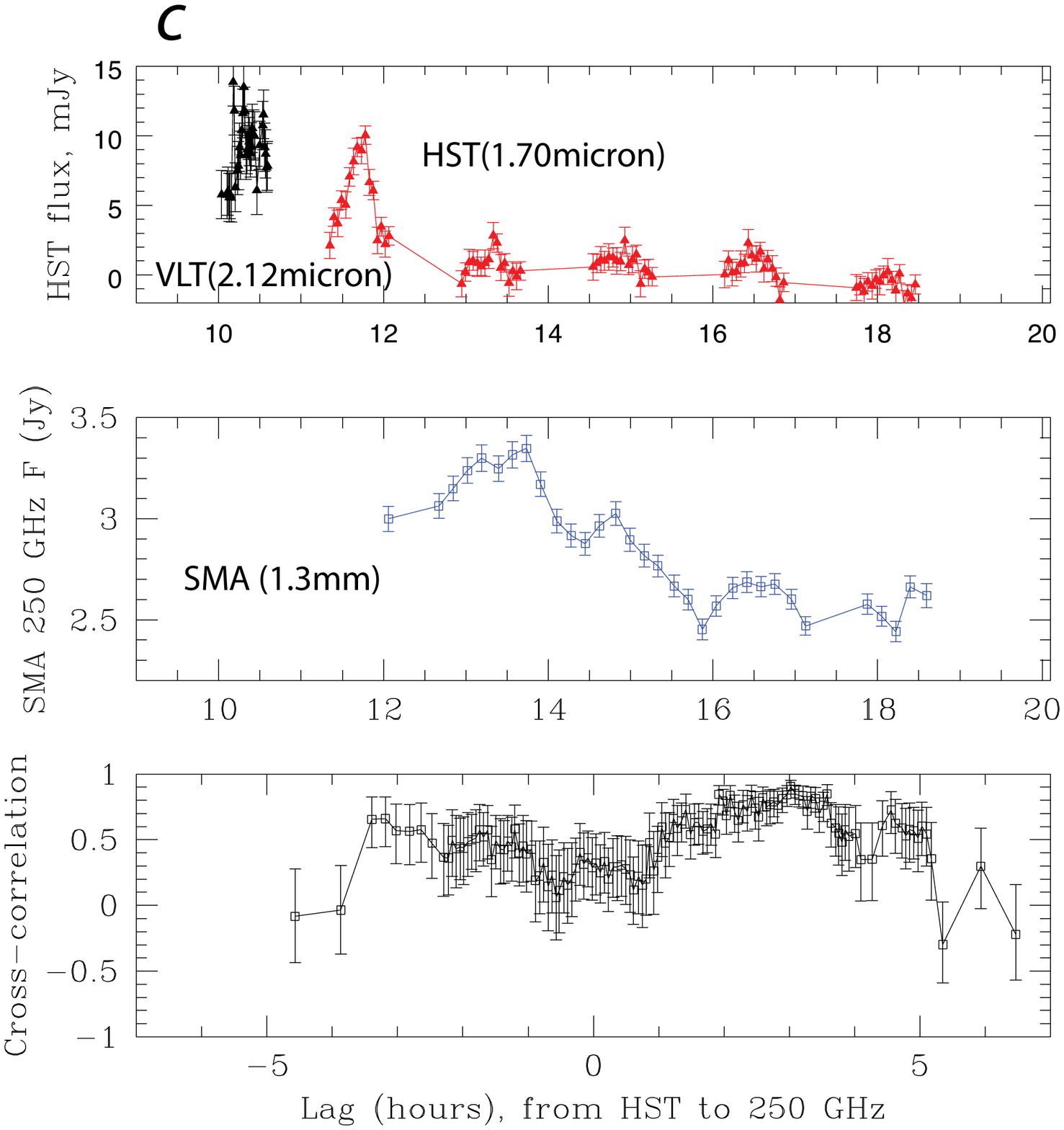}
\includegraphics[scale=0.4, angle=0]{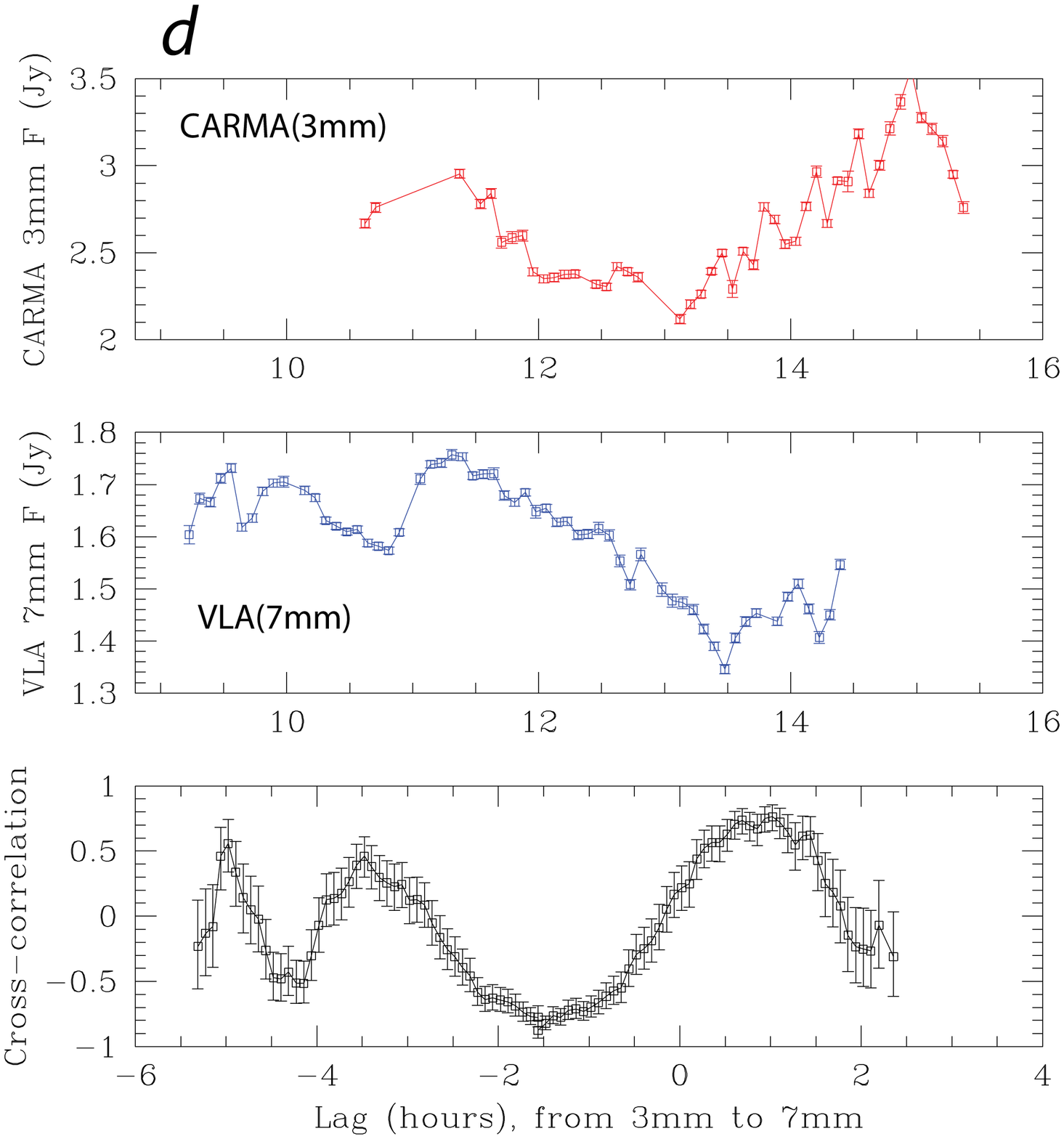}
\caption{ 
 (\textit{c - Left}) Similar to (a) 
except that NIR light curve is based on 
combining the data taken with the VLT and NICMOS whereas the 230 GHz data taken with 
the SMA on 2007, April 5.  The cross correlation is constrained to 
have a peak of 2.64$^{+0.5}_{-0.67}$ hours time delay. 
(\textit{d - Right})
The light curves shown in the top and center  panels 
are based on CARMA and VLA observations at 
94 GHz and 43 GHz  on 2007, April 
2, respectively. The cross correlation plot
shows  a peak with  1.02$^{+0.16}_{-0.31}$ hours of  time 
delay. One $\sigma$ error bars are given for all the time lags. 
}\end{figure}

\begin{figure}
\centering
\includegraphics[scale=0.5, angle=0]{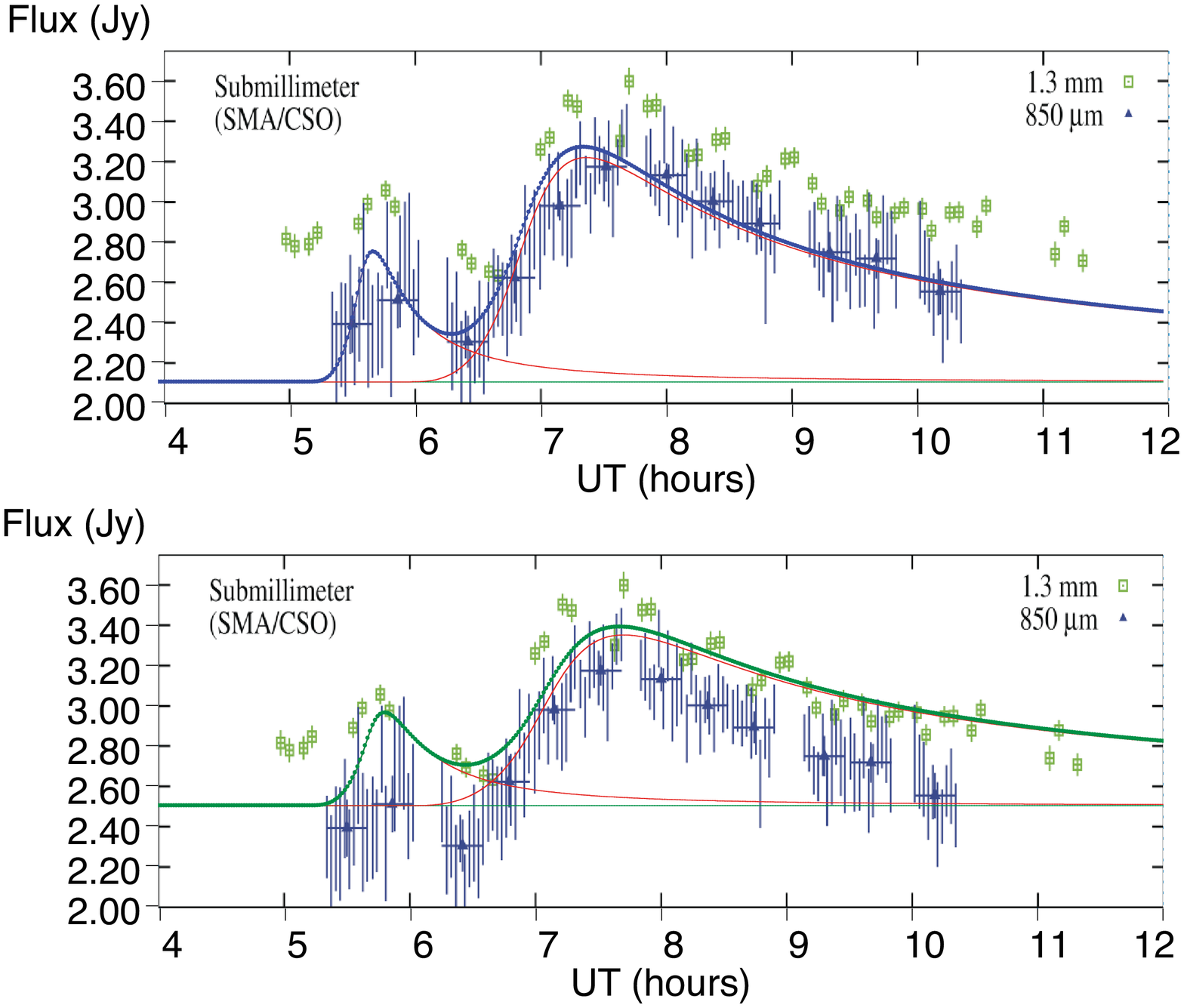}
\caption{
The SMA and CSO  data are taken on 2006 July 17  at 
1.3mm and 850$\mu$m, respectively (Marrone et al. 2008).  After background 
subtraction, 
the two weak and strong flares are fitted simultaneously, 
supporting   the  plasmon  model. 
The expanding blob model automatically generates the fit to the 
850$\mu$m data (top panel)  in blue and to the 
1.3mm data (bottom panel) in green.
}\end{figure}

\begin{figure}
\centering
\includegraphics[scale=0.5, angle=0]{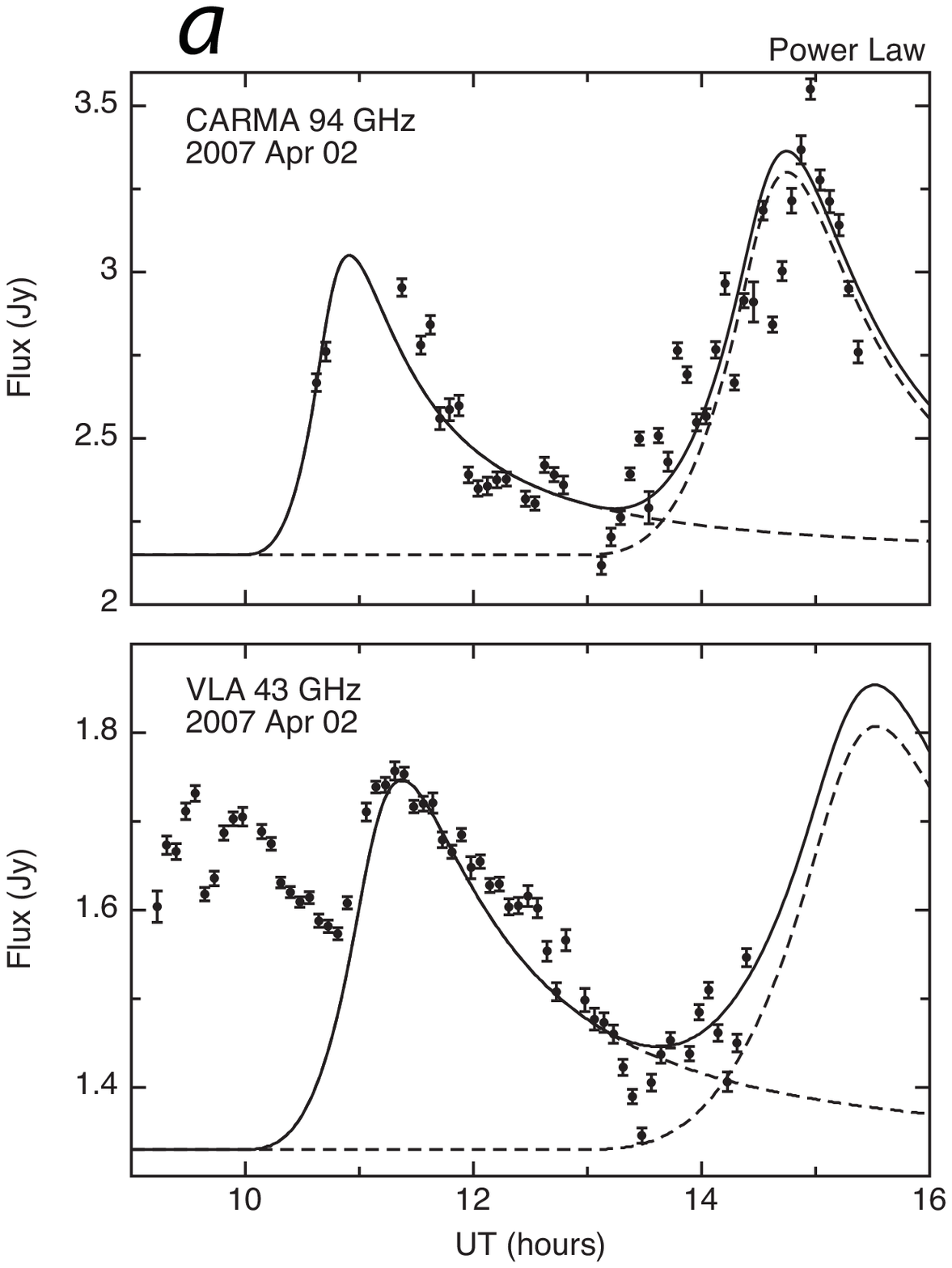}
\includegraphics[scale=0.5, angle=0]{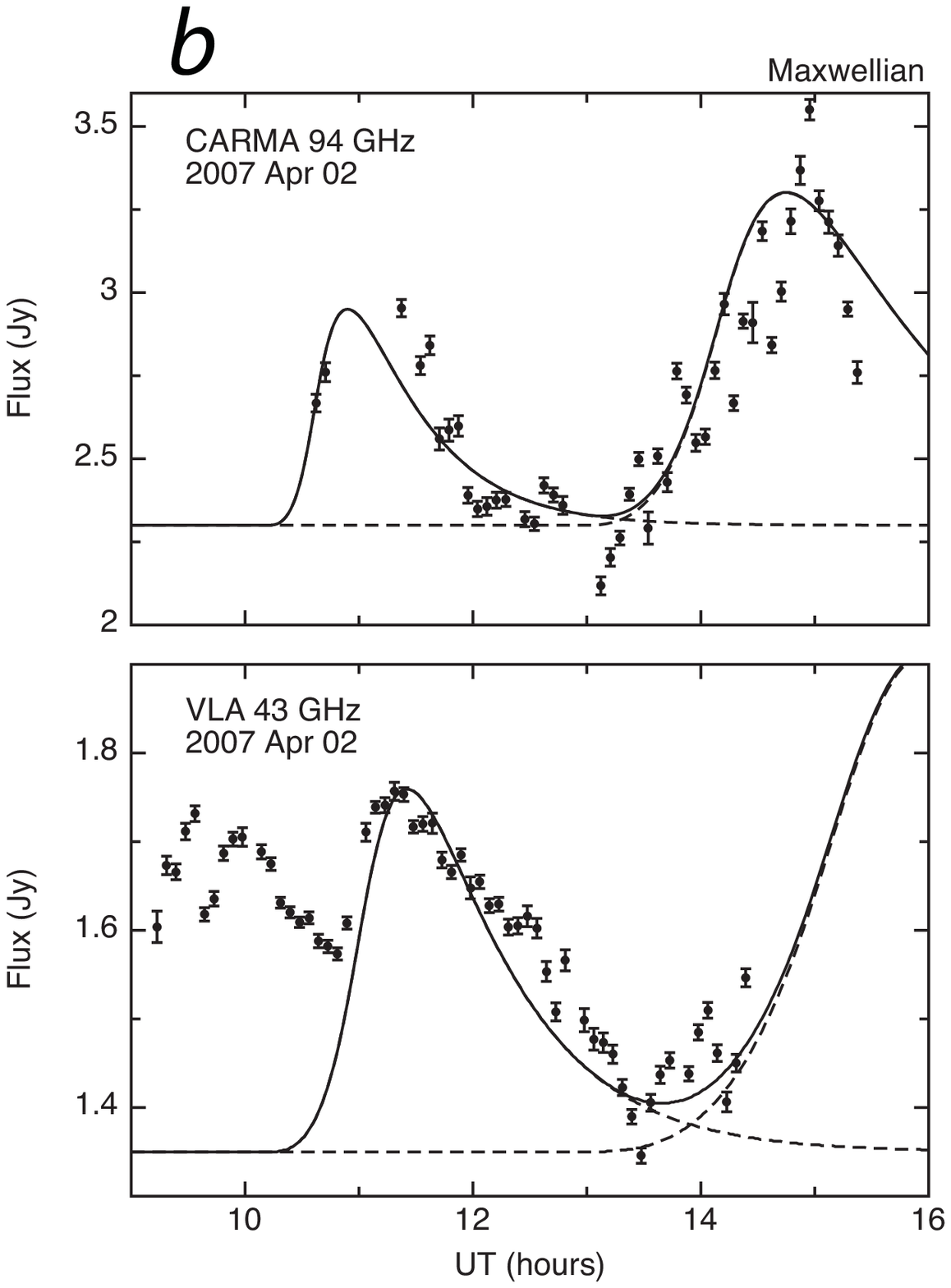}
\caption{
(\textit{a - Left})
A  fit to the CARMA  data at 94 GHz using two flares are shown in the top panel. 
The expanding blob model automatically generates the fit to the VLA data 
at 43 GHz, as shown in the bottom panel. The model has used a power law distribution of
electrons. 
(\textit{b - Right}) Similar to (a) except that a Maxwellian distribution of
particles is used to fit simultaneously  the CAMA and VLA light curves obtained on 2007 April 2.
}\end{figure}


\end{document}